\documentclass[a4paper,12pt]{JHEP3}
\pdfoutput=1 

\newcommand{\refeq}[1]{(\ref{#1})}

\usepackage{amsmath,epsfig}
\usepackage{amssymb,amsfonts}
\usepackage{latexsym}
\usepackage[latin1]{inputenc}
\usepackage{xcolor}

\usepackage{graphicx}
\usepackage{longtable}

\relax
\renewcommand{\theequation}{\arabic{section}.\arabic{equation}}

\def\be{\begin{equation}}
\def\ee{\end{equation}}

\newcommand{\ha}{{1 \over 2}}

\newcommand{\<}{\langle}
\renewcommand{\>}{\rangle}
\newcommand{\de}{\partial}
\newcommand{\bear}{\begin{eqnarray}}
\newcommand{\bea}{\begin{eqnarray}}
\newcommand{\eear}{\end{eqnarray}}
\newcommand{\eea}{\end{eqnarray}}
\newcommand{\fb}{\bar{\varphi}}

\def\hri#1#2{\href{http://arxiv.org/abs/#1}{[ArXiv:#1]#2}}
\def\hre#1#2{\href{http://arxiv.org/abs/#1/#2}{[ArXiv:#1/#2]}}

\newbox\pippobox

\def\II{\relax{\rm I\kern-.18em I}}

\def\m{\mu}
\def\n{\nu}

\def\g{\gamma}

\def\sp{\;\;\;,\;\;\;}

\def\f{\varphi}

\title{Holographic self-tuning of the cosmological constant}

\author{Christos Charmousis$^1$, Elias Kiritsis$^{2,3}$, Francesco Nitti$^2$\\
 ~\\
 $^1$
Laboratoire de Physique Th\'eorique, CNRS, Univ. Paris-Sud, \\ Universit\'e Paris-Saclay, 91405 Orsay, France

~\\
 $^2$
\href{http://www.apc.univ-paris7.fr}
{APC, Universit\'e Paris 7}, CNRS/IN2P3, CEA/IRFU, Obs. de Paris, Sorbonne Paris Cit\'e, B\^atiment Condorcet, F-75205, Paris Cedex 13, France (UMR du CNRS 7164).\\
 ~\\
 $^3$
 \href{http://hep.physics.uoc.gr/}
 {Crete Center for Theoretical Physics}, Institute for Theoretical and Computational Physics, Department of Physics, University of Crete
 71003 Heraklion, Greece
}

\abstract{
We propose a brane-world setup based on gauge/gravity duality in which the four-dimensional cosmological constant is set to zero by a dynamical self-adjustment mechanism. The bulk contains Einstein gravity and a scalar field. We study holographic RG flow solutions, with  the standard model brane separating an infinite volume UV region and an IR region of finite volume.
For generic values of the brane vacuum energy,  regular solutions exist such that the four-dimensional brane is flat. Its position in the bulk is determined dynamically by the junction conditions. Analysis of linear fluctuations shows that a regime of 4-dimensional gravity is possible at large distances, due to the presence of an induced gravity term. The graviton acquires an effective mass, and a five-dimensional regime may exist at large and/or small scales. We show that, for a broad choice of potentials,  flat-brane solutions are manifestly stable and free of ghosts. We compute the scalar contribution to the force between brane-localized sources and show that, in certain models, the vDVZ discontinuity is absent and the effective interaction at short distances is mediated by two transverse graviton helicities.}


\dedicated{\flushleft
Dedicated to the memory of Pierre Binetruy, a fine scientist and a leader.}

\preprint{CCTP-2017-3\\
ITCP-IPP 2017/14}

\begin{document}

\maketitle 

\section{Introduction and summary}

Effective quantum field theories for low-energy interactions are a general framework addressing observable physics from particle physics to cosmology.
While typically successful, they have so far failed to address the cosmological constant problem,  \cite{Weinberg:1988cp} (see also \cite{Padilla:2015aaa}, \cite{Pad} and \cite{Burgess}  for an updated review and references within). Indeed our main dynamical theory underlying cosmology, General Relativity (GR), and those of particle physics, namely quantum field theories in flat space-time, seem to be  incompatible when it comes to vacuum energy.

Experiments (such as the Lamb shift \cite{lamb} or the Casimir effect \cite{Bordag:2001qi}) indicate that any particle will give zero-point energy contributions to the vacuum energy, \cite{Polch}. These contributions scale with the fourth power of the cut-off, which can be  as high as the Planck scale, the generically assumed UV cut-off of any QFT. { On the other hand, vacuum energy couples to gravity as an effective cosmological constant, which by Einstein's equations gives rise to a non-zero space-time curvature.} If we assume the existence of supersymmetry broken at some scale $\Lambda_{SUSY}$, then  the cosmological constant is expected to be of order ${\cal O}(\Lambda_{SUSY}^4)$. Experiment states that such a scale must be quite larger than a TeV and therefore supersymmetry cannot solve the cosmological constant conundrum.

For illustration purposes, we may simply consider the contributions to zero point energy due to the electron: this provides a contribution to the  vacuum energy of order ${\cal O}(m_e^4)$. According to the principle of equivalence for GR any form of energy gravitates. Due to covariance, the vacuum energy gravitates as a cosmological constant.

Gravitationally, a positive cosmological constant will seed a  de Sitter space-time with a finite distance (curvature)  scale, the de Sitter horizon scale{\footnote{A negative cosmological constant instead would give finite life-time for the universe.}} (in the static frame). This scale is inversely proportional to the square root of the cosmological constant. Putting in the numbers for the vacuum energy due to the electron,  would tell us that the size of our Universe is comparable to the earth-moon distance, as Pauli was amused to note back in 1920 (see references within \cite{Padilla:2015aaa}). Needless to say that the Universe will become a lot smaller if we allow for heavier particles and higher UV scales or phase transitions in the Universe (which will also provide a cosmological constant due to the energies of the broken symmetry phase).

The experimental prediction, measured via gravity and cosmology, is the observed size of the accelerating universe, which gives a different answer. Given the present size of our observed universe, the observed vacuum energy  is of order ${\cal O}((10^{-3} eV)^4)$.

We are allowed to change/renormalize the value of the cosmological constant by a bare gravitational cosmological constant (cc) which can be added to the GR action. For this to work, our bare cc must be such that it exactly switches off QFT contributions to a renormalized value, the observed value of the cc. This involves an enormous fine-tuning which is the (first) cosmological constant problem in its ``classical'' formulation. This fine-tuning has to be done throughout the later history of the universe, for each time the vacuum energy appears, a bare value should be there to switch it off almost exactly. To this embarrassing fine-tuning  between theory and experiment one has to add the second problem of radiative instability of the vacuum: the cosmological constant will receive higher loop corrections to each order spoiling the fine tuning undertaken for the first problem. In many respects, this is a harder problem-one which has to be solved not only in the gravitational but also in the QFT sector (for recent progress see the sequestering proposals by \cite{Kaloper:2014dqa}).

The cosmological constant problem may be also pointing to a shortcoming  of GR and there has been some effort to approach the  problem from the viewpoint of modified gravity theories \cite{mg} in four space-time dimensions.
{One idea which has been proposed} is to introduce in the gravity sector some new degree of freedom, usually a scalar field, which can absorb vacuum energy contributions throughout the later evolution of the universe leaving space-time curvature unchanged.

Any mechanism by which the cosmological constant is adjusted dynamically by some extra degree of freedom is what is generally referred to as {\em self-tuning} (or {\em self-adjustment}) of the cosmological constant. More generically, we will refer to a model as  {\em self-tuning} if  flat four-dimensional space time is a solution to the gravitational field equations for generic values of the vacuum energy\footnote{For some ideas associated to a quasi-spontaneous breaking of conformal invariance see \cite{rat,sundrum}}.

Most recently,  the idea of self-tuning has been formulated in a subset of four-dimensional scalar-tensor theories \cite{Charmousis:2011ea}. In this setup, named {\em Fab Four}, the scalar field can eat up any cosmological constant  without fixing any of the parameters of the theory whilst space-time curvature remains flat. For a cosmological setup for example, the scalar field is time dependent for a locally flat Milne space-time.  The presence of integration constants allows zero curvature solutions whatever the value of the cosmological constant and without any fixing of coupling constants of the theory. In other words, the cosmological constant is fixed by the scalar field solution and not the theory, thereby realizing the self-tuning idea or self-adjustment mechanism.

In this work we propose a framework which implements the self-tuning mechanism using the brane-world idea, i.e. higher dimensions \cite{mg,reviews, k-review, maartens} and also, crucially, holography.

In the brane-world scenario \cite{Rubakov:1983bz,rs}, our four-dimensional universe ({\em brane}) is embedded in a higher dimensional {\em bulk}. Ordinary matter and gauge fields are constraint to propagate only on the brane, but gravity propagates in all dimensions and the brane interacts gravitationally with the higher-dimensional degrees of freedom.
The extra  dimensions may remain undetectable from present day experiments, if for example  their size  is sufficiently small or the bulk  is  sufficiently curved.

In the original brane-world model of Randall and Sundrum (RS) \cite{rs} with a single brane, the latter was embedded in a five-dimensional  anti de Sitter space-time. In order to have a flat brane solution to the field equations, a fine tuning was necessary  between the brane tension, interpreted as world-volume vacuum energy, and the (negative)  bulk cosmological constant. This was the brane-world version of the cosmological constant problem-flat solutions are not generic in the presence of vacuum energy. They are on the contrary very finely tuned. Brane-worlds were generalized in various directions and such generalizations are reviewed in \cite{mg,reviews,k-review,maartens,brax}.

It was a natural step to try and implement a brane-world version of the  self-tuning mechanism: the idea was that the brane vacuum energy due to matter  may curve the bulk, but leave the four-dimensional brane (our universe) flat. It was initially noted \cite{st1} (see also \cite{DeWolfe:1999cp,csaki}) that a non-trivial bulk scalar field could indeed relax the fine-tuning for the cosmological constant on the brane in a 5 dimensional brane-world setup. This idea was also implemented in 6 dimensional space-times, or  co-dimension 2 \cite{cod2}, for more generic gravity theories \cite{GB} or both \cite{Charmousis:2008bt}. Indeed the dynamical nature of the scalar introduced integration constant(s) that did allow for a flat brane solution without fine tuning of the brane tension. The 5 dimensional self-tuning solutions though had an important shortcoming: they had a naked singularity in the bulk space-time at a finite distance from the brane \cite{st1}, \cite{DeWolfe:1999cp}, \cite{csaki}. When this did not happen, the gravitational interaction on the brane was not four-dimensional, \cite{csaki}. Various other related setups were analyzed, leading eventually to instabilities or hidden fine tunings \cite{cod2}, \cite{GB}.

It  was also realized  that various brane-words in $AdS$ space-time have a holographic interpretation, \cite{APR,Verlinde,RZ,bb2}. This opened a new perspective on the relevant physics as it is mapped into QFT dynamics. The holographic correspondence provides a nontrivial map between gravity/string theory dynamics in the bulk and QFT dynamics at the boundary. Moreover it can be considered as a UV-complete definition of quantum gravity, \cite{smgrav}.
The study of holography for 20 years has revealed many novel features of QFT especially in the strongly-coupled regime,  as well as novel features of gravity and its connection to QFT thermodynamics  and hydrodynamics.
The rules of the game have been understood in many more contexts than the original N=4 sYM theory example and numerous successful checks have been done.

When it comes especially to cosmology, holography suggests several intriguing dual views encapsulated in the several versions of the de Sitter/(p)CFT correspondence, \cite{Strominger,BDM,MS,MS2,Anninos} which has been also extended to general cosmological flows, \cite{AFIM}. These look different from the
brane-world cosmology that is driven by moving branes{\footnote{The (equivalent) time dependent brane world perspective was undertaken in \cite{Binetruy:1999hy} and the connection to \cite{Kraus} was explained via Birkhoff' s theorem in the presence of branes in \cite{Bowcock:2000cq} }} and rolling vevs, \cite{Kraus,mirage} , but they may have a  deeper  connection.

In the rest of the introduction we  describe the structure and  holographic motivation for the brane-world self-tuning setup we study, and we present a summary of our results.

\subsection{Emergent gravity and the brane-world\label{holo}}

An important realization of the self-tuning setup is suggested by the holographic ideas on emergent gravity. This is a setup where the interactions of the Standard Model (SM),  including gravity, are generated by 4d conventional QFTs. For this to work, we need in the simplest setup three ingredients, \cite{smgrav}

\begin{itemize}

\item The gauge theories and other interactions of the SM.

\item A large-N, strongly coupled and stable 4d QFT$_N$ that will generate the gravitational sector (this may be a non-abelian gauge theory where $N$ is the number of colours).

\item A theory of bifundamental ``messengers" that will couple the QFT$_N$ to the SM by renormalizable interactions.
 Therefore the messengers must be charged under both gauge group of the SM  as well as the QFT$_N$ gauge group. They must have large masses, of order $\Lambda$ (the UV cutoff scale).

\end{itemize}

At energy scales $E\ll \Lambda$ the messengers can be integrated out and the SM is directly coupled to the operators of QFT$_N$. These operators involve the universal conserved stress tensor of QFT$_N$ as well as many other operators.
An appropriate linear combination of the stress tensors becomes the universal metric that will couple to the SM fields, and diffeomorphism invariance will be an emergent feature. This is were gauge/gravity duality comes into play.

{ Rather than using  the four-dimensional description above, we will now assume  the existence of a  holographically dual version of} the strongly coupled QFT$_N$ in terms of classical gravity and other interactions in a 5-dimensional bulk space with a (UV) near-AdS boundary. {  In this language, four-dimensional diffeomorphism invariance is manifest and is  a consequence of the overall energy conservation.}

The SM is weakly coupled at $E=\Lambda$ and therefore its coupling to QFT$_N$  follows the semi-holographic setup: it can be represented by a 4-brane embedded in the bulk geometry at the position corresponding to the cutoff scale induced by the messenger mass.

Therefore, in the gravitational description the setup is that of a SM-brane embedded in the QFT$_N$ bulk gravitational theory. The bulk fields of the gravitational sector couple to the SM fields on the brane.
An important ingredient of this coupling is the induced action for the bulk fields on the brane. This is generated by the SM quantum effects that will induce a non-trivial action for the bulk fields. Since the SM fields are localized on the brane, the same applies  to this induced gravitational action.

In general, bulk operators that are not protected by symmetries will obtain brane potentials that will scale as the the fourth power of the cutoff scale $\Lambda$. For the bulk operators that are protected by symmetries, like the graviton, possible conserved currents (giving rise to graviphotons) and the universal instanton density (giving rise to the universal axion) the corrections start at two derivatives, and scale as $\Lambda^2$, \cite{smgrav}.

The  framework of emergent gravity from 4d QFTs described above therefore can be modeled in the gravitational picture with a 4-d SM brane embedded in the bulk space-time generated by the QFT$_N$. In this paper we will simplify this effective description by keeping track of two basic bulk fields: the $metric$ as well as a single scalar. With this field content the action we will consider, up to two derivatives reads,
\be \label{A1-intro}
S = S_{bulk} + S_{brane },
\ee
where
\be\label{A2-intro}
S_{bulk} = M^{d-1} \int d^{d+1}x \sqrt{-g} \left[R - {1\over 2}g^{ab}\de_a\f\de_b \f - V(\f)\right],
\ee
\be\label{A3-intro}
S_{brane} = M^{d-1} \int d^d x \sqrt{-\g}\left[- W_B(\f) - {1\over 2} Z_B(\f) \gamma^{\mu\nu}\de_\mu\f\de_\nu \f + U_B(\f) R^{(\gamma)} \right]
\ee
where $g_{ab}$ is the bulk metric, $R$ is its associated Ricci scalar and $\gamma_{\mu\nu}$, $R^{(\gamma)}$ are respectively the induced metric and intrinsic curvature of the brane. We have kept the dimension $d$ above general although our main concern is for $d=4$. We expect  $M^{d-1}\sim N^2$.

\begin{figure}[h!]
\begin{center}
\includegraphics[width=12cm]{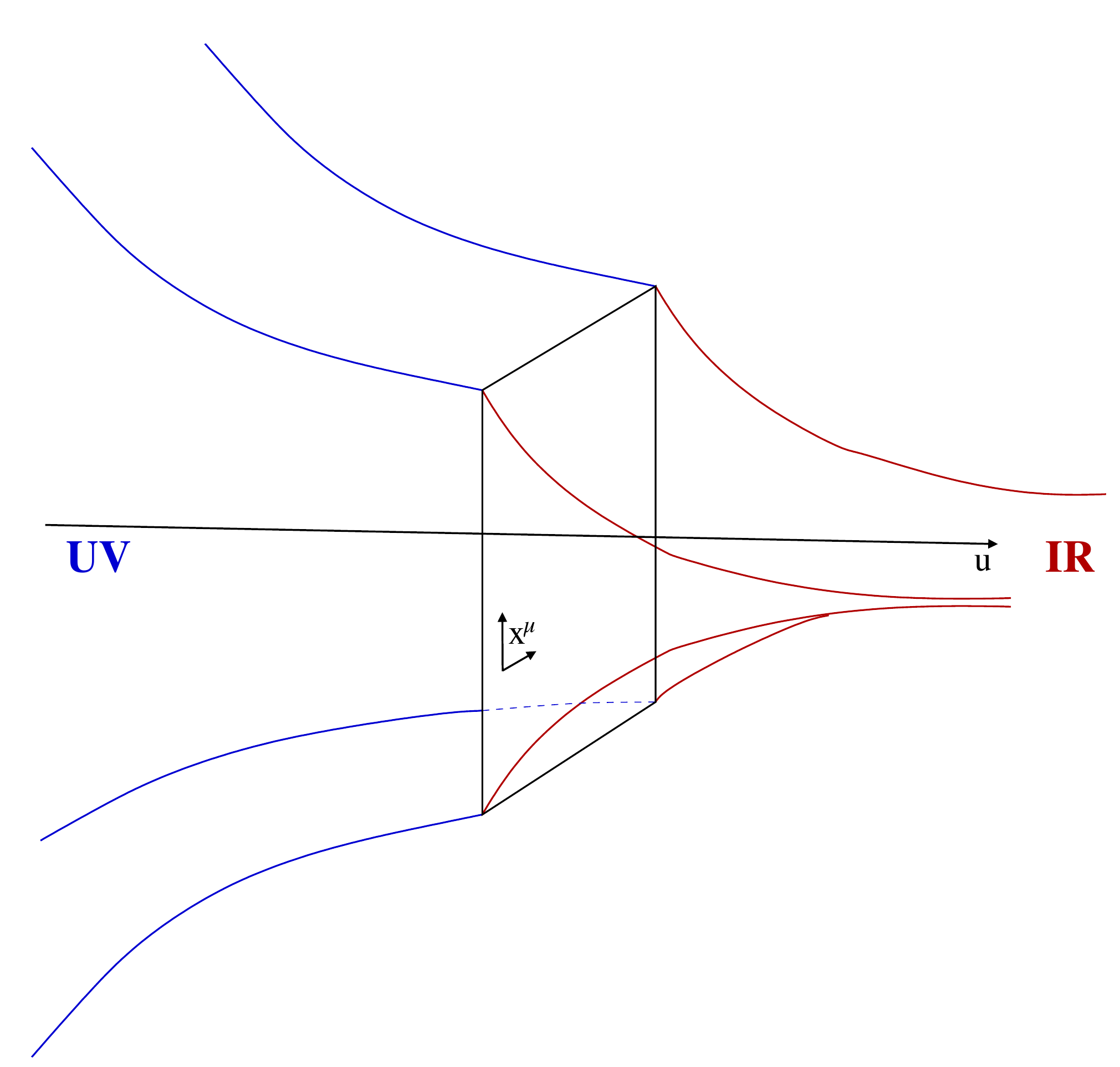}
\caption{Sketch of the Holographic RG-flow solutions, with the brane acting as the interface between an infinite volume UV region and a finite volume IR region. The coordinates $x^\mu$ are world-volume four-dimensional coordinates on the brane-world,  $u$ represents the holographic radial  direction.} \label{fig-intro}
\end{center}
\end{figure}

The  above action is the most general two-derivative action in Einstein-scalar theory which preserves the full group of bulk diffeomorphisms (including those transverse to the brane, since the latter is allowed to fluctuate).
 All we  assume initially for the bulk potential is that it has a maximum supporting a (stable) $AdS$ solution.   We will be interested in (fully backreacted) solutions in which the scalar field evolves in the bulk radial direction (transverse to the brane), interpolating between  an infinite volume asymptotic $AdS$ boundary region where $\f$ approaches the maximum of $V(\f)$, and a region  with asymptotically  vanishing volume element, with the brane separating the two. In the dual field theory language, the scalar corresponds to a relevant operator of the QFT$_N$, and the solution  to a renormalization group (RG) flow driven by this operator.  The large and small volume regions correspond respectively to the UV and IR of the RG flow. This structure is represented in Figure \ref{fig-intro}.
Although the overall volume of the bulk is infinite, our model allows  regimes in which gravity behaves as four-dimensional, as an effect  of the localized Ricci scalar term on the brane in equation (\ref{A3-intro}). This gives rise to a quasi-localized graviton resonance  as in the DGP model \cite{DGP}.

As we will eventually conclude, in this framework, and with the insights from the  holographic perspective,  it is possible to avoid all the drawbacks of previous brane-world self-tuning constructions.  Holography provides an important guideline in organizing the space of solutions.  Furthermore, the IR endpoint of the RG-flow can be singularity-free if the scalar field approaches another $AdS$ extremum (in this case a minimum of the potential). Moreover, some mild singularities are acceptable because they can be resolved.
 Furthermore, the holographic interpretation naturally requires an infinite volume region in the UV. As we will see, this is crucial for the self-tuning mechanism: any solution which has finite volume on both sides of the brane must necessarily be fine-tuned.

In general, self-tuning  models are severely constrained by Weinberg's no-go theorem \cite{Weinberg:1988cp}, which essentially states that in any local theory with dynamical gravity,  preserving local  Poincar\`e invariance, and whose solutions are determined by a local  action principle,  self-tuning cannot work\footnote{In the case of Fab Four,  the scalar breaks Poincar\'e invariance thus evading Weinberg's no-go theorem \cite{Padilla:2015aaa}}. The framework we present here  avoids this theorem, and this  has a clear interpretation in view of holography: each solution contains quantities (the ``vevs'' of the dual operators) which are not determined by extremizing a local action, but rather by a regularity condition which relates the UV and the IR, and has no classical analog in local field theories.

Given this input from holography it is now instructive to check our action ingredients from the brane world perspective. For a start, our brane will be an asymmetric one, separating an infinite volume UV region and a finite volume IR region. Asymmetric self-tuning models were studied early on by \cite{csaki}, but these did not include an induced gravity term.  Secondly, given that the overall volume of our brane model is infinite, there will not be a localized zero mode graviton fluctuation on the brane (as it is the case instead for the classic RS model \cite{rs}). This is in turn where the induced gravity term  plays an important positive role for the phenomenology of our model providing a quasi-localized graviton zero mode in the tensor fluctuations, by the same mechanism well-known in a flat bulk \cite{DGP}.

 The particular role of asymmetry \cite{Padilla:2004mc}, combined with the induced gravity term were realized in \cite{padilla} where dark energy models were constructed (but this time, without a bulk scalar). In the latter paper it was also realized, however,  that the positive role played by the induced gravity term in the tensor fluctuations, was negative{\footnote{literally opposite in sign!}} in the scalar fluctuations: there, it was found that the induced gravity term contributed to a scalar ghost whenever a spin 2 zero mode was {\em not} present in the spectrum. { This was because without a bulk scalar field,  dynamical scalar fluctuations only existed on the brane, but not in the bulk.} This is where the bulk scalar in our model plays an essential role:  it also contributes in the scalar sector allowing, as we will see, for the absence of scalar (but also tensor) fluctuation pathologies.

\subsection{Results and Outlook}

In this setup we consider solutions to the classical equations of motion
for $g_{\m\n}$ and $\f$ that correspond to Lorentz-invariant saddle points of the dual QFT$_N$, as described by the action $S_{bulk}$. The presence of the SM brane in the geometry is taken into account by the Israel matching conditions.

Our goal  in this paper is to first examine the existence of solutions to the bulk equations which are holographically acceptable (either with regular bulk geometries or with good IR singularities) having a flat induced metric on the brane.
This is the essence of the self-tuning mechanism: although there is a non-trivial vacuum energy (or cosmological term) on the brane, the metric of the brane universe is flat.

We find that holographically acceptable   solutions generically exist.  In these solutions, the brane is placed at a specific  equilibrium position $\f_0$ in the bulk, which is determined dynamically by solving  Israel's junction conditions. { We show that one can generically find  an acceptable equilibrium solution in the vicinity of a zero of $W_B$, for a generic  bulk potential $V(\f)$. Thus, the existence of  self-tuning  solutions is  generic in this framework.}

The next question we investigate is: is this equilibrium position stable? More specifically, are the fluctuations around this solution regular (not ghost-like) and stable (not tachyonic)? Connected to this question is also the following: what kind of interactions such fluctuations mediate on the brane world? Is gravity similar to observable gravity? Is the equivalence principle upheld?

We derive the fluctuation equations around the equilibrium brane position, for general bulk and brane potentials.  There are two sets of propagating modes. One is a spin-two mode associated to the 5d graviton. We find that the equations it satisfies are similar to the DGP scenario, \cite{DGP} with the important difference that in our case the bulk geometry is non-trivial.

We calculate the propagator that controls the interaction of sources on the SM-brane. This propagator is DGP-like at short enough distances but is a massive propagator at long distances\footnote{This behavior was seen before in a DGP framework with a codimension higher than 1 thick brane, \cite{ktt,ktt2}.} The  reason for this is the behavior of the bulk to bulk propagator on the brane. At short enough distances it vanishes, with the same behavior as in flat space. But at long enough distances it asymptotes to a constant that is determined by the bulk geometry. It is this different behavior that is responsible for the mass at long distances.

The framework presented here  has a rich  gravitational phenomenology, displaying several different potential signals of long- and short-distance modified gravity.  The graviton propagation  is four-dimensional at both short and long distances,  and also has a mass. Depending on parameters, a five-dimensional phase may appear at intermediate distances.
The effective four-dimensional gravitational coupling constant  is controlled by the induced Einstein term on the brane, and the mass of the graviton is  controlled by the same quantity and by the geometry around the equilibrium position. We  lay out the conditions for constructing specific models in which the modified gravity regime falls outside the scales probed by current observations. This includes having an arbitrarily light graviton in a technically natural way.

The analysis of the scalar fluctuations is more involved\footnote{For an earlier discussion of scalar fluctuations in brane-worlds with a bulk scalar, see \cite{radion}}. 
There is a single gauge invariant scalar fluctuation in the bulk, but two invariant ones on the brane. We derive the dynamics of the scalar fluctuations and we formulate it as matrix Sturm-Liouville problem. This formulation enables us to derive sufficient conditions for the fluctuations to be manifestly regular (not-ghostlike) and stable (not tachyons).  We also construct the brane-to-brane scalar propagator, which takes the form of a matrix coupling two kinds of sources: the trace of the stress tensor, and the scalar ``charge''.

We do not address here a full discussion of the phenomenology of the scalar sector. This is an important aspect, because it leads to constraints from fifth-force and violation of the equivalence principle.   Moreover, it is important to investigate how the non-linearieties of the theory modify the gravitational effects  beyond one-graviton exchange, as these can lead to stringent constraints on scalar-tensor theories (as was analyzed by \cite{kaloper} in the context of ``Fab Four''-like theories). However neither the linearized  scalar-mediated interaction nor the non-linear effects are universal, and they can manifest themselves at different scales in a model-dependent way. Thus this discussion  must  be carried out in specific models and is beyond the scope of the present paper.

Our results are encouraging but constitute only the tip of the iceberg.
There are several further tasks that must be accomplished before this setup is physically acceptable.

\begin{itemize}

\item A detailed analysis on the dependence of the observable parameters (4d Planck scale, mass of the graviton) from the inputs (nature of bulk QFT, UV couplings and the  induced brane cosmological constant) must be made in order to assess which ingredients provide a physical answer.

\item The massive graviton has, generically, a vDVZ discontinuity, \cite{VDVZ}. Finding the associated Vainshtein scale, \cite{vainshtein}, is important in order to understand the viability of the setup. It is important to note that the theory of the massive graviton is an effective theory near the equilibrium position and for this reason is not subject to the standard constraints on massive graviton theories. Such constraints are stringent if the theory only a contains a massive 4d graviton and no other gravitational degrees of freedom, \cite{dgt}.  On the other hand, consistent theories containing massive gravitons like KK theory and string theory/holography have appropriate couplings to avoid such direct constraints, \cite{MS,kn}.   Similar considerations have also been addressed in  scalar-tensor theories of the   ``Fab Four'' type in \cite{kaloper}. In that work it was shown that,  requiring non-linearities to screen  extra scalar modes around spherically symmetric solutions, together with the validity of effective field theory at the observed scales, puts non-trivial constraints.  In the present context, to answer the same questions one would have to analyze solutions with spherically symmetric brane sources, and investigate how the non-linear scale interplays with the other bulk and brane scales, and it is not easy to ``guess'' whether the constraints will invalidate the framework. This is an important but complex study, and will be left for a future work.

\item It is interesting that this setup always provides for a massive graviton on the brane. It has been observed that the cosmological evolution driven by a massive graviton is similar to an effective cosmological constant $M_P^2\Lambda\sim m_0^2M_P^2$, \cite{k-review} which is the right size to explain the observable cosmological constant. Whether there is a connection between these two observations remains to be seen by analyzing the full cosmology of the theory.

\item Although the conditions for ``healthy'' scalar fluctuations have been derived, more details need to be known about the forces mediated by the scalar excitations. The fact that there are two possible scalar excitations on the brane indicate that there are generically two charges associated to the scalar interaction. The nature of the scalar force, its range and its couplings to observable matter matter must be elucidated, as a function of the inputs: the localized action and the bulk dynamics.

\item The existence of a flat 4d-space-time solution which accommodates a large brane vacuum energy while allowing for reasonable gravitational interactions, does not fully solve the  problem. One should investigate how one arrives at such a solution { dynamically.} For this one first needs to investigates alternative solutions with maximal symmetry but where the induced brane-world metric is positively or negatively curved.
    The final step is to derive the full time-dependent evolution equations for the system brane+bulk.

\item The issue of radiative corrections to the framework we discuss in this paper is important. The bulk gravitational theory has both higher derivative corrections (that are controllable at strong coupling according to AdS/CFT intuition) and loop corrections that are controllable at large $N$. The induced brane action for the bulk fields is expected to be generated by brane-field quantum effects and all such effects are assumed to be included in the brane potential  two-derivative terms. There can be higher derivatives corrections that we have neglected here. They will provide corrections to the matching conditions that are nor IR relevant. In the worst case scenario they can affect the scale $m_4$ that controls the onset of massive brane gravity.

\item The full time-dependent dynamics of the system must be derived and analyzed. This is tantamount to analysing the cosmological evolution of the setup. In particular this is important in order to verify the naive expectation where the brane starts at early times in a bulk position near the boundary and far away from the ``equilibrium" position $\f_0$. The ensuing evolution towards this equilibrium position  can be mostly driven by the brane cosmological constant giving therefore a period of brane inflation.

    Approaching $\f_0$ the effective cosmological constant becomes smaller and smaller and the brane evolution is driven more and more by the energy densities on the brane. These expectations are reasonable and should be verified. An interesting open problem is to assess what can act as dark energy in this setup. { Several possibilities can be investigated already within this framework, due to the presence of  scalar modes (including the brane position) which may act as quintessence or leave a residual cosmological constant if the brane is slightly displaced from its equilibrium position.}

\item It is important to stress that the brane cosmological constant is not a fixed potential of the bulk fields  but also depends in general on brane-field order parameters (examples for the SM are the Higgs field or chiral symmetry condensates). This intertwines interestingly with the self-tuning mechanism and in principle allows both an accommodation of phase transitions into the relaxation mechanism but also the possibility that the solutions to the CC Problem and the Electroweak hierarchy problem are intimately connected.

\item  The fact that gravity is generically 5 dimensional off the brane world indicates that there may be a period in the evolution of the (brane) universe where there is an exchange of energy between the SM-brane and the bulk, \cite{ktt2}. Such an effect can affect the cosmology on the brane\footnote{This phenomenon has been investigated in phenomenological brane setups in \cite{bb1,bb2,bb3}.}.

\end{itemize}

This paper is structured as follows.

Section 2 presents the model, the vacuum solutions,   and the self-tuning mechanism arising from Israel's junction conditions. We give a review of  the geometry of holographic RG flows and what makes for holographically acceptable singularities. We show that self-tuning junctions are generically present for a wide variety of brane and bulk potentials, and we give concrete examples with and without an IR singularity.

In Section 3 we lay the ground for the analysis of linear perturbations around vacuum solutions, and identify the relevant bulk and brane perturbations, as well as  their gauge transformations. After fixing the gauge we derive the linearized bulk field equations and junction conditions for  physical scalar and tensor perturbations.

Section 4 is dedicated to the analysis of tensor modes, and in particular to the calculation of the tensor-mediated interaction mediated  between sources localized on the brane. We compute the tensor brane-to-brane propagator  and discuss its different regimes, and discuss the associated phenomenology at different scales.

In Section 5 we analyze scalar perturbations. We write the gauge-fixed linearized  junction conditions in terms of a single scalar perturbation, and show that the bulk equation plus junction conditions  can be written in terms of a vector-valued  Sturm-Liouville problem with  Robin boundary conditions. We discuss the stability of the background solutions and  give sufficient conditions for the absence of ghosts/tachyons. We compute the scalar brane-to-brane propagator which enter the scalar-mediated interaction between brane-localize sources, and we speculate on a class of models free of the vDVZ problem.

Several technical details are left to the appendix.  In Appendix A we give a classification of the different possible types of junctions; Appendix B relates the boundary values of the fields at the brane with the asymptotic behavior in the UV, in particular the  UV coupling for the relevant operator driving the flow  in the dual field theory. In Appendix C we review Weinberg's no-go theorem and describe how it is avoided in our framework. Appendix D contains the technical details of the linear perturbations around the vacuum. Finally, in Appendix E we give details about the large- and small- momentum asymptotics of the bulk Green's function, which is one of the  ingredients entering in the brane-to-brane propagator.

\section{The Self-tuning theory}

We consider a scalar-tensor Einstein theory in a $d+1$-dimensional bulk space-time parametrized by coordinates $x^a\equiv (u, x^\mu)$. We consider a $d$-dimensional brane embedded in the bulk parametrized by coordinates $x^\mu$ . The most general 2-derivative  action to consider reads,
\be \label{A1}
S = S_{bulk} + S_{brane }
\ee
where,
\be\label{A2}
S_{bulk} = M^{d-1} \int d^{d+1}x \sqrt{-g} \left[R - {1\over 2}g^{ab}\de_a\f\de_b \f - V(\f)\right] + S_{GH},
\ee
\be\label{A3}
S_{brane} = M^{d-1} \int d^d x \sqrt{-\g}\left[- W_B(\f) - {1\over 2} Z_B(\f) \gamma^{\mu\nu}\de_\mu\f\de_\nu \f + U_B(\f) R^{(\gamma)} \right]+\cdots,
\ee
where $M^{d-1} \equiv (16 \pi G_5)^{-1}$ is the bulk Planck scale,  $g_{ab}$ is the bulk metric, $R$ is its associated Ricci scalar and $\gamma_{\mu\nu}$, $R^{(\g)}$ are respectively the induced metric and intrinsic curvature of the brane while $V(\f)$ is some bulk scalar potential. $S_{GH}$ is the Gibbons-Hawking term at the space-time boundary (e.g. the UV boundary if the bulk is asymptotically $AdS$).

 The ellipsis in the brane action involves higher derivative terms of the gravitational sector fields ($\f,\gamma_{\m\n}$) as well as the action of the brane-localized fields (the ``Standard Model'' (SM), in the case of interest to us).
 $W_B(\f), Z_B(\f)$ and $U_B(\f)$ are scalar potentials which are generated by the quantum corrections of the brane-localized fields (that couple to the bulk fields, see \cite{smgrav}). As such, they are localized on the brane. In particular, $W_B(\f)$ contains  the brane vacuum energy, which takes contributions from the brane matter fields.  All of $W_B(\f), Z_B(\f)$ and $U_B(\f)$ are cutoff dependent and generically, $W_B(\f)\sim \Lambda^4$, $Z_B(\f)\sim U_B(\f)\sim \Lambda^2$ where $\Lambda$ is the
UV cutoff of the brane physics as described here. Its origin was motivated in subsection \ref{holo}.

\subsection{Field equations and matching conditions}

The  bulk field equations depend only on $V(\f)$ and are given by:
\bea
&& R_{ab} -{1\over 2} g_{ab} R = {1\over 2}\de_a\f\de_b \f -  {1\over 2}g_{ab}\left( {1\over 2}g^{cd}\de_c\f\de_d \f + V(\f) \right),  \label{FE1}\\
&& \de_a \left(\sqrt{-g} g^{ab}\de_b \f \right)- \sqrt{-g} {\de V \over \de \f} =0 \label{FE2}
\eea
The brane, being codimension-1, separates the bulk in two parts, denoted by ``$UV$'' (which contains the conformal $AdS$ boundary region or more generally, in non-asymptotically  $AdS$ solutions,  the region where the volume form becomes infinite ) and ``$IR$'' (where the volume form eventually vanishes, and may contain the $AdS$ Poincar\'e horizon, or a (good) singularity, or a black hole horizon etc, as we will discuss in section \ref{int-IR}). We will take the coordinate $u$ to increase towards the IR region.

Denoting $g^{UV}_{ab}, g^{IR}_{ab}$ and $\f^{UV}, \f^{IR}$  the solutions for the metric and scalar field on each side of the brane, and by $\Big[ X\Big]^{IR}_{UV}$ the jump of a quantity $X$ across the brane, Israel's junction conditions are:
\begin{enumerate}
\item Continuity of the metric and scalar field:
\be\label{FE3}
\Big[g_{ab}\Big]^{UV}_{IR} = 0,   \qquad \Big[\f\Big]^{IR}_{UV} =0
\ee
\item Discontinuity of the extrinsic curvature and normal derivative of $\f$:
\be\label{FE4}
\Big[K_{\mu\nu} - \gamma_{\mu\nu} K \Big]^{IR}_{UV} =   {1\over \sqrt{-\gamma}}{\delta S_{brane} \over \delta \gamma^{\mu\nu}}  ,  \qquad \Big[n^a\de_a \f\Big]^{IR}_{UV} =- {1\over \sqrt{-\gamma}}{\delta S_{brane} \over \delta \f} ,
\ee
where $K_{\mu\nu}$ is the extrinsic curvature of the brane, $K = \gamma^{\mu\nu}K_{\mu\nu}$  its trace, and $n^a$ a unit normal vector to the brane,  oriented towards the $IR$.
\end{enumerate}
Using the  form of the brane action, equations (\ref{FE4}) are given explicitly by:
\bea
&&
\Big[K_{\mu\nu} - \gamma_{\mu\nu} K \Big]^{IR}_{UV} = \Bigg[\ha W_B(\f) \gamma_{\mu\nu} + U_B(\f) G^{(\gamma)}_{\mu\nu} - {1\over 2}Z_B(\f) \left(\de_\mu \f \de_\nu \f - \ha \g_{\mu\nu} (\de \f)^2 \right)  \nonumber \\
&& \qquad + \left(\gamma_{\mu\nu}\gamma^{\rho\sigma}\nabla^{(\g)}_\rho \nabla^{(\g)}_\sigma - \nabla^{(\g)}_\mu\nabla^{(\g)}_\nu\right) U_B(\f) \Bigg]_{\f_0(x)}, \label{FE5}\\
&& \Big[n^a\de_a \f\Big]^{IR}_{UV}  = \left[{d W_B \over d \f} - {d U_B\over d \f} R^{(\g)} + \ha {d Z_B \over d \phi}(\de \f)^2 - {1\over \sqrt{\g}}\de_\mu \left( Z_B \sqrt{\g} \g^{\mu\nu}\de_\nu \f \right) \right]_{\f_0(x)},  \label{FE6}
\eea
where $\f_0(x^\mu)$ is the scalar field on the brane.

\subsection{The Poincar\'e-invariant ansatz} \label{sec FE}

We consider the case where the bulk space-time has $d$-dimensional Poincar\'e invariance, so that the solution would be dual to the ground state of a Lorentz-Invariant QFT.
The brane will be embedded at specific radial distance $u_0$ so that the induced metric is a flat $d$-dimensional Minkowski metric. In the domain-wall (or Fefferman-Graham) gauge,  the metric and scalar field (on each side of the brane)  are:
\be\label{FE7}
ds^2 = du^2  + e^{2A(u)} \eta_{\mu\nu}dx^\mu dx^\nu \, , \qquad \f = \f(u),
\ee
 We use the notation:
\be
{d\over d u} = \dot{}\,\,  ,
\ee
and we  denote by $(A_{UV}(u), \f_{UV}(u)) $ and $(A_{IR}(u), \f_{IR}(u)) $ the bulk solution in the UV and IR regions, respectively.
The brane sits  at a fixed value $u_0$ and we define:
\be\label{FE8}
A_0 \equiv A(u_0) , \qquad \f_0 \equiv \f (u_0).
\ee
Only $\f_0$ (not $u_0$) is a gauge-invariant quantity\footnote{By gauge invariant we mean invariant under bulk diffeomorphisms.}.  The induced metric on the brane is $\g_{\mu\nu} = e^{2A_0}\eta_{\mu\nu}$.

With the ansatz (\ref{FE7}),  the field equations (\ref{FE1}-\ref{FE2}) become equivalent to:
\be\label{FE8-a}
d(d-1) (\dot{A})^2  - {1\over 2} (\dot{\f})^2 = - V(\f), \qquad  \ddot A = -{1\over 2(d-1)}(\dot{\f})^2.
\ee
These equations can be cast in a first order form by defining  UV and IR superpotential functions $W_{UV}(\f)$ and $W_{IR}(\f)$ \cite{DeWolfe:1999cp}, such that:
\bea
&& \dot{A}_{UV}(u) = -{1\over 2(d-1) } W_{UV}(\f(u)) , \qquad \dot{\f}_{UV}(u) = {d W_{UV} \over d\f} (\f(u)) \, , \label{FE10}\\
&& \dot{A}_{IR}(u) = -{1\over 2(d-1) } W_{IR}(\f(u)) , \qquad \dot{\f}_{IR}(u) = {d W_{IR} \over d\f} (\f(u))\, . \label{FE11}
\eea
The scalar functions $W_{UV, IR}$ are both solutions to the (gauge-invariant) superpotential equation:
\be \label{superpot}
-{d \over 4(d-1)} W^2 + {1\over 2}\left( {d W\over d\f}\right)^2 = V .
\ee
The choice of $W$  determines the geometry  on each side,  up to the choice of an initial condition $(A_*, \f_*)$ ,  which only affects a shift in $A$  (i.e. an overall choice of scale of the $d$-dimensional theory).
These boundary conditions have a clear interpretation in the boundary QFT dual to the bulk gravitational theory: $A_*$ sets the scale of the boundary Minkowski metric of the dual QFT while $\f_*$ determines the UV coupling constant of the scalar operator $O(x)$ dual to $\f$. On the other hand $W$ is invariant under bulk diffeomorphisms.
The superpotential equation (\ref{superpot}) implies an inequality
\be
\left|W(\phi)\right|\geq B(\phi)\equiv 2\sqrt{-{d-1\over d}V(\phi)}
\label{B}\ee
which also defines the function $B(\phi)$ that acts as a lower bound on the space of solutions of the superpotential equation, \cite{multirg}.

The continuity conditions (\ref{FE3}-\ref{FE4}) simply state that $A(u)$ and $\f(u)$ are continuous across the brane:
\be\label{FE12}
A_{UV}(u_0) = A_{IR} (u_0) = A_0,  \qquad \f_{UV}(u_0) = \f_{IR} (u_0) = \f_0.
\ee
Therefore, only one initial condition $(A_*, \f_*)$ must be imposed, for example in the UV. The interpretation of these initial conditions is holographically clear and it will be discussed in greater detail in subsection \ref{scales}.

The non-trivial matching conditions are the ones imposed on the first derivatives, (\ref{FE5}-\ref{FE6}). Indeed, the extrinsic curvature and normal derivatives of $\f$ are given by:
\be \label{FE13}
K_{\mu\nu} = \dot{A} e^{2A} \eta_{\mu\nu}, \qquad K_{\mu\nu} - \gamma_{\mu\nu} K = -(d-1) \dot{A} e^{2A}\eta_{\mu\nu}, \qquad  n^a\de_a \f = \dot{\f}.
\ee

The junction conditions can be cast in a  gauge-invariant form using the superpotentials $W$ on each side of the brane: making use of the expressions (\ref{FE10}-\ref{FE11}) for $\dot{A}$ and $\dot{\f}$, as well as (\ref{FE13}), equations  (\ref{FE5}-\ref{FE6}) simply become statements about the jump in the superpotential and its derivative across the brane \cite{csaki}:
\bea
&& \left. W_{IR} -  W_{UV} \right|_{\f_0} = W_B(\f_0)  \label{match1}\\
&& \nonumber \\
&& \left. {d W_{IR} \over d\f}  -  {d W_{UV} \over d\f} \right|_{\f_0}  ={d W_B \over d\phi}(\f_0) \label{match2},
\eea
To summarize, the full system of bulk and brane field equations boils down to two bulk equations, (\ref{superpot}),  relating the two super potentials $W_{IR}$ and $W_{UV}$  to the bulk potential $V$,  and two matching conditions relating the bulk and brane superpotentials together, (\ref{match1}), (\ref{match2}). Before explaining the logic we will use in picking the relevant solutions to these equations (which will be the subject of section \ref{selfie}), we make a digression on holography and the properties and interpretation of the UV and IR parts of the geometry.

\subsection{Holographic interlude} \label{interlude}

In a bulk theory allowing for a holographic interpretation, not all bulk geometries are on the same footing. Below we summarize the structure  of the ``UV'' (i.e. large volume) and ``IR'' (small volume) regions of the bulk geometry, and their interpretation in the holographic dictionary.

\subsubsection{UV region} \label{int-UV}

 First of all, we will consider only UV-complete solutions, i.e. those containing  a region which extends all the way to  an asymptotically $AdS$ boundary  where $e^A \sim e^{-u/\ell} \to +\infty$.  The presence of such a region on one side of the interface is a crucial ingredient of the self-tuning mechanism: this is because, as we will discuss below in more detail, generic solutions in this region flow to the UV fixed point independently of the particular value $W_{UV}(\f_0)$ which solves equations (\ref{match1}-\ref{match2}). Therefore, we do not need to fine-tune the UV side of the solution.

The asymptotic UV region  usually corresponds to the scalar field asymptoting to  a maximum of the scalar potential. A given potential may  allow for several UV fixed points, but one can restrict the boundary conditions of the gravitational theory to pick one of them: indeed,  the choice of UV boundary conditions is part of the definition of the holographic theory. This includes not only the choice of the UV extremum,  but also the boundary conditions on scalar fields, the boundary induced metric, etc.

 Even with these restrictions, there always exist an {\em infinite} number of solutions to the superpotential equation in the UV, which satisfy all the correct boundary conditions at leading order in a near-boundary expansion, and differ by subleading terms.

In fact,  there exists an arbitrary integration constant $C_{UV}$ to the superpotential equation which parametrizes  a continuous family of nonequivalent solutions which get closer and closer to each other as one   approaches the extremal point of $V$.  For a recent detailed discussion of the solutions to the superpotential equation, see \cite{multirg}. All these solutions asymptote the same $AdS$ geometry, and they are all regular close to the boundary of $AdS$.

To be more explicit, such a ``UV'' $AdS$ solution is realized near a maximum of bulk potential (say at $\f = 0$),
\be \label{int1}
V(\f) \simeq -{d(d-1) \over \ell^2} + {m^2 \over 2} \f^2 + \ldots
\ee
The constant term fixes the asymptotic $AdS$ length $\ell$, and the mass term fixes the dimension of the corresponding operator\footnote{We assume here what is called ``standard'' form of the holographic dictionary. For operator dimensions $\Delta$ such that $d/2-1<\Delta <d/2$,  there is an ``alternative'' definition of the theory which is obtained by replacing $\Delta \leftrightarrow (d-\Delta)$ in equation (\ref{int2}). We will not discuss this case further.} by:
\be \label{int2}
\Delta = {d\over 2}\left(1+ \sqrt{1 + {4 m^2 \ell^2\over d^2}}\right).
\ee
We assume $-d^2/4< m^2 <0$ so that $2<\Delta <d$ and the operator is relevant.
The  superpotentials corresponding to (\ref{int2})  all have the  form, for $\phi\simeq 0$:
\be\label{int3}
W_{UV}(\f) = {2(d-1)\over \ell} + {d-\Delta\over 2} \f^2 + \ldots + C_{UV} \f^{d\over d-\Delta} \Big(1 + \ldots\Big) + {\cal O}\left(\f^{2d\over d-\Delta}\right).
\ee
where the dots indicate analytic higher order terms, and $C_{UV}$ is an undetermined constant. Solving for  the scalar field  and scale factor via equations (\ref{FE10}),  one finds:
\be \label{int4}
e^{A(u)} \simeq e^{-u/\ell}, \quad \f(u) \simeq g_0  \, \left(\ell \,  e^{u/\ell} \right)^{d-\Delta}  + {C_{UV} \ell \over (2\Delta -d)}\, g_0^{\Delta\over (d-\Delta)} \,  \left(\ell \, e^{u/\ell}\right)^\Delta, \quad u \to -\infty,
\ee
where $g_0$ is one more integration constants  which, importantly, does not appear in the superpotential. In the equation above, the factors of $\ell$ are inserted to absorbe the appropriate mass dimensionality of $g_0$ and $C_{UV}$ ($d-\Delta$ and one, respectively) while keeping $\f(u)$ dimensionless.
  
We now describe how the solution above is interpreted in the holographic dictionary.
\begin{itemize}
\item The scale factor diverges as $\f \to 0$, signaling an asymptotically $AdS$ region with conformal boundary at $u=-\infty$, where the scalar reaches the ``UV fixed point'' $\f=0$.
\item Both leading and subleading terms in the scalar field vanish as we go to the boundary ($u\to -\infty$), signaling that the fixed point is an attractor.
\item The constant $g_0$ controlling the leading term in the scalar field near-boundary expansion represents the  coupling of the dual operator $O(x)$ associated to $\f$ in the dual field theory in the far UV. In other words the UV CFT is deformed by a term of the form:
\be \label{int4-ii}
S_{source}  = \int d^d x\,  g_0 O(x).
\ee
Notice that $g_0$ does not appear in the superpotential, but rather is generated by the  boundary conditions one imposes at the AdS boundary (extreme UV limit of the dual field theory).
\item Similarly, one has to fix asymptotically the boundary conditions for the leading term in the metric. With a generic ansatz of the form,
\be \label{int4-iii}
ds^2 \to du^2 + e^{-2u/\ell}\left(\gamma^{(0)}_{\mu\nu} + \ldots\right) , \qquad u \to -\infty,
\ee
$\gamma^{(0)}_{\mu\nu}$ represents the  metric of the space where the UV CFT is defined , and it is also fixed by boundary conditions at the $AdS$ boundary.  In particular, in the solution we are considering, the CFT lives in flat space-time with Minkowski metric. It is crucial that neither $g_0$ nor $\gamma^{(0)}_{\mu\nu}$ are fields with respect to which we have to extremize the gravitational action, but they are part of the definition of the dual field theory.
\item
The subleading term in the near-boundary expansion of $\f$ in equation (\ref{int4})  is controlled by $C_{UV}$. In the dual field theory, this term corresponds to the  vacuum expectation value of the operator $O$:
\be\label{int5}
\< O \> = C_{UV} \ell  \,\, g_0^{\Delta\over (d-\Delta)}
\ee
\end{itemize}
We see explicitly from equations  (\ref{int3}) and (\ref{int4})  that   the integration constant $C_{UV}$  enters only at subleading order, and all superpotentials get closer and closer to each other as $\f \to 0$, independently of $C_{UV}$. In other words, no matter  what initial conditions we pick for $W_{UV}$  away from $\f=0$, the solution is  attracted to the same asymptotically AdS boundary at $\f=0$. This also implies that the initial condition $W(0)=2(d-1)/\ell$ is ill-defined because it does not fix the  solution.

\subsubsection{IR region} \label{int-IR}
The situation is conceptually  different in the IR.
The difference between the UV and the IR is that not all solutions reaching the IR are regular, and  not all of them are acceptable, but only  those  obeying some restrictions. The others are to be considered as ``spurious'' solutions of Einstein's equations which  are unphysical from the holographic point of view, i.e. they do not correspond to a true state (saddle-point)  in the dual field theory.

More specifically, a solution is acceptable in the IR if it belongs to one of the two classes below:
\begin{itemize}
\item IR-regular  solutions;
\item ``Good'' IR-singular solutions (near the boundary of field space).
\end{itemize}

Below we will explain what characterizes these two classes. The crucial point is that, as we will explain, independently of the choice of $V(\phi)$, there is at most a {\em finite number} of IR-acceptable solutions of the superpotential equation\footnote{We exclude from the discussion here the case of flat directions in the dual QFT.}.

 Before we delve into the classification of IR solutions we note that,  on the IR side of the brane, eventually $e^A \to 0$. Indeed, by definition, going towards the IR  the  scale  factor is decreasing. We will discard the presence of  an IR brane (or ``hard wall'' in the holographic lore) cutting off the small volume part of the geometry, and at which  $A$ reaches a finite limit.   This case suffers from the same problems as in non-computable singularities that we will discuss below (i.e. it is neither regular nor acceptable).   On the other hand, assuming there is no hard wall,  the scale factor  cannot approach a non-vanishing constant asymptotically,  without the theory violating the null energy condition (this can be seen using $\ddot A = -(\dot\phi)^2 < 0$).

\paragraph{Regular solutions.}
These are solutions in which the curvature invariants are all finite as $e^A \to 0$. In practice, in a co-dimension-one setup, the only asymptotic behavior compatible with regularity is:
\be \label{sing1}
e^{A(u)} \sim e^{-u/\ell_{IR}} \qquad u \to + \infty
\ee
where we approach the Poincar\'e  horizon of an $AdS$ space-time with curvature radius $\ell_{IR}$. This corresponds to both $V(\phi)$ and $W(\phi)$ approaching a finite constant:
\be \label{sing2}
V \to -{d(d-1) \over \ell_{IR}^2}, \qquad W \to {2(d-1) \over \ell_{IR}^2}.
\ee
Since the asymptotic geometry approaches  $AdS$, we are again in the presence of  an asymptotically conformal theory. However, now this  is in the interior of the space, where the scale factor approaches zero asymptotically. Therefore this CFT is found in the IR limit.  The actual IR limit of the scalar field,  $\phi_{IR}$,  may be finite (in which case the dual  theory flows to an IR conformal fixed point at a finite value of the coupling), or it can be  infinite ( runaway $AdS$ behavior, \cite{superp}).

An important difference with respect to the case of a UV fixed point is that solutions $W(\phi)$ reaching an IR fixed point are {\em isolated points} in the space of solutions of the superpotential equations and are not part of a continuous family. In other words,  an infinitesimal deformation (in the space of solutions) leads to missing the fixed point and flowing elsewhere while typically becoming singular in the process.

\paragraph{Acceptable singular solutions.}

Putting aside the AdS IR asymptotics described above, all other cases $e^A \to 0$, lead to an IR naked singularity where we have additionally that $\phi \to +\infty$. However, some singularities may be acceptable from holographic arguments or gravitationally if  there exists a way of resolving them. The presence of a (classical curvature) singularity in holography may be interpreted as follows:
\begin{enumerate}

\item The solution does not describe a semiclassical saddle point. These are what are customarily called ``bad" singularities in holography, \cite{Gubser}.

\item The singularity appears because we have not included all possible relevant degrees of freedom. If we include them then the singularity is resolved. Examples of such resolutions exist both by re-including KK states of the bulk theory, \cite{sing1,sing2}, or in more complicated contexts stringy states, \cite{sing3,sing4}. Such resolvable singularities are usually called ``good singularities" in holography, \cite{Gubser}.

\end{enumerate}

A criterion for a ``good" (i.e. resolvable) singularity was proposed by Gubser  in \cite{Gubser}.
 It postulates that the solution admits an infinitesimally small deformation which may cloak the singularity behind a regular black hole horizon. In this way, ``heating up'' slightly the theory, in principle, cloaks the naked singularity without a drastic change to the solution. For a concrete example, take the case of our bulk action (\ref{A2}) with a specific Liouville potential $V(\phi) \sim \exp b \phi$. In this case the general solution with the relevant planar $(d-1)$-symmetry is known \cite{cc1}. It is found that  black hole solutions exist only for $b < \sqrt{2d \over d-1}$,  otherwise solutions have an uncloaked naked singularity. This agrees with the postulated criterion (\ref{sing3}) as we will see in a moment.

There is additional evidence concerning solutions with ``good" singularities. The calculation of correlators involves the solution of the fluctuation equations with appropriate boundary conditions.
There are two possibilities:
\begin{itemize}

\item The behavior of correlators at finite energy does not depend on the resolution of the singularity.
    This case is realized if in the associated Sturm-Liouville problem only one of the two linearly independent solutions is normalizable at the singularity. In this case the boundary condition (normalizability) fixes the correlator uniquely. In mathematical terms, an equivalent statement is that the corresponding radial Hamiltonian is essentially self-adjoint. We will call such ``good" singularities {\em computable} or {\em IR-complete}. Such a singularity resolution was encountered early on in higher co-dimension brane world models \cite{roberto}, in \cite{horowitz},  and in the holographic context in \cite{gravitons}.

\item The behavior of correlators at finite energy {\em does}  depend on the resolution of the singularity.
    This case is realized if in the associated Sturm-Liouville problem {\em both} linearly independent solutions are normalizable at the singularity. In this case one needs an extra boundary condition, (which is supplied by the singularity resolution).
In mathematical terms, an equivalent statement is that the corresponding radial Hamiltonian has an infinity of self-adjoint extensions determined by extra boundary conditions at the singularity.
Therefore, without an explicit resolution of the singularity the correlators cannot be computed. We will call such ``good'' singularities {\em non-computable} or {\em IR-incomplete}.Examples of such cases are described in detail in \cite{superp,multirg}.

\end{itemize}

 There are many examples where IR-completeness fails, but the most straightforward is the example of a hard wall, i.e. an IR-brane that cuts-off the small volume region of the geometry: in this case, all solutions of bulk  wave equation are trivially normalizable at the wall, but different boundary conditions lead to very different spectra.  IR-incompleteness does not necessarily mean that the holographic model is unphysical: rather it hints that  with the present ingredients we do not have enough information to compute any observable without embedding it into a more complete theory (a higher dimensional  bulk geometry or a string theory). For convenience, to make sure the five-dimensional description we are using  is predictive, we require IR-completeness. This puts a  restriction on the asymptotics of the bulk potential at infinity as we discuss below.

It can be shown (see e.g. \cite{multirg}) that the IR-completeness criterion always implies Gubser's criterion. Both  imply a bound on the growth rate of the potential in the IR. We suppose for definiteness that the potential grows asymptotically as:
\be \label{sing2-i}
V(\f) \sim -V_\infty \exp b \f, \qquad b>0, \; V_\infty >0.
\ee
It is useful to define the quantity:
\be \label{Q}
Q  = \sqrt{d\over 2(d-1)}.
\ee
Then:
\begin{enumerate}
\item Gubser's criterion requires
\be \label{sing3}
b < 2Q
\ee
\item  IR-completeness requires:
\be\label{sing4}
b < 2\sqrt{{d+2 \over 6(d-1)} }
\ee
\end{enumerate}
One can check that (\ref{sing3}) is stronger than (\ref{sing4}) for any $d>1$.

This is not the end of the story: the conditions in equations (\ref{sing3}-\ref{sing4})  are not sufficient as they refer to the potential but not to the solution itself. Indeed, whether or not the singularity is acceptable according to one or the other criterion depends on the behavior of the superpotential at infinity which characterizes the solution.  Analyzing the superpotential equation one finds two types of asymptotic solutions corresponding to the potential  behaving as in equation (\ref{sing2-i}):
\begin{enumerate}
 \item A continuous family, whose asymptotic behavior is independent of the parameter $b$ in (\ref{sing2-i}):
\be \label{sing5}
W_C(\f) \simeq C \exp Q \f , \qquad \f \to +\infty,
\ee
where $C$ is an arbitrary positive constant.

\item An isolated solution with a milder asymptotic behavior:
\be \label{sing6}
W_*(\f) \simeq W_\infty \exp {b\over 2} \f, \qquad \f \to +\infty,
\ee
The parameter $W_\infty$ in this case is fixed, and given by:
\be
W_\infty = \sqrt{ 8 V_{\infty} \over (2Q)^2 - b^2}.
\ee
Due to equation (\ref{sing3}), one  sees  that the  special solution $W_*$ in equation (\ref{sing6})   grows more slowly than any of the solutions in the continous family\footnote{More generally, the special solution is characterized by  the property  property $W_* (\f) \sim \sqrt{V(\f)}$ as $\f\to +\infty$.},  (\ref{sing5}).
\end{enumerate}
It turns out  that {\em only the isolated solution} $W_*(\f)$ with special asymptotic behavior (\ref{sing6}) satisfies Gubser's criterion and,  in case (\ref{sing4}) is satisfied, also IR-computability\footnote{In particular, even if we relax IR-computability, there is still only one  solution satisfying Gubser's criterion.}.

\begin{figure}[h!]
\centering
\includegraphics[width=0.7\textwidth]{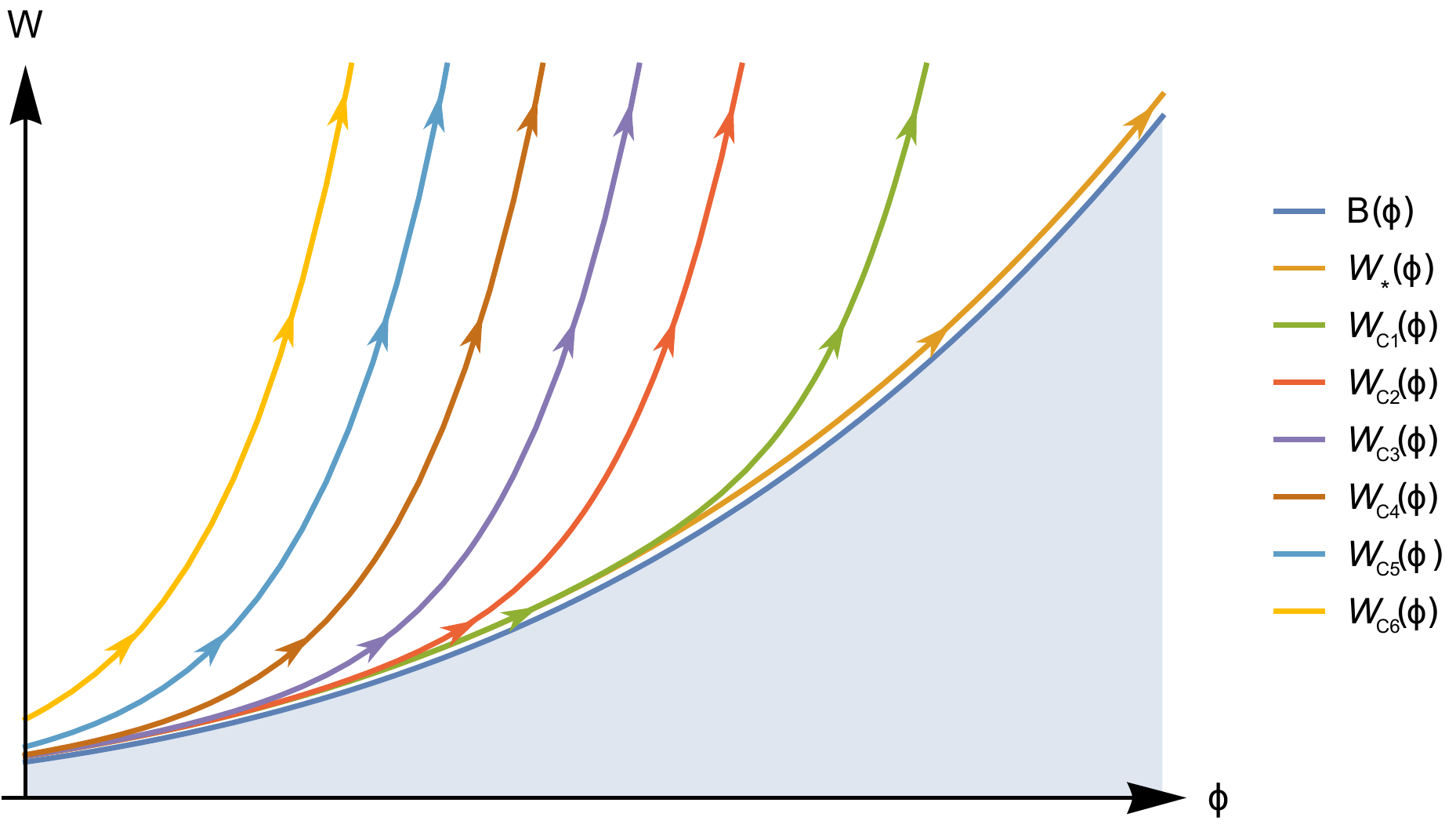}
\caption{Schematic view of the solutions of the superpotential equation that reach $\phi \to +\infty$. The curve $W_*$ represents the special solution. The curves labeled $C_1$ through $C_6$ are different curves belonging to the continuous family, and they grow exponentially faster than $W_*$. The function $B$ defines the boundary of the forbidden region as defined in equation (\protect\ref{B}).}
\label{fig:sing2}
\end{figure}

A sketch  of the superpotential solutions in the IR is given in Figure \ref{fig:sing2}.

The conclusion therefore is that there is  at most {\em a finite number} of solutions that have an acceptable behavior in the IR: those that reach an IR fixed point at a minimum of $V$, plus the  special singular solution (\ref{sing6}) that extends to infinity in field space. However all of them may flow to the same UV. Therefore, imposing IR-regularity picks one or a few of the many solutions which flow to the same asymptotic $AdS$ boundary.

\subsection{The self-adjustment mechanism} \label{selfie}

In the context of brane-worlds, the self-tuning paradigm \cite{st1}  appeared shortly after the original Randall-Sundrum (RS) model with one or two branes \cite{rs} embedded in AdS space-time.

In the original RS model, one has to fine-tune the tension of the brane with the bulk cosmological constant in order for the embedded brane geometry to remain flat. Any detuning of the brane tension, resulted in an effective inflation (or deflation) of the brane, therefore to an effective cosmological constant acting on the brane, \cite{KR}. The fine tuning of the brane tension, in the context of the brane-world, is the translation of the well-known fine tuning of the cosmological constant. Therefore this is the brane-world version of the big cosmological constant problem \cite{Weinberg:1988cp}.

It was noted that including a scalar field in the bulk geometry would introduce extra freedom in the solutions, which relieved the brane to bulk fine-tuning condition \cite{st1}. Unfortunately, this simple resolution did not come without problems of its own. We will scrutinize this in more detail, as it is central to our discussion.

The equations at the heart of the self-tuning mechanism are the matching conditions for the superpotential we wrote at the end of subsection (\ref{sec FE}):
\bea
&& \left. W_{IR} -  W_{UV} \right|_{\f_0} = W_B(\f_0)  \label{match1-i}\\
&& \nonumber \\
&& \left. {d W_{IR} \over d\f}  -  {d W_{UV} \over d\f} \right|_{\f_0}  ={d W_B \over d\phi}(\f_0) \label{match2-i},
\eea
where both $W_{UV}(\f)$ and $W_{IR}(\f)$ are solutions of the superpotential equation,
\be \label{superpot-i}
-{d \over 4(d-1)} W^2 + {1\over 2}\left( {d W\over d\f}\right)^2 = V .
\ee
There are two possible ways of looking at these equations and  what they constrain, and these result in the ``old'' self-tuning solutions (which generically lead to bad IR singularities) and holographically-motivated self-tuning, where the IR singularity disappears or if it still persists,  it is of a good kind.

\paragraph{Old self-tuning.}

This is the case discussed in most general terms  by Csaki {\it et al.} in \cite{csaki},  who consider essentially the same class of actions as the one here, apart from the induced Einstein and kinetic terms on the brane. They arrive at the same equations (\ref{match1-i}-\ref{match2-i}) and do a counting of integration constants which can be summarised as follows:
\begin{enumerate}
\item Each of the two functions $W^{IR}$ and $W^{UV}$ solves the first order equation (\ref{superpot-i}), and therefore each one is determined modulo an integration constant  $C^{UV}$, $C^{IR}$;
\item For a generic  brane potential   $W^B(\f)$, the two matching conditions (\ref{match1-i}-\ref{match2-i}) fix  $C^{UV}$, $C^{IR}$ for {\em any} generic value of $\f_0$ (\ref{FE12}).
\end{enumerate}
In other words $\f_0$ is a free integration constant which self tunes with an arbitrary value of the brane tension while the brane remains geometrically flat. The effective cosmological constant of the brane is zero\footnote{By effective we mean the one measured on the brane by measuring the expansion of the brane universe.} independently of the brane-world tension and of the couplings in the action. This indeed gives self-tuning solutions, but the heart of the problem is that a generic value of $C^{IR}$ will result in a solution in the confinous family (\ref{sing5}), which has an unacceptable IR singularity. This was the case for all explicit solutions considered  in \cite{csaki} (when $V$ is given by a  cosmological constant or an IR exponential potential), which indeed are in the continuous branch (\ref{sing5}). On the other hand, the special solution (\ref{sing6}) was not considered.   Moreover, $\f_0$ appears as a modulus of the solution.

\paragraph{Holographic self-tuning}

 According to the rules of holography, summarized in the previous subsection, the perspective on the choice of integration constants is different:
\begin{enumerate}
\item The $IR$ constant  $C^{IR}$ should be fixed by demanding that the IR singularity is absent (Poincar\'e horizon in $AdS$) or it is of the good type.

Typically there is only  one such  solution to eq (\ref{superpot-i}) (or at most a discrete set).  According to this point of view,  the solution $W^{IR}$  should be fixed once and for all to be the regular one, or the one with special asymptotics (\ref{sing6}) in the case of an IR-exponential potential\footnote{Or more generally, for an arbitrary potential, the one with large-$\f$ asymptotics $W_*(\f)\sim \sqrt{V(\f)}$.}, before we impose the matching conditions (\ref{match1-i}-\ref{match2-i}).

\item Once $W^{IR}$ is fixed by regularity , equation (\ref{match1-i}-\ref{match2-i}) will determine 1) the integration constant $C^{UV}$ in the UV superpotential ;  2) {\em the brane position} in field space, $\f_0$.

\end{enumerate}

Therefore, demanding IR regularity fixes the brane position. This is  desirable, as it is likely that the old self-tuning setup still leaves some light radion-like modes in the spectrum, since the brane position (in $\f$ space)  looks like a modulus. On the other hand, since the {\em good} values of $C^{IR}$ are (generically) at most  discrete,  in the holographic version there is no continuous deformation parameter in the space of solutions to the superpotential equation plus matching conditions.

Notice that $C^{UV}$ contains information about the v.e.v of the operator dual to $\f$, so the presence of the brane is changing  the v.e.v's with respect to the solution with no brane in which $W^{UV} = W^{IR}$.   On the other hand, we are still free to choose the UV sources, which are encoded in the UV boundary conditions of equations (\ref{FE10}) for the metric and the scalar field.


We will now rewrite the matching conditions in a way which makes the philosophy explained above algebraically manifest. Equations (\ref{match1-i}) and (\ref{match2-i}) can be rewritten in such a way that one can obtain a solution without knowing anything about the behavior of the UV superpotential: the latter can be eliminated from the equation and one is left with a single equation that determines $\f_0$ from $V(\f)$ and the IR solution alone.

In fact, we solve  (\ref{match1-i}-\ref{match2-i})  for $W^{UV}(\f_0)$ and $\de_{\f}W^{UV}(\f_0)$,
\be\label{sol1}
W^{UV}(\f_0) = W^{IR}(\f_0) -  W^B(\f_0) \qquad \de_\f W^{UV}(\f_0) = \big(\de_\f W^{IR} -  \de_\f W^B\big)(\f_0),
\ee
and then  use the superpotential equation (\ref{superpot-i}) for $W^{UV}$ to arrive at an equation for $\f_0$ only:
\be\label{sol4}
-{Q^2 \over 2}\Big( W^{IR}(\f_0) -  W^B(\f_0)\Big)^2 \,+\,  {1\over 2} \left({d W^{IR} \over d\f} -  {d W^B \over d\f}\right)_{\!\!\f_0}^2 =  V(\f_0),  \qquad Q \equiv \sqrt{d\over 2(d-1)}.
\ee
This equation contains as input only the model data (the bulk and brane potentials) and the IR solution, which is fixed by IR regularity. Once we solve it for $\f_0$, we can obtain $W^{UV}(\f_0)$  from (\ref{sol1}). This  in turn can then be used as an initial condition for the UV superpotential equation, and it fixes $W^{UV}(\f)$  completely\footnote{The sign ambiguity in $dW^{UV}/d\f$ is fixed  by  the sign of $d(W^{IR} - W^B)/d\f$ at the interface.}.\\

We pause here to clearly explain   on how the adjustment mechanism of the cosmological constant is realized in this model and what we mean by self-tuning.

A self-adjustment mechanism is one in which it is not necessary to pick special values of the parameters of the model in order to obtain a solution to the field equations in which the 4d metric is Minkowski. Here, the model parameters are the bulk and brane potentials, and these include the 4d vacuum energy (which  is contained in the $\f$-independent part of $W_B(\f)$).

 The standard CC problem can be stated by saying that for {\em generic} values of the  parameters (in particular,  of the 4d vacuum energy plus bare cosmological term), flat 4d space is not a solution.

In our model on the other hand, for arbitrary potentials $V$ and $W^B$ in the bulk and brane, such that they allow a solution of equation (\ref{sol4}), the ``UV'' side of the geometry  adjusts itself dynamically, for given $C_{UV}$ and $\f_0$, so that flat space is a solution for the 4d metric. As we discussed in section \ref{interlude}, generic initial conditions for $W_{UV}$ at $\f_0$ (as long as  $W_{UV}>0$, as we will discuss in section \ref{types}) connect to the same large volume AdS region, therefore any of these $W_{UV}$ gives rise to a regular geometry satisfying the same boundary conditions. From the boundary field theory perspective, they differ only in the vacuum expectation value of the operator dual to $\f$, which is encoded in the integration constant which fixes $W_{UV}$. This integration constant, (or in the field theory language,  the v.e.v. of the operator in the UV CFT), is the extra parameter which is not fixed by the  bulk field equations and which is responsible for  the self-adjustment to flat space. Stated differently, the UV geometry adjusts itself in order to be glued to the regular IR solution through the interface, {\em whatever} the parameters at the interface are. We will give another picture of this mechanism in purely 4d terms in Appendix \ref{nogo} when we discuss Weinberg's no-go theorem.

This is not the end of the story, however. As we will see in the next sections, even though we may have solutions generically,  not all solutions are acceptable:  as we will see in the next section,  the solution which we have called  $W_{UV}$  does not always connect to an asymptotically $AdS$ region, but sometimes it actually describes a different IR geometry,  in which case the self-tuning mechanism does not work.
Furthermore, one has to impose some phenomenological and consistency requirement:  the existence of 4d gravity at the observed distance scales, which will be discussed in section \ref{sec pheno}, and the stability of the solution (absence of ghosts and tachyons) (see section \ref{sec scalar}).   These requirements  may restrict the range of allowed parameters in the potentials, but do not introduce a need for tuning independent parameters\footnote{This,  on the contrary,  is the case in the RS model, in which two a priori independent constants - the bulk cosmological constant and the brane tension - need to be related to each other.}. In any case, this has now become a problem in  model building (finding a set of bulk and brane potentials such that there is enough room for realistic physics) rather than a conceptual fine-tuning or naturalness problem.\\

We end this section by discussing the relation between the integration constants of the bulk gravitational equations and the parameters which define the  dual field theory.

For a given choice of bulk and brane potentials  in the action (\ref{A2}-\ref{A3}),  any specific bulk solution of the form (\ref{FE7})  depends on two sets of integration constants:
\begin{itemize}
\item[(a)] The two integration constants entering $W_{UV}$ and  $W_{IR}$ by IR  regularity, and the the equilibrium value $\f_0$.

\item[(b)] The integration constants that determine the scalar field and metric profile by integrating equations (\ref{FE10}-\ref{FE11}). These include the scale factor at the interface, $A_0$,  as well as  the brane position $u_0$.
\end{itemize}

As we have discussed,  the integration constants in the first set are all fixed\footnote{Up to possible discrete choices: if multiple solutions are allowed, these however are always isolated points in the space of all possible choices of integration constants.} once the action is given: either by regularity ($W_{IR}$), or dynamically via equations (\ref{sol1}) ($W_{UV}$ and $\f_0$).
On the other hand, integration constants in the second set (b) are  not fixed by any quantity entering in the action: rather, they are completely fixed by the choice of the UV coupling $g_0$, which is part of the definition of the dual boundary theory:
\be \label{scales5c-sec}
A(u)  \to - {u \over \ell_{UV}}\qquad \f(u) \to  g_0 \exp \left({\Delta_- u\over \ell_{UV}}\right)  , \quad u \to -\infty,
\ee
where $\Delta_- = (d-\Delta)$, $\Delta$ being the dimension of the operator dual to $\f$ in the UV field theory.

As it is shown in Appendix \ref{scales}, one can find a simple relation between the integration constants at the brane, and the UV coupling $g_0$:
\be\label{scales16-sec}
g_0 = e^{\Delta_- A_0} \left(e^{\bar{{\cal A}}(\f_0)} /\ell_{UV}\right)^{\Delta_-}  \,,
\ee
where  $\bar{{\cal A}}(\f_0)$ depends only of integration constants in class (a) (and depends on neither  $A_0$ nor $g_0$):
\be
\bar{{\cal A}} (\f_0) \equiv {1\over \Delta_-} \log \f_0 +  {1\over 2(d-1)} \int^{\f_0}_{0} \left( {W_{UV} \over W_{UV}'} - {2(d-1) \over \Delta_- \f}\right).
\ee

\subsection{Consistent self-tuning solutions} \label{types}

As we have seen in section \ref{selfie}, once the IR solution $W_{IR}$ and the brane potential $W_{B}$ are fixed,  the interface position $\f_0$ and the UV superpotential $W_{UV}$ are determined by the two equations (\ref{sol1}). In this section we  summarize the different possible qualitative behaviors at the intersection, depending on the sign and the size of the brane potential at the interface,  $W_B(\f_0)$. A detailed discussion is given in Appendix \ref{types-app}.

\begin{figure}[h!]
\begin{center}
\includegraphics[width=7.3cm]{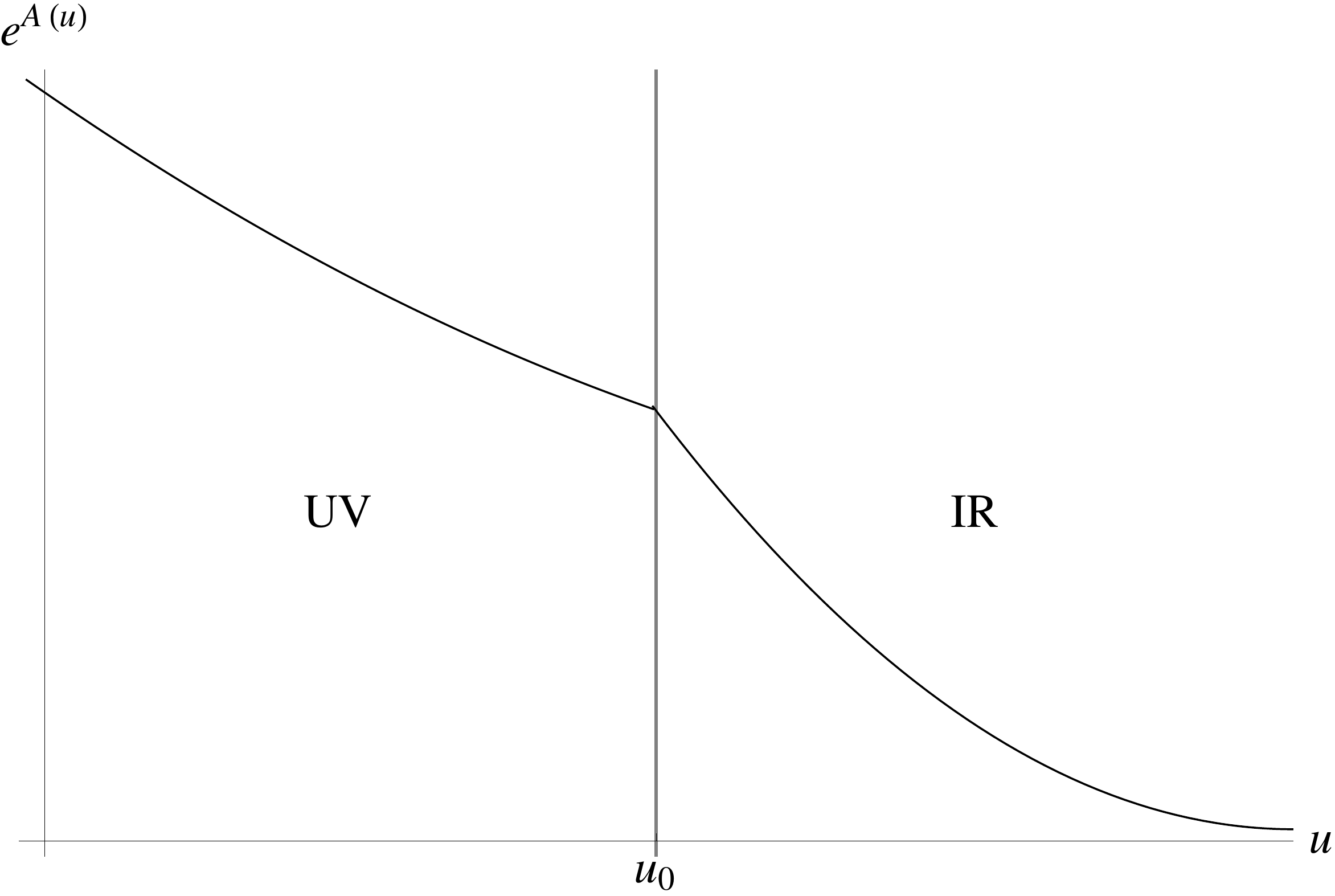} \includegraphics[width=7.3cm]{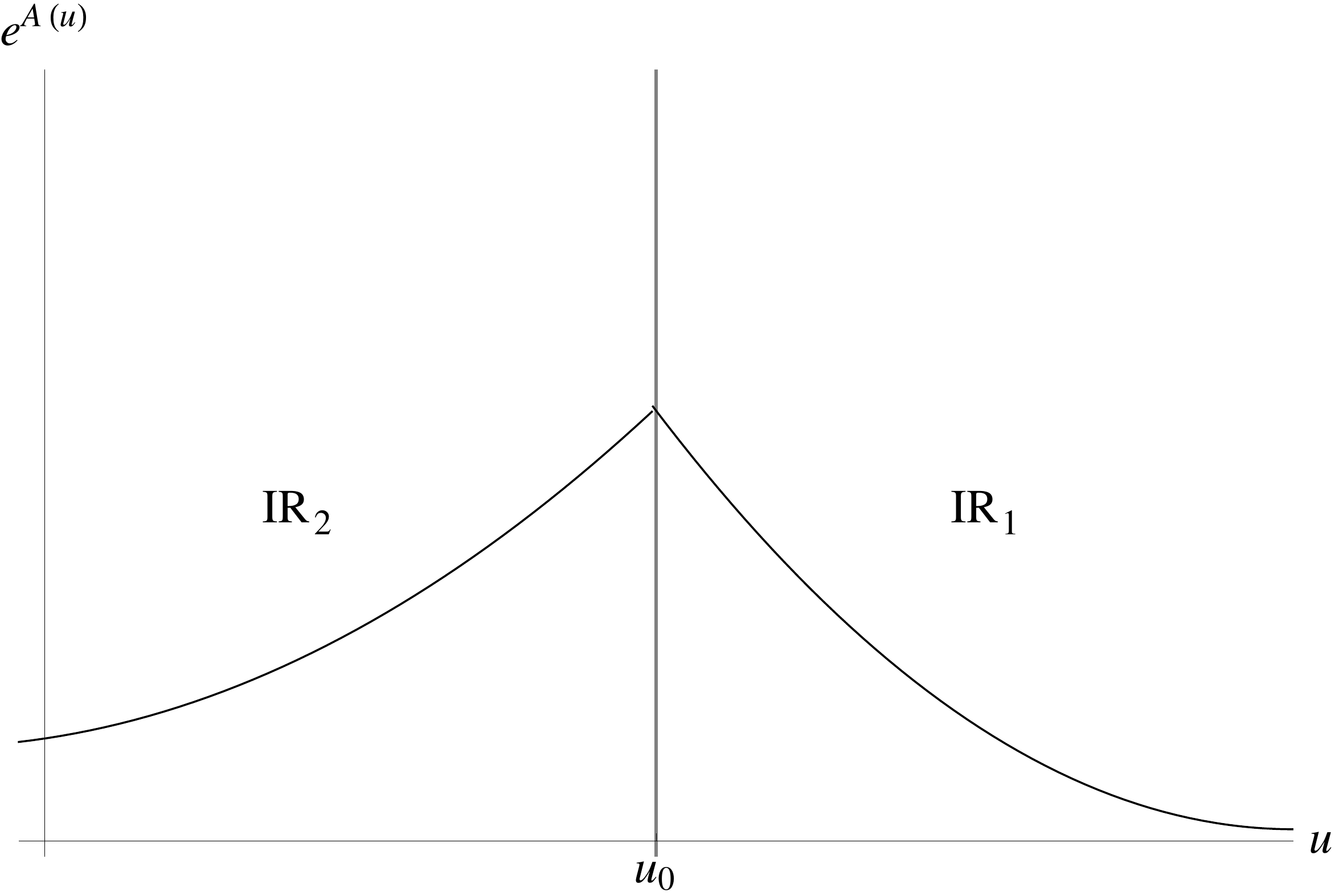} \\
(a) \hspace{7cm} (b)
\caption{The behavior of the scale factor $e^A$ as  a function of the holographic coordinate $u$, in the two cases  $W_{UV}>0$ (left) and $W_{UV}<0$ (right). (a) In case $A$,  the junction  connects an IR and a UV region. (b) In case $B$, the junction is a maximum of the scale factor and connects two IR regions.} \label{Scalefactor}
\end{center}
\end{figure}

The first qualitative distinction comes from the sign of $W_{UV}(\f_0)$, which is given by the sign of $W_{IR}-W_B$ at the interface:
\begin{enumerate}
\item[A.] $W_{UV}(\f_0)>0$.
In this case the scale factor is monotonic as we cross the interface, like in figure \ref{Scalefactor} (a). The solution on the left of the interface ($u<u_0$ in our conventions) flows to the UV asymptotic region, and approaches the $AdS$ boundary where $e^A \to +\infty$, {\em independently} of the precise value of $W_{UV}(\f_0)$ (recall our discussion in section \ref{int-UV} about the asymptotically $AdS$ UV region being an attractor).

Notice that since $W_{UV}(\f_0)$ must be in the allowed region, i.e. $|W_{UV}(\f_0)| > B(\f_0)$,  defined in (\ref{B}).  Positivity of  $W_{UV}(\f_0)$ automatically implies that $W_B(\f_0) < W_{IR} - B(\f_0)$.

\item[B.] $W_{UV}(\f_0)<0$.
In this case on the other hand the interface is a local maximum for the scale factor, which decreases on both sides of the brane, as in figure \ref{Scalefactor} (b).  The junction connects two solutions of the ``IR'' type, and no asymptotic UV boundary region. In order to be acceptable,  the solution on both sides has to coincide with the ``special'' IR solution which is either regular or have a good singularity. Since the special IR solution is unique, this happens only if the brane potential satisfies the condition $W_B(\f_0)= 2W_{*}(\f_0)$, i.e. the junction is symmetric\footnote{More generally, there may be a finite number of special IR solutions, in which case the bulk may be  asymmetric  across $u_0$ but still requires one of the tunings $W_B(\f_0) = W_1^*(\f_0)+W_2^*(\f_0)$, where $W_1^*$ and $W_2^*$ are the two IR-special solutions on the left and on the right.}. This is an extra condition which requires fine-tuning of the brane potential against the bulk. This is the generalization of the usual RS fine tuning of the bulk vs. brane cosmological constant.
\end{enumerate}

We therefore arrive at the following statement: \\
{\em  A solution to the matching condition implements the self-tuning mechanism only if :}
\be \label{ex1}
W_B(\f_0) < W_{IR}(\f_0) - B(\f_0).
\ee

On the other hand, as we will see, one of the results from the analysis of scalar fluctuations in section \ref{sec scalar} is the following:\\
{\em  A sufficient condition for stability  is:}
\be \label{ex1-b}
W_B(\f_0) >0.
\ee
However, one can relax (\ref{ex1-b}) and check stability on specific models case-by-case.

Notice that the r.h.s. of equation (\ref{ex1}) is necessarily positive since $W(\f_0)<B(\f_0)$ is in the forbidden region. Therefore, the safest possibility is that both (\ref{ex1}) and (\ref{ex1-b}) are satisfied at the same time.%

We will now show that for a  very broad class of brane potentials $W_B(\f)$, solutions of the junction conditions exist which  can self-tune away any large amount of vacuum energy and satisfy   $0<W_B(\f_0) < W_{IR}(\f_0) - B(\f_0)$.

First, we expect the function $W_B(\f)$  to scale (in units of the UV $AdS$ radius) as
\be\label{ex7}
W_B(\f) = {\Lambda^4 \over M^3} w(\f)
\ee
where  $\Lambda$ is the cut-off on the brane, $M$ the 5-dimensional Planck scale and $w(\f)$ is  a dimensionless function independent of $\Lambda$  (we may expect a  mild dependence on  $\Lambda$ in $w(\f)$  but for simplicity here  we ignore this possibility). For a given class of functions  $w(\f)$, self tuning is successful if:
\begin{enumerate}
\item  a  solution $\f_0 >0 $ to equation (\ref{sol4})  exists without assuming $\Lambda$ to be small or without tuning the parameters in $w(\f)$,
\item the solution satisfies the condition (\ref{ex1}). If in addition it satisfies also  condition (\ref{ex1-b}), the solution  is manifestly ghost-free.
\end{enumerate}

We now argue that, for the two conditions above to be satisfied, it is enough to ask that $w(\f)$ have a zero at some finite $\f=\bar{\f}$, with positive derivative:
\be \label{ex8}
w(\bar{\f}) = 0, \qquad w'(\bar{\f}) >0.
\ee
 As we will show, under these  conditions:
\begin{enumerate}
\item we can always  find an equilibrium position, which sits in the vicinity of $\bar{\f}$,  for any value of $\Lambda$ in   a continuous range
\be \label{ex9}
0 < \Lambda < \Lambda_{max}
\ee
in which both conditions (\ref{ex1}-\ref{ex1-b}) are satisfied.
\item The maximum allowed value $\Lambda_{max}$ scales like $[V(\bar\f)]^{1/8}$. In particular, for a potential which has an  exponential behavior at large $\f$, $\Lambda_{max}$ scales  {\em exponentially} with the position of the zero of $w(\f)$:
\be \label{ex10}
V(\f) \to -V_\infty \exp b\f \quad \Rightarrow \quad \Lambda_{max} \propto \exp {b \bar{\f} \over 8}
\ee
\end{enumerate}

With  a small amount of  tuning in the parameters of $w(\f)$ (which may even be  natural, as we will discuss below) we can achieve $\bar{\f} \gg 1$ and  self-tune away a large vacuum energy.

The argument goes as follows. Suppose $\bar{\f}$ is such that $w(\bar\f) = 0$.  If we go arbitrarily close to $\f = \bar{\f}$, no matter the value of  $\Lambda$ we reach a region where $W_B\ll W_{IR}$. In this region, we can define a function $\epsilon(\f)$ such that:
\be\label{ex11}
W_{B} (\f) = W_{IR} (\f) \epsilon (\f) , \qquad \epsilon (\f) \ll 1,
\ee
Inserting this expression in equation (\ref{sol4}) and linearizing in $\epsilon(\f)$, we find:
\be\label{ex12}
V(\f)\epsilon(\f)  + {1\over 4} \left({d \over d\f} W^2_{IR}\right) \epsilon'(\f) =0,
\ee
where we have used the fact that $W_{IR}$ solves the superpotential equation
 (\ref{superpot-i}). We now expand equation (\ref{ex12}) with respect to $\f$, close to $\bar{\f}$.  Since $\epsilon(\bar{\f}) = 0$ by assumption,
\be\label{ex13}
\epsilon(\f) = \epsilon_1 (\f - \bar{\f}) + {\cal O}\left((\f - \bar{\f})^2\right),  \qquad \epsilon_1 = {\Lambda^4  \over M^3} {w'(\bar{\f}) \over  W_{IR}(\bar{\f })}.
\ee
 Equation (\ref{ex12}) becomes, in this approximation, a linear equation in $\f$ which is solved by setting:
\be \label{ex14}
\f_0 = \bar{\f} - {\de_\f (W^2_{IR}) \over 4V}\Big|_{\f=\bar{\f}}.
\ee
Because we have taken $V<0$ , and because $W_{IR}$ is a monotonically increasing function, then $\f_0 > \bar{\f}$, and $W_B(\f_0) >0$ by our assumption (\ref{ex8}).
Moreover, at $\f_0$, $W_{IR} - W_B \simeq W_{IR} (1 - \epsilon(\f_0) ) > B(\f_0) $ as long as  $\epsilon(\f_0)$ is small. Thus $W_B(\f_0)$ is in the range which satisfies equations (\ref{ex1}-\ref{ex1-b}).

Thus we have shown that an acceptable equilibrium solution exists for any value of   $\Lambda$,  {\em provided} $\epsilon(\f_0) \lesssim 1$. This condition however does depend on $\Lambda$, since $\epsilon = W_B / W =  \Lambda^4 w / W$. On the other hand the value of the equilibrium position (\ref{ex14}) is independent of $\Lambda$, since we have assumed that $\bar{\f}$ itself does not depend on $\Lambda$.  Thus, if $\Lambda$ is too large, the value $\f_0$ in equation (\ref{ex14})  will fall outside the region where $\epsilon$ is small, and the argument will break down. We can put a bound on how large a  $\Lambda$ we can sustain by using the linearized approximation for $\epsilon$ around $\bar{\f}$, where:
\be \label{ex15}
\epsilon(\f_0) \simeq \epsilon_1 (\f_0 - \bar{\f}) = {\Lambda^4 \over M^3} {w'(\bar{\f}) \over W_{IR}} {\de_\f (W^2_{IR}) \over (-4V)}\Big|_{\f=\bar{\f}}.
\ee
Thus,  the condition $\epsilon \lesssim 1$ translates into a condition for $\Lambda$. If we suppose $\bar{\f} \gg 1$, enough to be in the asymptotic exponential region for the potential, and use equations (\ref{ex5}-\ref{ex6}), the condition is:
\be\label{ex16}
\Lambda \lesssim \Lambda_{max}  = C \, {\exp ({b \bar{\f}/8}) \over  (w'(\bar{\f}))^{1/4}} , \qquad C \equiv \left({M^6 V_\infty \over 8}\, {(2Q)^2 -b^2 \over b^2}\right)^{1/8}.
\ee
Thus $\Lambda_{max}$ can be made very large if the zero of $w(\f)$ sits at large $\f$.

\subsection{Concrete examples} \label{concrete}
To end this section, we will present two concrete examples where the self-tuning mechanism is at work. They are presented for illustrative purposes and we will not try to develop them into fully phenomenologically acceptable models.  Constructing a model that satisfies all the stability constraints and leads to an acceptable phenomenology is beyond the scope of this paper, although all the phenomenological requirements will be however explicitly specified in sections 4 and 5,  and will further constrain model-building beyond what we discuss in the two examples below.

Here we will limit ourselves to showing that stable, self-tuning models with arbitrary large $4d$ vacuum energy do exist.

In Section 6.1 we present an example  based on a polynomial bulk potential. In this case the IR is non-singular, but at the equilibrium solution the brane has negative tension. Thus equation (\ref{ex1-b}) is violated and one cannot exclude the presence of ghost scalar modes, and further analysis is required.

In Section 6.2 on the other hand we will show a class of ``safe'' self-tuning  models that satisfies both requirements (\ref{ex1}-\ref{ex1-b}) for a very large range of parameters, including very large contributions to the (bare) vacuum energy. These are based on the existence of a solution close to a zero of $W(\f)$, discussed in the previous subsection\footnote{The main difficulty in constructing  a manifestly ghost-free self-tuning model, i.e. one that satisfies $0< W(\f_0) < W_{IR}(\f_0) - B(\f_0)$, is that the special solution $W_{IR}(\f)$  has the same scaling behavior as  $B(\f)$ for large $\f$, and all solution lying between $W_{IR}$ and $B(\f)$ follow very closely $W_{IR}$, then eventually  stop (bounce) before reaching infinity. Thus, the strip  $0< W_B(\f_0) < W_{IR}(\f_0) - B(\f_0)$ is very narrow. However,  the class of models we present in section \ref{concrete-2} admits generically at least one solution falling in this strip.}

\subsubsection{Case study I: an IR-regular model} \label{concrete-1}

To construct a model which is free of bulk singularities,   we consider  a potential with three extrema at $\f=0, \pm v$:
\be \label{ex2}
V(\f) = -12 + {1\over 2} \bigg(\f^2 - v^2\bigg)^2 - {v^4\over 2},
\ee
Each extremum supports an $AdS$ fixed point, and we will be interested in flows that start in the UV at $\f=0$ and end in the IR at $\f=v$.  We have set the to unity the UV $AdS$ radius. For concreteness we also set $v=1$.  The regular IR solution is the one that flows to the $AdS$ fixed point  at $\f=1$. We take the brane potential to be:
\be\label{ex3}
W_b(\f) = \omega \exp[\gamma \f].
\ee

\begin{figure}[h!]
\begin{center}
\includegraphics[width=12cm]{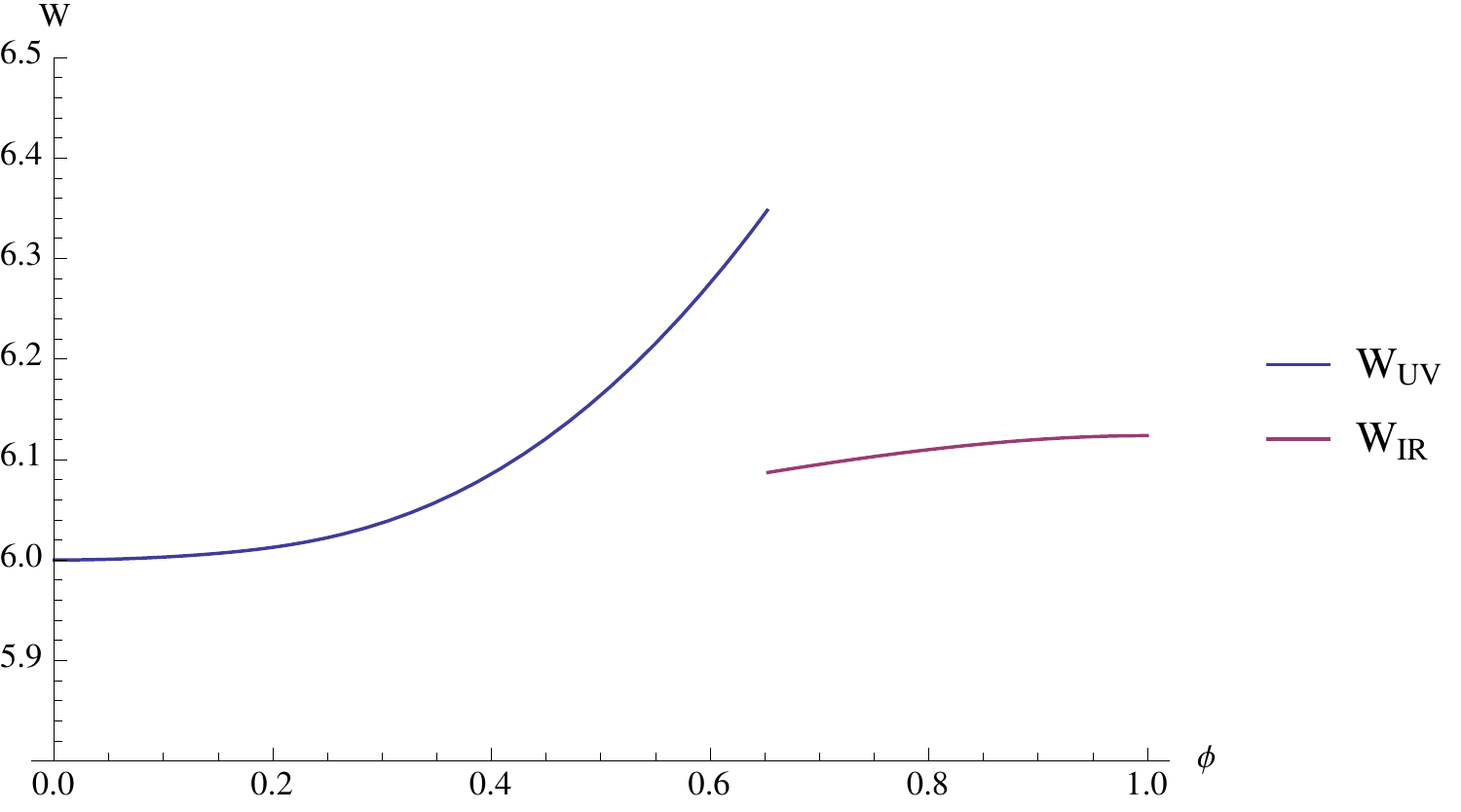}
\caption{The superpotential from the UV ($\f=0$) to the IR ($\f=1$). The UV-IR matching is at $\f_0 = 0.65$. On the vertical axis,  the superpotential is measured in units of the UV $AdS$ radius.} \label{superfig}
\end{center}
\end{figure}

\begin{figure}[h!]
\begin{center}
\includegraphics[width=12cm]{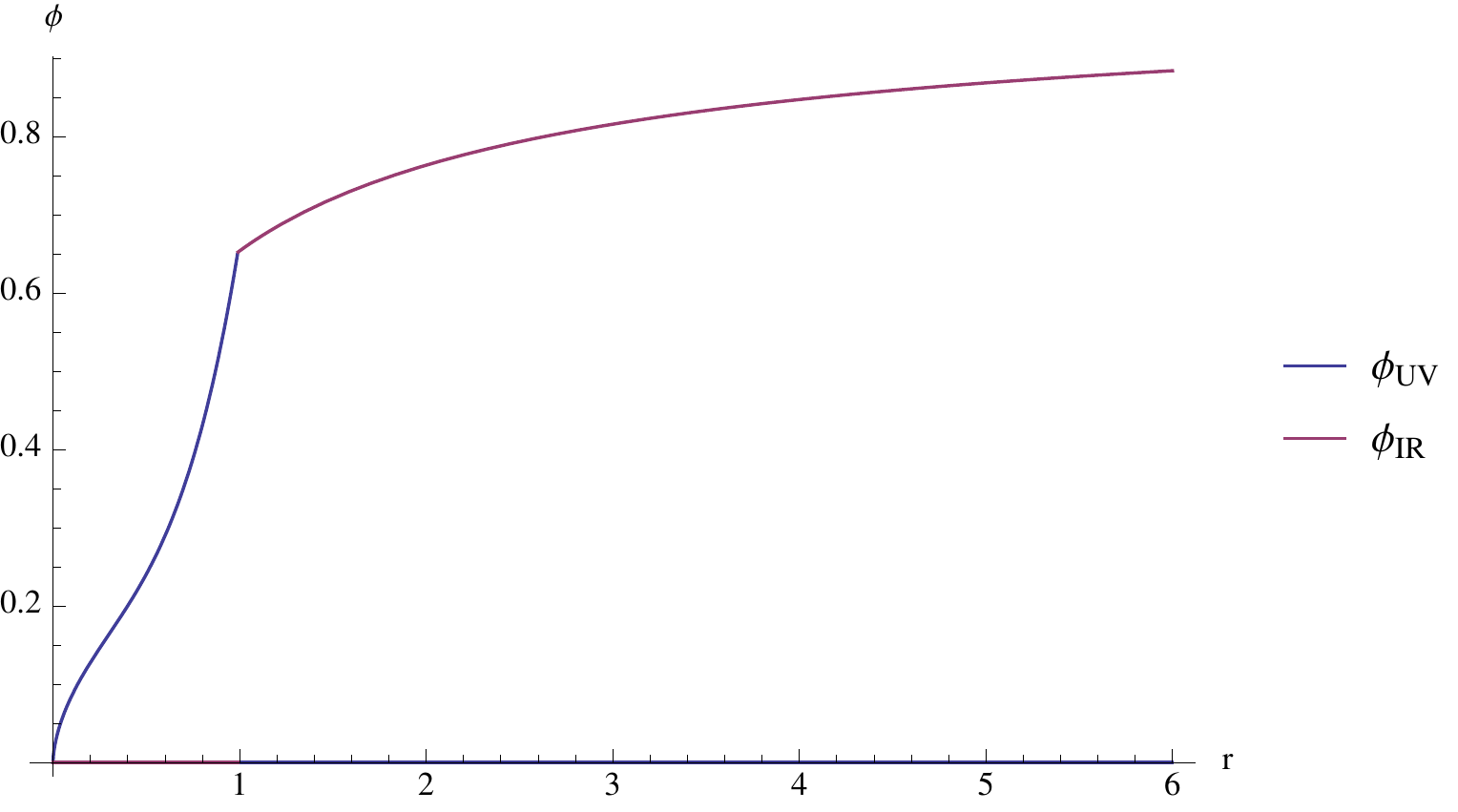}
\caption{The scalar field profile as a function of the holographic radial coordinate. On the horizontal axis,  we have used the conformal coordinate $r = \int e^{-A}du$. In this coordinate,  the $AdS$  boundary is at $r=0$, and the (regular) IR horizon,  is reached as $r\to +\infty$, where $\f(r)$ approaches the fixed point value  $\f_{IR}=1$.   The solution is obtained by fixing the  initial condition at the interface, $A(\f_0) = 0$. The position of the interface is  $r_0 = 0.99$.} \label{scalar}
\end{center}
\end{figure}

Depending on the values of $\omega$ and $\gamma$ one may find solutions to the matching conditions in the interval $0< \f < 1$. For instance, we use the following values:
\be\label{ex4}
\omega = -0.01, \; \gamma = 5 \qquad \Rightarrow \qquad \f_0 = 0.65.
\ee
The superpotential is displayed in Figure \ref{superfig}, and the scalar field profile in figure (\ref{scalar}).
In order for the matching condition (\ref{sol4}) to have a solution in the range $(0,1)$, with this model we need $w_B$ negative, which is not manifestly ghost-free. Thus, although it is free of bulk singularities, before using this model for phenomenology one has to check the absence of ghosts explicitly by computing the spectrum of scalar fluctuations.  This problem will be absent in the class of models discussed in the next subsection.

\subsubsection{Case study II: a class of stable self-tuning models} \label{concrete-2}

The main problem with the model in the previous section is that $\f$ has a finite range between the UV and the IR, and all solutions overshooting $\f=1$ have bad singularities. Hence, the equilibrium condition (\ref{sol4}) may not have solutions in this range, or the solution may violate the condition (\ref{ex1}). Here we present a class of models which does not have this shortcomings, and such that the junction is self-tuning and manifestly ghost free for a wide range of parameters.

We investigate a model in which $\f$ has infinite range: the UV fixed point is still at $\f=0$ but the IR is reached as  $\f \to +\infty$. The potential we choose is: 
\be \label{ex5}
V(\f) = -12 - \left({\Delta(4-\Delta)\over2 } - {b^2\over 4}\right) \f^2 - V_1 \sinh^2 {b\f \over 2}
\ee
where we have set $\ell_{UV}=1$ and $d=4$.
This potential is monotonically decreasing,  behaves at large $\f$  as $-(V_1/4)\exp(b\f)$, and it supports solutions with  acceptable singularities for $b< 2\sqrt{2/3}$ since in $d=4$, $Q = \sqrt{2/3}$.

The coefficient  of the quadratic term was chosen so that $m^2 = \Delta(\Delta-4)$ and the dimension of the dual operator is manifest. For definiteness, we choose:
\be \label{ex18}
b = {Q\over 4}= {1\over4}\sqrt{2\over 3} , \quad \Delta = 3, \quad V_1 = 1,
\ee
but  their precise values are not important (as long as $b< 2Q$,  $2< \Delta<4$ and $V_1>0$.  With such potential, all solutions to the superpotential equations that extend to $+\infty$ are singular, and the special solution with a good singularity has asymptotics:
\be\label{ex6}
W_{IR} \simeq \sqrt{2\over (2Q)^2 - b^2} \exp {b\f \over 2}, \qquad \f \to +\infty.
\ee

As an example which follows  the logic described in section \ref{types},  consider a polynomial function $W_B(\f)$ which has at least one zero at $\bar{\f} >0$, with positive derivative\footnote{For simplicity, here we have suppressed the bulk Planck scale $M$ compared to equation (\ref{ex7}). It can be reistated by letting $\Lambda^4 \to \Lambda^4/M^3$. $\Lambda$ and $M$  are dimensionless since we have set $\ell_{UV}=1$.
}:
\be\label{ex17}
W_B(\f) = \Lambda^4 \left[-1 - {\f \over s} +  \left({\f \over s}\right)^2  \right].
\ee
The parameter $s$ controls the position of the zero. Based on the discussion in Section \ref{types}, we expect to find a stable  solution in the vicinity of $\bar{\f}$, for all values of $\Lambda$ up to a maximum value  on $s$.

\begin{figure}[h!]
\begin{center}
\includegraphics[width=10cm]{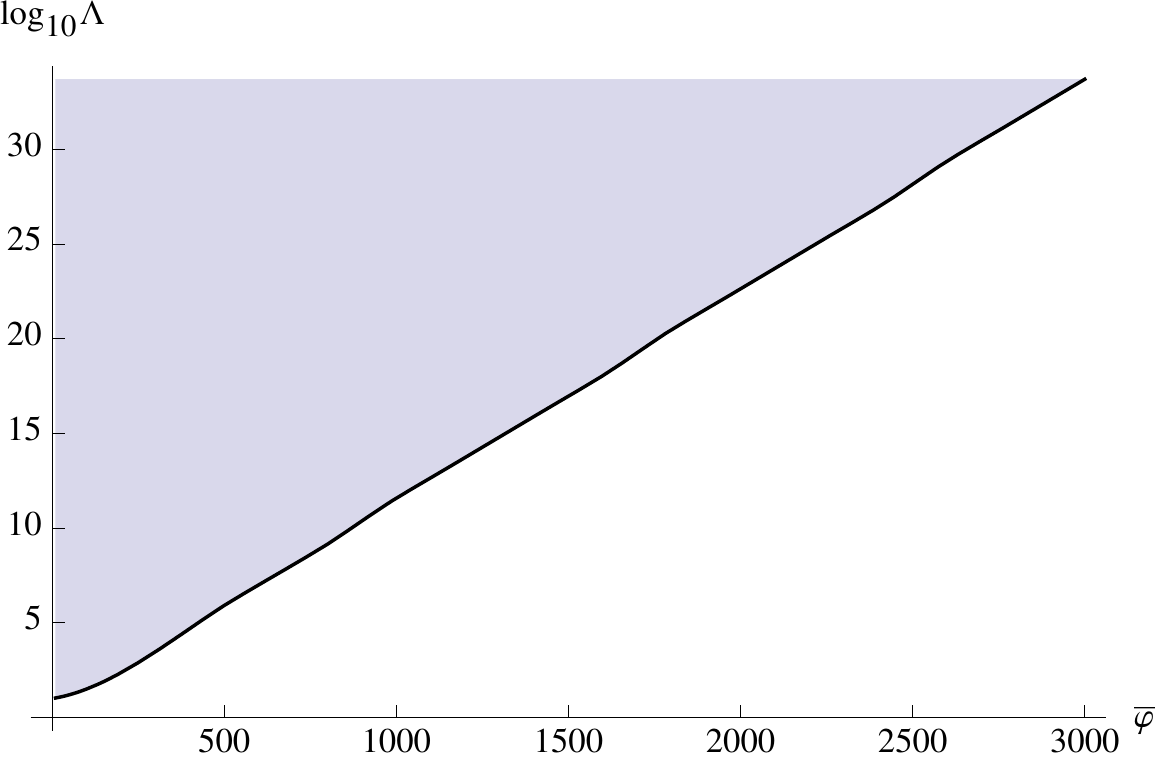}
\caption{The  unshaded part of the graph is the region of parameter space $(\Lambda, \bar{\f})$ where a flat, manifestly stable (i.e. with $W_B(\f_0) >0$)  solution to the junction condition exists. This graph was obtained numerically using the bulk potential of the form (\protect\ref{ex5}) with parameters given in equation (\protect\ref{ex18}), and brane potential of the form (\protect\ref{ex17}).} \label{lambda}
\end{center}
\end{figure}

This is indeed what happens: we have solved numerically the superpotential equation for $W_{IR}$,  imposing IR asymptotics as in (\ref{ex6}),
and solved the equilibrium condition for $s$ in a range between 10 and  2000, for which  $\bar{\f}$ ranges between 15 and 3200. We always find an acceptable (stable) equilibrium position close to $\bar{\f}$, for values of $\Lambda$ outside the shaded region in figure \ref{lambda}. We see from that figure that, for a given value of $\bar\f$, stable flat solutions exit for large ranges of  the vacuum energy scale  $\Lambda$: the latter  can     be as large as $10^{30}$ for $\bar{\f}\sim 2500$ (which a tuning of $s$ of only 1 in  $10^{3}$).

\section{Linear perturbations around flat solutions}

In order to assess the phenomenological viability of the framework we developed in the previous section, it is crucial to address two important points:
\begin{enumerate}
\item The model must reproduce the standard four-dimensional gravitational interaction between matter sources, at least in a broad enough range of distance scales to be compatible with observation.
\item The vacuum equilibrium solution (flat defect at a fixed bulk position $u_0$) must be stable.
\end{enumerate}
In this and the following two sections  we will address these points at the level of linearized perturbations around the vacuum solution.

Although the analysis of bulk linear perturbations in general Einstein-dilaton theories is known (see for example \cite{peloso} for  a discussion), to be self-contained  we briefly reproduce it in this work. On the other hand, a full treatment of linear fluctuations in an asymmetric brane-world with a general brane action like the one in (\ref{A3}) has not previously appeared (to the best of our knowledge). The detailed  calculations are presented in appendix \ref{app pert}. Here we report the main definitions and final results.

\subsection{Bulk perturbations}

To set up the study of linear perturbations, it is convenient to  work  in conformal coordinates,  such that:
\be
ds^2 = a^2(r)\left(dr^2 + \eta_{\mu\nu}dx^\mu dx^\nu\right) \sp  a(r)dr =du\sp  a(r) = e^{A(u(r))}\;.
\label{coord} \ee
 A prime will denote derivative with respect to $r$, a dot with respect to $u$. Einstein's equations and the junction conditions in these coordinates can be found in Appendix \ref{app pert}.

We  introduce perturbations of  the metric and scalar field,  on each side of the brane, in the form:
\be\label{pert1}
ds^2 = a^2(r)\left[(1+2\phi) dr^2 + 2A_\mu dx^\mu dr + (\eta_{\mu\nu} + h_{\mu\nu}) dx^\mu dx^\nu\right], \quad \varphi = \bar{\varphi}(r) + \chi
\ee
where the fields $\phi, A_\mu, h_{\mu\nu}, \chi$ depend on $r,x_\mu$ and are treated  as small quantities.
We  further decompose the 5 dimensional bulk modes into tensor, vector and scalar perturbations with respect to the 4 dimensional diffeomorphism group,
\be\label{pert3}
A_\mu = \de_\mu B + A^T_\mu, \quad h_{\mu\nu} = 2\eta_{\mu\nu}\psi + 2\de_\mu\de_\nu E + 2 \de_{\left(\mu\right. } V^T_{\left.\nu \right)} + \hat{h}_{\mu\nu}
\ee
with $\de^\mu A_\mu^T = \de^\mu V^T_\mu = \de^\mu \hat{h}_{\mu\nu} = \hat{h}^\mu_\mu=0$.  All indices $\mu,\nu$ are raised and lowered with the flat Minkowski metric $\eta_{\mu\nu}$.

Therefore, we have one bulk tensor $\hat{h}_{\mu\nu}$, two bulk transverse vectors $(A_\mu^T, V_\mu^T)$, five bulk scalars $(\phi, \psi, \chi, B, E)$ (plus one brane scalar, describing brane bending as  we will see later). At the linearized level, general coordinate transformations $(\delta r,\delta x^\mu) = (\xi^5,  g^{\mu\nu}\xi_\nu)$  act as  gauge transformations, under which:
\bea
&& \delta \psi = - {a'\over a} \xi^5 \qquad \delta \phi =-(\xi^5)'  - {a'\over a} \xi^5 \nonumber \\
&& \nonumber \\
&& \delta B = -\xi' - \xi^5, \qquad \delta E = - \xi,  \qquad  \delta \chi = - \fb' \xi^5, \label{pert4}  \\
&&  \nonumber \\
&& \delta A_\mu^T = - (\xi^T_\mu)', \qquad \delta V_\mu^T = - \xi_\mu^T 
\nonumber \\
&& \nonumber\\
&& \delta \hat{h}_{\mu\nu} = 0
\eea
where we have introduced a decomposition of the diffeomorphism parameter $\xi_\mu$ in its transverse and longitudinal components, i.e.  $\xi_\mu = \xi_\mu^T + \de_\mu \xi $ with $\de^\mu \xi_\mu^T=0$.

The tensor mode $\hat{h}_{\mu\nu}$ is gauge-invariant, and gauge symmetry plus constraints allow to eliminate the two  vectors and four of the bulk scalars. The remaining physical bulk scalar can be identified with the gauge-invariant combination:
\be\label{pert3a}
\zeta = \psi - {1 \over z} \chi,
\ee
where $z(r)$ is the background quantity:
\be \label{pert3b}
z \equiv {a \bar\f'\over a'}.
\ee
The bulk gauge-invariant fluctuations satisfy the second order equations:
\bea
&&\hat{h}''_{\mu\nu} + (d-1){a'\over a} \hat{h}'_{\mu\nu} + \de^\rho\de_\rho  \hat{h}_{\mu\nu} = 0 \label{pert4-1} \\
&& \nonumber \\
&& \zeta'' + \left[(d-1){a'\over a} + 2 {z'\over z}\right] \zeta'  + \de^\rho\de_\rho  \zeta = 0 .\label{pert4-2}
\eea
These two equations must be solved independently on each side of the interface, and the solutions must be glued using the linearized junction conditions, to the discussion of which we now turn.

\subsection{Brane perturbations and first junction condition}

We consider the perturbations in the brane position, which adds one more scalar perturbation which is localized on the interface and couples to the bulk scalar modes via the perturbed Israel matching conditions (\ref{FE5}-\ref{FE6}).

The brane is described by an embedding $X^A(\sigma^\alpha)$ where $X^A= (r, x^\mu)$ and $\sigma^\alpha$ are world-volume coordinates. We choose the gauge $\sigma^\alpha = x^\mu \delta^{\alpha}_\mu$, so the embedding is completely specified  by the radial profile $r(x^\mu)$. We consider a small deviation from the equilibrium position $r_0$:
\be\label{pert2}
r(x^\mu) = r_0 + \rho(x^\mu)
\ee
The  brane scalar mode $\rho$  represents brane bending. This is an additional scalar mode which only has dynamics in the tangential directions to the brane, and no bulk dynamics.

The fluctuations on the brane are the brane bending mode plus those induced by the bulk perturbations. In particular, the induced metric and scalar field on the brane are given by:
\be\label{jun1}
\gamma_{\mu\nu} = a_0^2 \left(\eta_{\mu\nu} + h_{\mu\nu} + 2{a'_0\over a_0} \eta_{\mu\nu}\rho \right),  \quad \f  = \bar{\f}_0 + \chi + \bar\f'_0 \rho
\ee
Above,  and  in what follows, a subscript $0$ refers to background quantities evaluated at the unperturbed brane position $r_0$. The last term in each of the two equations above, comes from expanding the background solution around the equilibrium position $r_0$, using equation (\ref{pert2}).

The first junction conditions (\ref{FE3})  (i.e. the continuity conditions for the scalar field and induced metric),  are given in terms of the scalar and tensor perturbations by:
\be \label{jun3}
\Big[\psi +{a'\over a} \rho\Big]^{UV}_{IR} = 0, \quad  \Big[\chi + \fb_0'\rho \Big]^{IR}_{UV} =0, \quad \Big[E \Big]^{UV}_{IR} = 0,   \qquad \Big[\hat{h}_{\mu\nu}\Big]^{IR}_{UV} =0.
\ee
Notice that if $\rho\neq 0$, neither $\psi$ nor $\chi$ are continuous and neither is the gauge-invariant variable $\zeta$ defined in equation (\ref{pert3a}).  In fact, its jump equation is also gauge-invariant, as we will discuss below, after introducing gauge-invariant brane quantities.

\subsection{Gauge fixing   and second junction condition}
Under the linearized diffeomorphisms (\ref{pert4}) , the brane-bending mode transforms as:
\be \label{pert2-a}
\delta \rho(x) = \xi^5(r_0,x).
\ee
It is useful to introduce the new bulk scalar variables:
\be\label{jun2}
\hat{\psi}(r,x) = \psi + {a'(r)\over a(r)} \rho(x),  \quad \hat{\chi}(r,x) = \chi + \fb'(r) \rho(x),
\ee
These variables are continuous across the interface, as it is clear from equation (\ref{jun3}), and their value at the brane is gauge invariant, as can be seen by combining  the transformations (\ref{pert4}) and (\ref{pert2-a}):
\be \label{jun4}
\Big[\hat\psi\Big]^{UV}_{IR} =  \Big[\hat\chi \Big]^{IR}_{UV} =0 , \qquad \delta \hat\psi(r_0) = \delta \hat\chi(r_0)= 0 .
\ee
Therefore, on the brane we have one gauge-invariant induced tensor mode $h_{\mu\nu}(r_0)$ and {\em two} gauge-invariant scalar modes $\hat{\psi}(r_0), \hat{\chi}(r_0)$: we can think of one of them as induced from the bulk mode $\zeta$, and the other as being the invariant  version of the brane-bending mode. The jump condition  for $\zeta$ can be read-off from its definition (\ref{pert3a}) and  equations (\ref{jun2}-\ref{jun4}):
\be \label{jun5}
\Big[\zeta\Big]^{UV}_{IR} = \Big[{a'\over a \fb'}\Big]^{UV}_{IR} \hat{\chi}(r_0).
\ee
The fact that both $\zeta$ and its derivative (as we will see shortly) are discontinuous across the brane makes the dynamics of scalar perturbation more complex to treat  than for tensor modes. The latter  are  continuous with discontinuous derivatives, a situation which can be  described in the standard way  with a $\delta$-function potential at the interface.

To proceed further, it is convenient to fix the gauge completely.  If one wishes, one could choose the gauge $\rho=0$, in which the brane sits at its unperturbed position\footnote{The brane  is not flat though, since the induced metric still receives contributions from the bulk fluctuations, as it is clear from equation (\ref{jun1}).} and the fields $\psi$ and $\chi$ are continuous. This however does not make the bulk variable zeta continuous, because equation (\ref{jun5}) is gauge-invariant.

On the other hand, it is useful to work in a gauge which maximally simplifies the equations  in the bulk. We make the gauge choice:
\be\label{jun6}
\chi = B = 0.
\ee
In this gauge one can solve the bulk constraint equations for $\phi$  and $E$ in favor or $\psi$, and be left with the only bulk fluctuation $\psi$, which coincides with $\zeta$.

The full derivation of the remaining  junction conditions is detailed in appendix \ref{app pert}. As a result, we summarize below the bulk equations and matching conditions for tensor and scalar modes, in the absence of brane sources (these will be added in the next subsections). Below we specialize to the physical value $d=4$. Here and in the rest of the paper we define:
\be \label{uzero}
U_0 \equiv U_B(\f_0), \quad Z_0 \equiv Z_B(\f_0).
\ee
\paragraph{Tensor modes}
We have,
\bea
&&\hat{h}''_{\mu\nu} + 3{a'\over a} \hat{h}'_{\mu\nu} + \de^\rho\de_\rho  \hat{h}_{\mu\nu} = 0 \label{jun7} \\
&& \nonumber \\
&& \Big[\hat{h}_{\mu\nu}\Big]^{IR}_{UV} =0, \qquad a_0 \left[\hat{h}'_{\mu\nu}\right]^{IR}_{UV} =  -  U_0 \de^\rho \de_\rho \hat{h}_{\mu\nu}(r_0). \label{jun8}
\eea
where $U_0 \equiv U_B(\f_0)$.
Equation (\ref{jun7}) must be solved independently on the UV and IR sides of  the interface, and the results matched using the junction conditions (\ref{jun8}). The continuity condition was obtained in the previous subsection. For the derivation of the second junction condition,  see  Appendix \ref{app tensor}.
\paragraph{Scalar modes}
With the gauge choice (\ref{jun6}), the bulk equations and junction conditions can be written in closed form in terms of  the variable  $\psi$ and the brane-bending mode $\rho$ (for a detailed  derivation, see  Appendix \ref{app scalar}):
\bea
&&
\psi'' + \left(3{a'\over a} + 2{z'\over z}\right) \psi' + \de^\mu\de_\mu \psi = 0, \label{jun9}\\
&& \nonumber \\
&&\nonumber \\
&&\Big[ \fb' \rho \Big]_{UV}^{IR} = 0\,  , \qquad  \Big[\psi\Big]_{UV}^{IR} = -\left[{1\over z}\right]_{UV}^{IR} \fb' \rho \, ; \label{jun10}\\
&& \nonumber \\
&& \nonumber \\
&&
\left[{z^2 \over 6} \psi'\right]_{UV}^{IR} =  \left( {2U_0 \over a_0} - \Big[{a \over a'}\Big]_{UV}^{IR}\right)  \de^\mu \de_\mu \left(\psi + {a'\over a} \rho\right) +  {1\over a_0}\left({dU_B\over d\f} \right)_0\bar\f'\de^\mu\de_\mu \rho ; \label{jun11}\\
&& \nonumber \\
&&
\Big[z \psi'\Big]_{UV}^{IR} =- 6 {dU_B\over d\f}(\f_0)  \de^\mu \de_\mu\left(\psi + {a'\over a} \rho\right) + \left( {Z_0 \over a_0} \de^\mu \de_\mu - \tilde{{\cal M}_b}^2 \right) \fb'\rho\, ; \label{jun12}
\eea
where:
\be
  z \equiv {a\fb' \over a'} , \qquad  \tilde{{\cal M}_b}^2 = a_0 \left({d^2 W_B \over d\f^2}(\f_0) - \left[ {d^2 W \over d\f^2 }\right]^{IR}_{UV} \right)\, . \label{jun13}
\ee
All quantities on the right-hand side are evaluated at the unperturbed brane position. Notice that the perturbations enter on the right hand side only in the continous combinations $\hat{\psi}(r_0)$ and $\hat{\chi}(r_0)$, defined in equation (\ref{jun2}). On the other hand, the junction conditions above imply that both $\psi$ and $\psi'$ are discontinuous at the interface.

To simplify the junction conditions, we can  eliminate $\rho$ on the right-hand side of equations (\ref{jun11}-\ref{jun12}), in favor of the jump in $\psi$, by solving    equation (\ref{jun10}):
\be\label{scalar13.5-sec}
 \qquad \fb'(r_0) \rho = -{[\psi] \over [1/z]}.
\ee
Using this result, it is easy to show from the definition (\ref{jun2})  that:
\be\label{scalar13.6-sec}
\hat{\psi}(r_0) = {[z\, \psi] \over [z]}.
\ee
In the two equations above,  and in the ones that follows, we use the notation $[\;\;]$ as a shorthand for $[\;\;]^{IR}_{UV}$ to indicate the jump of a quantity across the interface.

Using these  results,  equations (\ref{jun11}-\ref{jun12}) become relation between the left and right functions and their derivatives:
\bea
&& \left[z \psi'\right]= -\left({6\over a_0} {d U_B \over d\f}\Big|_{\f_0}\right) \Box {[z \, \psi] \over [z]} - {1\over a_0}\left( Z_0 \Box  - a_0^2 {\tilde {\cal M}}^2 \right){[ \psi] \over [z^{-1}]} \label{sm1-sec3} \\
&&  \left[z^2 \psi'\right] =  6 \left( 2 {U_0 \over a_0}- \left[a\over a'\right]  \right) \Box {[z \,\psi] \over [z]} \label{sm2-sec3} - \left({6\over a_0} {dU\over d\f}\Big|_{\f_0}\right) \Box {[ \psi] \over [z^{-1}]}
\eea

An efficient  way to deal with this kind of boundary value problem will be developed in Section \ref{sec scalar}.

\section{Tensor perturbations and induced gravity}

We now concentrate on the analysis of the tensor modes, whose dynamics are described by equations (\ref{jun7}-\ref{jun8})  and investigate to what extent these can reproduce the standard observed gravitational interaction between brane matter sources.

In theories with extra dimensions, one way to obtain four-dimensional gravity is to have normalizable  massless (i.e. satisfying  $\de^\mu\de_\mu \hat{h}_{\rho\sigma}= 0$)  modes. However, the system (\ref{jun7}-\ref{jun8}), generically,  does not admit such normalizable massless modes, unless the infrared is IR-incomplete \cite{gravitons,superp}. This is intuitively clear  because the  volume of the bulk is infinite on the UV side. This is unlike what happens e.g. in the RS model, where the asymptotic region containing the AdS boundary is cut-off, and the volume is finite, allowing for zero-modes\footnote{For the sake of completeness, we note that one {\em can} have normalizable zero modes in infinite volume if one accepts  IR-incomplete singularities \cite{gravitons}. However, even in this case one needs to impose fine-tune boundary conditions at the singularity. We will not consider this possibility further.}.

An alternative way that four-dimensional gravity on a brane can arise from a higher-dimensional theory is through the DGP mechanism of brane-induced gravity \cite{DGP}, in which the gravitational interaction is the result of the superposition of infinitely many bulk modes which give rise to a long-lived quasi-localized resonance on the brane. In this case, the standard gravitational interaction is reproduced at short distances.

In this model we have the right ingredients for the  induced gravity mechanism to be at work: indeed, the junction conditions for tensor modes (\ref{jun8}),
are    of the same form as in the DGP model in flat space\footnote{In the DGP scenario in flat bulk space, $U_0$ must be hierarchically large compared with the bulk Planck scale. This is difficult to achieve in controlled frameworks like string theory, \cite{ktt,a}.}. Moreover, at distances shorter than the bulk curvature scale, the second term in the bulk equation (\ref{jun7}) can be neglected, therefore we can expect the DGP mechanism to work as in flat space. In the rest of this section we will explicitly confirm this expectation, and discuss in detail how the gravitational interaction is reproduced (and modified) at different scales.

To investigate the gravitational interaction between  brane sources, we must introduce brane-localized matter, which we assume couples  to the induced metric and possibly to the dilaton field at the brane:
\be\label{ind3-ii}
S_m = \int d^dx \, \sqrt{\gamma} {\cal L}_m(\gamma_{\mu\nu}, \psi_i,\f_0)
\ee
where $\psi_i$ denotes collectively the matter fields (which we assume to be trivial in the vacuum). The matter stress tensor is then:
\be\label{ind3-iii}
T_{\mu\nu}(x) = -{2\over \sqrt{\gamma}} {\delta S_m \over \delta \gamma^{\mu\nu}(x)}.
\ee
We assume the matter stress tensor to be conserved in the vacuum, $\de^\mu T_{\mu\nu}=0$. The junction conditions including the stress tensor as a source are (see Appendix \ref{app tensor}):
\vspace{0.5cm}
\be \label{ind3-vii}
\left[\hat{h}_{\mu\nu}\right]^{IR}_{UV}=0 \sp \left[\hat{h}'_{\mu\nu}\right]^{IR}_{UV} =  -  e^{-A_0}U_0 \de^\rho \de_\rho \hat{h}_{\mu\nu} - e^{-A_0}{1\over M^3}\hat{T}_{\mu\nu},
\ee
\vspace{0.5cm}
where the source is given by
\be\label{ind3-viii}
\hat{T}_{\mu\nu} = T_{\mu\nu} - {1\over 3}\eta_{\mu\nu} T +  {1\over 3}{\de_\mu\de_\nu\over \de^2} T, \qquad T \equiv \eta^{\mu\nu}T_{\mu\nu},
\ee
and it is conserved and traceless, $\de^\mu \hat{T}_{\mu\nu} = \eta^{\mu\nu}\hat{T}_{\mu\nu} = 0$.

The field equation  (\ref{jun7}) and the matching condition (\ref{ind3-vii}) can be obtained by varying the following quadratic action:
\bea\label{ind3-iv}
S && =  - {1\over 4} M^{3} \int d^dx dr e^{3A(r)}\left[ (\de_r \hat{h})^2 + (\de_\mu \hat{h})^2 \right]  \,\,  -  \, \, {1\over 4} M^{3}U_0 e^{2A_0} \int_{r=r_0} d^dx  (\de_\mu \hat{h})^2 \nonumber \\
&& -{1\over 2} \,\,  e^{2A_0} \int_{r=r_0} d^dx \, \hat{T}(x) \hat{h}(x),
\eea
in which we are temporarily suppressing the tensor indices. The overall coefficient in equation (\ref{ind3-iv}) was fixed by matching it to the    $h\Box h$ term in the quadratic expansion of the Einstein-Hilbert terms in the action\footnote{For  a  tensor perturbations $\hat{h}$ around flat space we have, to quadratic order and up to total derivatives,
$$ \sqrt{-g} R \simeq -{1\over 2} \Box \hat{h} - {1\over4} \hat{h} \Box \hat{h}. $$  } (\ref{A2}-\ref{A3}), or equivalently by matching the linearised matter coupling as defined by equation (\ref{ind3-iii}).

The field equation resulting from the action (\ref{ind3-iv}) is:
\be\label{ind3-v}
\de_r \left(e^{3A(r)} \de_r \hat{h}\right) + \Big[e^{3A(r)} +  U_0 e^{2A_0} \delta(r-r_0) \Big] \de_\mu \de^\mu \hat{h} =  \delta(r-r_0) {e^{2A_0} \over M^{3}} \hat{T}
\ee
In this way we have  written  in compact form both the bulk field equation and the junction conditions in a single equation.
In order to solve it, we use the same procedure followed in \cite{DGP} in flat space and in \cite{ktt2} in AdS: we  define a scalar Green's function $G(r,x;r',x')$, such that:
\bea \label{ind4}
&&\left[\de_r e^{3A(r)} \de_r  + \Big[e^{3A(r)} +  U_0 e^{2A_0} \delta(r-r_0) \Big] \de_\mu \de^\mu \right] G(r,x;r',x') \nonumber \\ && = \delta(r-r_0) \delta^{d}(x-x').
\eea
Then, the solution of equation (\ref{ind3-v}) is given by:
\be \label{ind4-ii}
\hat{h}_{\mu\nu}(x,r) =  {e^{2A_0} \over M^{3}} \int d^dx'\,   G(r,x;r_0,x')  \hat{T}_{\mu\nu}(x'),
\ee
where we have reinstated the tensor indices. The interaction mediated between brane-localized sources is found by inserting the expression (\ref{ind4-ii}) back into the action (\ref{ind3-iv}):
\be \label{ind4-iii}
S_{int} = -{e^{4A_0}\over 4 M^3} \int d^4x \, d^4x' \,  G(r_0,x; r_0, x')\left(T^{\mu\nu}(x) T_{\mu\nu}(x') - {1\over 3} T(x) T(x')\right)
\ee
where we have used the transverse and traceless property of $\hat{h}_{\mu\nu}$. Notice that the tensor structure is appropriate for the exchange  of a massive tensor mode. In the next subsection we will describe in detail the generic qualitative behavior of the brane-to-brane Green's function appearing in equation (\ref{ind4-iii}).

Notice that equation (\ref{ind4-iii}) does not yet describe the gravitational interaction measured by flat-space observers on the brane, since the induced metric on the brane is $\gamma_{\mu\nu} = e^{2A_0}\eta_{\mu\nu}$. Therefore, the distance measured by  the $x$-coordinate is not the one measured on the brane with the standard  flat metric $ds^2=dx\cdot dx$. To translate equation  (\ref{ind4-iii}) into standard coordinates, one must write it in a covariant way and carefully keep track of all the constant warp factors. This final step will be performed in the next section, where we identify, among other things, the effective four-dimensional Planck scale.

\subsection{The bulk and brane propagators}

To find the Green's function appearing in equation (\ref{ind4-iii}),  we Fourier transform $G(r,x^\mu; r_0, 0)$ with respect to $x^\mu$ to $\tilde{G}(r, p; r_0) $,   change  $\de^\mu \de_\mu \to -p^2$ in equation (\ref{ind4}) and look for  a solution of the form:
\be\label{ind5}
\tilde{G}(r, p; r_0) =  D(p,r)B(p)
\ee
where  $D(p,r)$ is the bulk Green's function, and it solves the equation:
\be\label{ind6}
\left[\de_r e^{3A(r)} \de_r   - e^{3A(r)} p^2 \right] D(p,r)= -\delta(r-r_0)e^{3 A_0} \,\, .
\ee
This equation must be solved imposing normalizable boundary conditions at the UV and IR ends of the radial direction, so that the perturbation (\ref{ind4-ii}) represents a {\em state} in the theory (an excitation above the vacuum) and not a change of the theory itself by a UV deformation. IR normalizability on the other hand amounts to a regularity requirement for the perturbation.

Inserting the ansatz (\ref{ind5}) into (\ref{ind4}),  and using (\ref{ind6}),  we find an algebraic equation for $B(p)$, whose solution is:
\be\label{ind7}
B(p) = -{ e^{-3A_0} \over 1 \,  +\,   \left[e^{-A_0} U_0\, D(p,r_0)\right] \, p^2}.
\ee
Inserting the above result in equation (\ref{ind5}), we obtain the  brane-to-brane propagator in momentum space:  
\be\label{ind8}
\tilde{G}(r_0,p;r_0) = -e^{-3A_0}{ D(p,r_0) \over 1 + [ e^{-A_0} U_0\, D(p,r_0)] p^2}
\ee
Notice that, if there exists a regime in which
\be \label{ind8-0}
e^{-A_0} U_0 p^2 D(p,r_0) \gg 1,
\ee
then
the brane-to-brane propagator  is approximately $4$-dimensional, i.e. $\propto 1/p^2$, indicating that it is possible to recover the standard four-dimensional interaction.

Below,  we analyze the general features of the propagator  (\ref{ind8}), but we  postpone the discussion of the physical scales observed on the brane (these include the effective four-dimensional Planck scale and the crossover scale) to the next subsection.

The detailed behavior of the brane-to-brane Green's function  is determined by the function $D(p,r)$ evaluated at $r_0$. We will show below that the inequality (\ref{ind8-0}) is always satisfied at large enough $p^2$, and always violated at small $p^2$, regardless of the details of the bulk theory. We will be focusing on  the Euclidean propagator, therefore taking $p^2>0$.

.

To gain some insight on the behavior of $D(p,r_0)$, we may look at the  small and large momentum limit of equations (\ref{ind9-i}-\ref{ind9-ii}). The transition scale between these two regimes is determined by the bulk curvature at the interface,  ${\cal R}_0 \approx W(\phi_0)$: for $p\gg{\cal R}_0  $ we can neglect the derivative of $A(r)$ and treat  equations (\ref{ind9-i}-\ref{ind9-ii}) as in flat space-time; for $p\ll {\cal R}_0 $ the curvature term dominates, and we can expand the solution as a power series in $p^2$. Below we discuss these two regimes in more detail.

The detailed discussion of the behavior of $D(r_0, p)$ for large and small momenta compared to ${\cal R}_0$ is carried out in Appendix \ref{app-bulk}, where it is shown that:%

\be\label{scales-t}
D(r_0, p) \simeq \left\{\begin{array}{ll} \displaystyle{1\over 2p} & \quad  p \gg {\cal R}_0, \\
& \\
d_0 +  d_2 p^2 + d_4 p^4 + \ldots & \quad p \ll {\cal R}_0, \end{array} \right. 
\ee
The coefficients $d_i$ are explicitly computed in the Appendix \ref{app-smallp}, and are given by:
 \be\label{ind15}
d_0 =  e^{3A_0} \int_0^{r_0} dr' e^{-3A_{UV}(r')}.
\ee
\be\label{ind15-2}
d_2  =  - e^{3A_0} \int_0^{r_0} dr' e^{-3A_{UV}(r')}\int_{r'}^{r_0} dr'' e^{3A_{UV}(r'')}\int_0^{r''} dr''' e^{-3A_{UV}(r''')}
\ee
\be\label{ind15-3}
\!\!\!\!\!\!d_4 =  e^{3A_0} \int_0^{r_0} dr' e^{-3A_{UV}(r')}\int_{r'}^{r_0} dr'' e^{3A_{UV}(r'')}\int_0^{r''} dr''' e^{-3A_{UV}(r''')}\int_{r'''}^{r_0} e^{3A_{UV}(r^{iv})}\int_0^{r^{iv}} dr^v e^{-3A_{UV}(r^v)}
\ee
One may be worried that  the expansion (\ref{ind14}) could  break down at some finite order (or even at leading order) due to  non-analyticity as $p^2\to 0$: after all,  in flat space, $D(p,r_0) \sim p^{-1}$ as $p^2\to 0$. However, as shown in Appendix \ref{secreg}, this cannot be the case
and  at least the first two coefficients  (\ref{ind15}-\ref{ind15-2}) are  always well defined in holographic theories.
More specifically,  the small-$p$ expansion is analytic if the bulk spectrum of normalizable eigenmodes  has a mass gap.   Otherwise,  the expansion breaks  down and some non-analytic terms generically appear. However, as we show in Appendix \ref{secreg}, this happens at an order higher than $p^4$ (except in the case of a regular $AdS$ IR fixed point, where the first non-analytic term is of the order  $p^4 \log p$).

\begin{figure}[h!]
\begin{center}
\includegraphics[width=14cm]{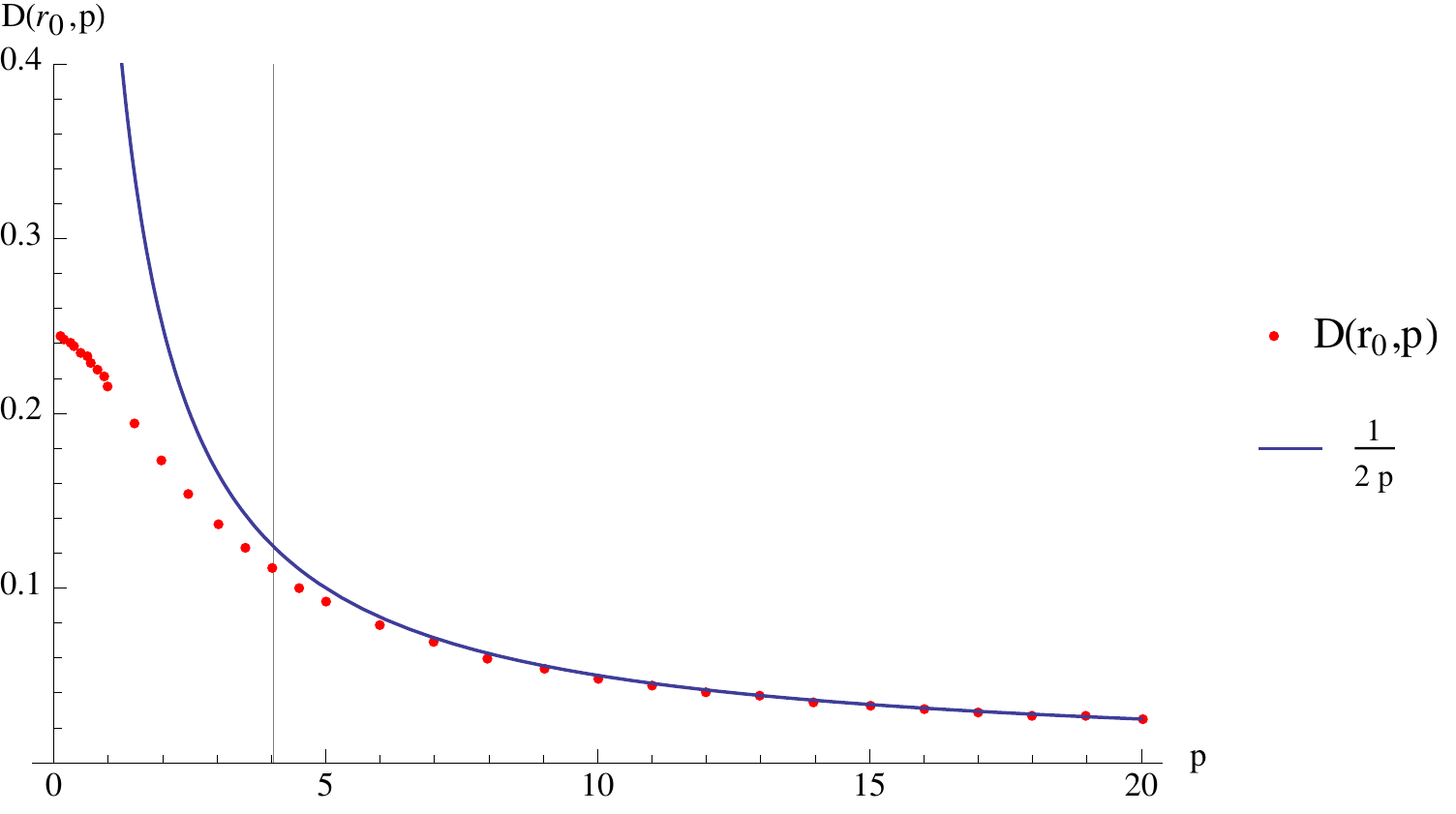}
\caption{The function $D(r_0,p)$ as a function of momentum, compared with $1/2p$. This graph is obtained  numerically from the specific example with $AdS$ UV and IR asymptotics presented in section \protect\ref{concrete-1}. The scales on both the horizontal and the  vertical axis are in units of the UV $AdS$ length.  The transition scale $1/r_t$ (solid line) is about 4 (in UV-AdS units).} \label{Dpfig}
\end{center}
\end{figure}

The behavior of the bulk propagator, obtained numerically in the specific example discussed in section \ref{concrete-1},  is  sketched in  figure \ref{Dpfig}.

Having determined the properties of the bulk spin-2 Green's function,
the tensor mode brane propagator can  be  obtained from equation
(\ref{ind8}). However, in order to relate that expression to the actual gravitational interaction measured by brane observers, we have first to translate it in physical coordinates on the brane. Indeed,  equations  (\ref{ind4-iii}) and   (\ref{ind8}) are expressed in terms of coordinates $x^\mu$ (and the associated momenta $p_\mu$), but in these coordinates the induced metric on the brane differs by the Minkowski metric by a scale factor $e^{2A_0}$:
\be
\gamma_{\mu\nu} = e^{2A_0}\eta_{\mu\nu}
\ee
In order to rewrite the results in a transparent way in physical coordinates, it is convenient to first write equation (\ref{ind4-iii}) in a manifestly covariant way, and then change the embedding coordinates of the interface\footnote{We might have as well done a coordinate transformation in the whole bulk, but this would have affected the asymptotic metric and would change the definition of the UV sources in a non-universal (i.e. solution-dependent) way. Therefore, we prefer to keep working in the coordinates (\ref{FE7}) in the bulk  and rescale the brane coordinates only} to:
\be \label{brane1}
y^\mu  = e^{A_0}x^\mu,  \quad q_\mu = e^{-A_0}p_\mu , \quad x^\mu p_\mu = y^\mu q_\mu.
\ee

Introducing the appropriate factors of $e^{A_0}$, the manifestly covariant form  of equation (\ref{ind4-iii}) is:
\be\label{brane2}
S_{int} = -{1\over 2 M^3} \int d^4x \sqrt{\gamma} \int d^4x'\sqrt{\gamma} \,  G(r_0,x; r_0, x')\left(\gamma^{\mu\rho}\gamma^{\nu\sigma}  - {1\over 3}\gamma^{\mu\nu}\gamma^{\rho\sigma}\right)T_{\mu\nu}(x) T_{\rho\sigma}(x'),
\ee
where:
\be\label{brane3}
G(r_0,x; r_0, x') = \int {d^4p \over (2\pi)^4} \tilde{G}(r_0,p; r_0) e^{i x^\mu p_\mu},
\ee
and  $\tilde{G}(r_0,p; r_0)$ is given in equation (\ref{ind8}). Performing the change of coordinate (\ref{brane1}) and writing everything in momentum space, we obtain the interaction between two brane stress-tensors at physical momenta $q^2$ in physical brane units:
\be\label{brane4}
S_{int} = \int {d^4 q \over (2\pi)^4} G_4(q) \left(T_{\mu\nu}(q)T^{\mu\nu}(-q) -  {1\over 3} T(q) T(-q) \right),
\ee
where now the {\em physical} brane-to-brane propagator in momentum space is:
\be\label{brane5}
G_4(q) = -{e^{A_0} \over 4M^3} {\tilde{D}(q, r_0) \over 1+ q^2 e^{A_0} U_0 \tilde{D}(q, r_0)}, \quad  \tilde{D}(q, r_0) =  D(e^{A_0} q, r_0).
\ee
In deriving equation (\ref{brane5}) one must carefully  take into account  an extra factor $e^{4A_0}$ arising from the measure in the Fourier transform in terms of $q$ rather than $p$.

There are several  crossover scales, which we discuss below,    governing the  qualitative behavior of the full brane-to-brane propagator $G_4(q)$.

\begin{enumerate}
\item {\bf The {\em crossover scale} $m_t$}:
With respect to  physical momenta, the crossover  (\ref{scales-t}) between  large  and small momentum behavior of the bulk propagator is now given by

\be\label{scales-t2}
\tilde{D}(q,r_0) \simeq \left\{\begin{array}{ll} \displaystyle{e^{-A_0}\over 2q} & \quad q \gg m_t \\
& \\
d_0 + q^2 \,e^{2A_0} d_2 + \ldots & \quad q \ll  m_t \end{array} \right. ,  \qquad  m_t \approx e^{-A_0}{\cal R}_0.
\ee
where ${\cal R}_0 \approx W(\f_0)$ is the bulk curvature scale close to the interface.

We define the associated distance scale $r_t\equiv 1/m_t$.

\item {\bf The {\em DGP scale} $r_c$}:
\be\label{scales-rc}
r_c \equiv {U_0 \over 2};
\ee
This scale arises within the large-$q$ regime of the bulk propagator, $q\gg m_t$.  In this regime, using the first line of equation (\ref{scales-t2}),    equation (\ref{brane5}) can be approximated by:
\be\label{scales2}
\tilde{G}_4(q) \simeq - {1\over 4 M^3} \, {1 \over 2 q + 2 r_c q^2}  \qquad q  \gg m_t
\ee
This is DGP-like \cite{DGP}, with the scale $r_c$ given in (\ref{scales-rc}). This scale sets the transition between a five-dimensional regime and a four-dimensional one:
\be \label{brane-6}
\tilde{G}_4(q) \simeq \left\{ \begin{array}{ll} \displaystyle{-{1\over 2M_p^2} \, {1 \over  q^2}}  & \qquad q  >  1/r_c \\ & \\ \displaystyle{-{1 \over  2M^{3}} \, {1 \over  q}}  & \qquad   q < 1/r_c \end{array}\right.
\ee
with the four-dimensional effective Planck scale  given by:
\be\label{brane7}
M_p^2 =  4 r_c \, M^{3} = 2 M^3 U_0
\ee
We stress that the crossover in equation (\ref{brane-6}) takes place only if $r_c < r_t$, since the approximation equation (\ref{scales2}) holds only for $q> 1/r_t$.

The dimensionless ratio between the measured 4d Planck scale and the DGP scale is given in terms of the parameters of the model as:
\be\label{brane7b}
r_c M_p = \left ({M U_0 \over 2} \right)^{3/2}
\ee

In particular, notice that it only depends on $\f_0$ and not on the integration constant $A_0$, therefore equation (\ref{brane7b}) is independent on the UV coupling $g_0$.

\item {\bf The {\em graviton mass  scale} $m_g$}, which we  define below.  Below this energy scale,  the brane propagator becomes approximately constant as  a function of $q$. As we will see momentarily, this scale plays the role of an effective graviton mass.

The scale $m_g$ arises in the small-$q$ regime, $q\ll m_t$, in which we can use the small momentum expansion (second line of equation (\ref{scales-t2}))  for the bulk propagator. Stopping at $O(p^2)$ in this expansion, we obtain from equation  (\ref{brane5}):
\be\label{scales3}
\tilde{G}_4(q) \simeq - {1\over 2 \tilde{M}^2_p} \, {1 \over  m_g^2 + q^2 + O(q^4)}  \qquad q  \ll m_t
\ee
The graviton mass scale $m_g$ and the effective Planck scale  $\tilde{M}_p$ in this regime are given by:
\be\label{brane8}
m_g^2 = {e^{-A_0}\over U_0 d_0\left(1 - {e^{A_0}d_2\over U_0 d_0^2}\right)}, \qquad \tilde{M}_p^2 = {2M^3 U_0}\left(1 -{ e^{A_0} d_2 \over U_0 d_0^2}\right).
\ee

\item {\bf The {\em massive gravity crossover scale} $m_4$}
This scale governs the regime in which the brane propagator can be approximated by a massive graviton one. For this to be the case, the  $O(q^4)$ terms  in the expression (\ref{scales3}) must be  negligible compared to the $O(q^2)$ term: in such a regime,  the graviton exchange mimics  the interaction mediated by  a  massive graviton with mass $m_g$.
Using the small-$q$ expansion from equation (\ref{scales-t2}) in the definition (\ref{brane5}), we find:
\be
\!\!\!\!\!\!G_4(q) =- {1\over 4M^3 U_0} \left( 1 + {e^{-A_0}\over d_0 U_0} + q^2 \left(1 - {e^{A_0} d_2 \over d_0^2 U_0} \right) + q^4\left(e^{3A_0} {d_2^2 \over d_0^3 U_0} - e^{3A_0} {d_4 \over d_0^2 U_0}\right) + O(q^6) \right)^{-1}.
\ee
Demanding that the $q^4$ term be negligible with respect to the $q^2$ term, we find that  the ``massive graviton'' approximation (\ref{scales3}) holds at momenta:
\be\label{brane9}
q \ll  m_4, \quad m_4^2 \equiv e^{-2A_0}{(d_0 d_2 - e^{-A_0}d_0^3 U_0) \over (d_2^2 - d_0 d_4) }.
\ee

\end{enumerate}

To analyze the overall qualitative behavior of graviton exchange as a function of the distance  scale, we must distinguish two situations: $r_t > r_c$ and $r_t < r_c$. As we will see, the behavior in these  situations is  DGP-like and massive-gravity-like, respectively.
\begin{itemize}
\item {\bf $r_t > r_c$} (figure \ref{figscales1})

\begin{figure}[h!]
\begin{center}
\includegraphics[width=14cm]{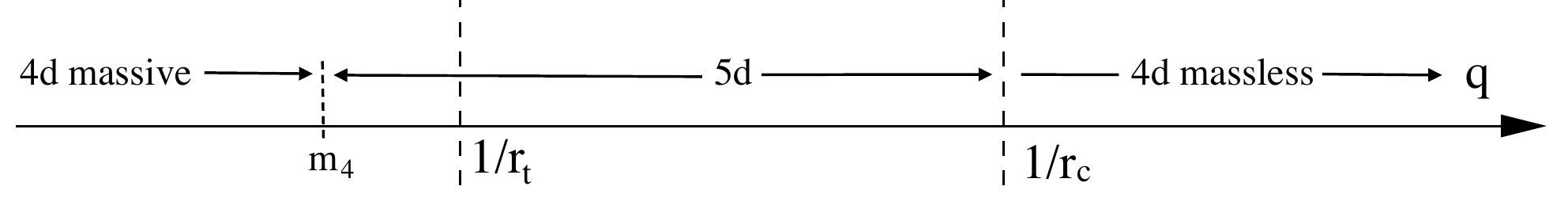}
\caption{Graphical representation of the different regimes (depending on momentum) of the brane-to-brane propagator, in the case $r_t > r_c$. In this example, we have chosen to set  the massive gravity transition scale $m_4< 1/r_t$, therefore it is only below $m_4$  that we have a four-dimensional single-particle-like propagator.} \label{figscales1}
\end{center}
\end{figure}

In this case we can broadly distinguish three regimes:
\begin{enumerate}\item  $q > 1/r_c$:
 the interaction mimics four-dimensional massless gravity\footnote{By massless here we refer to the $1/q^2$ behavior of the propagator, {\em not} to the tensor structure, which for tensor modes is always the one of a massive graviton, see equation (\ref{brane4}).} with Planck scale given in by (\ref{brane7});
\item $1/r_t < q < 1/r_c$:\\
 this   is an intermediate  DGP-like five-dimensional  regime approximated by equation (\ref{scales2});
\item $q < 1/r_t$ the interaction matches onto a massive-graviton type exchange, with mass $m_g$ given in (\ref{brane8}). In particular, it mimics a massive graviton for all momenta $q \ll m_4$ (see equation (\ref{brane9} )
\end{enumerate}

\item  {\bf $r_t < r_c$}  (figure \ref{figscales2})\\

\begin{figure}[h!]
\begin{center}
\includegraphics[width=14cm]{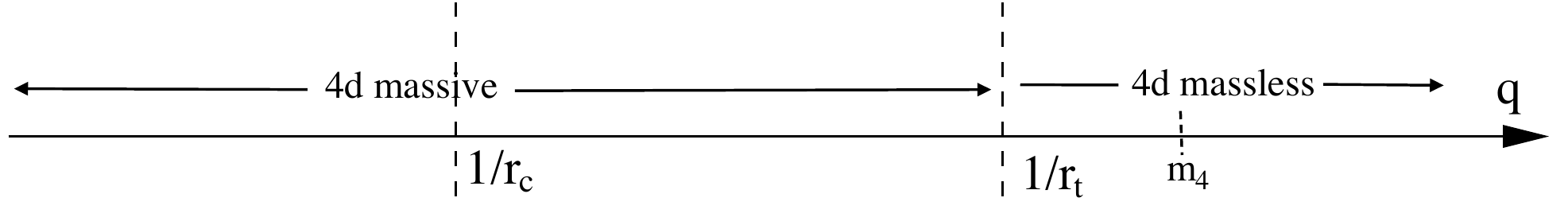}
\caption{Graphical representation of the different regimes (depending on momentum) of the brane-to-brane propagator, in the case $r_t < r_c$. In this example, we have chosen to set  the massive gravity transition scale $m_4>1/r_t$, thus the transition across $q\simeq 1/r_t$ goes directly to a four-dimensional massive propagator.} \label{figscales2}
\end{center}
\end{figure}

In this case, there is no DGP-like regime, and the relevant transition  scales are  $m_4$ (defined in equation (\ref{brane9})) and $1/r_t$:
\begin{enumerate}
\item $q> 1/r_t$ :\\
  we have massless 4d propagation with Planck scale given again by  equation (\ref{brane7}).
\item $m_4< q< r_t$:\\
Higher derivative corrections (resulting from higher powers of $q^2$ in the propagator) are important.
\item $q< m_4$:\\
the behavior is that  of a  massive graviton propagator with mass $m_g$ and Planck scale $\tilde{M}_p$ given in equation (\ref{brane8}).
\end{enumerate}
Notice that if $m_4 > 1/r_t$, we have a four-dimensional single-particle behavior  in the entire range of momenta. This situation is  schematically represented in figure \ref{figscales2}. Also,  if $m_g \ll 1/r_t < m_4$, the propagator can still be  well approximated by a  massless propagator all the way down to momenta $q \simeq m_g$.
\end{itemize}

\begin{figure}[h!]
\begin{center}
\includegraphics[width=14cm]{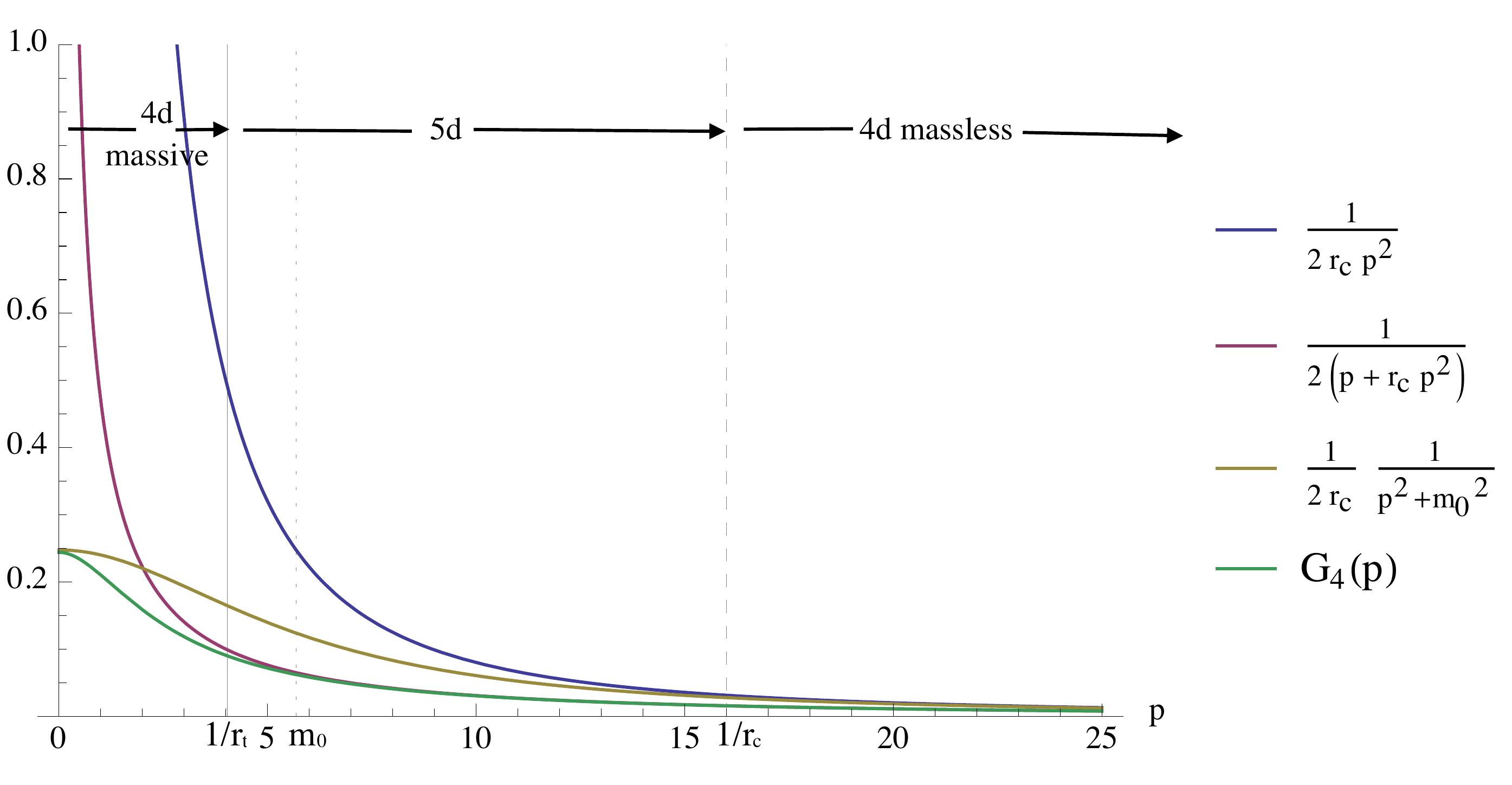}
\caption{The effective brane-to-brane propagator for $r_t>r_c$. The vertical lines are from left to right: $1/r_t$ (solid),  $m_g$ (dotted), $1/r_c$ (dashed). All quantities are measured in boundary $AdS$ units.} \label{fig1}
\end{center}
\end{figure}

\begin{figure}[h!]
\begin{center}
\includegraphics[width=14cm]{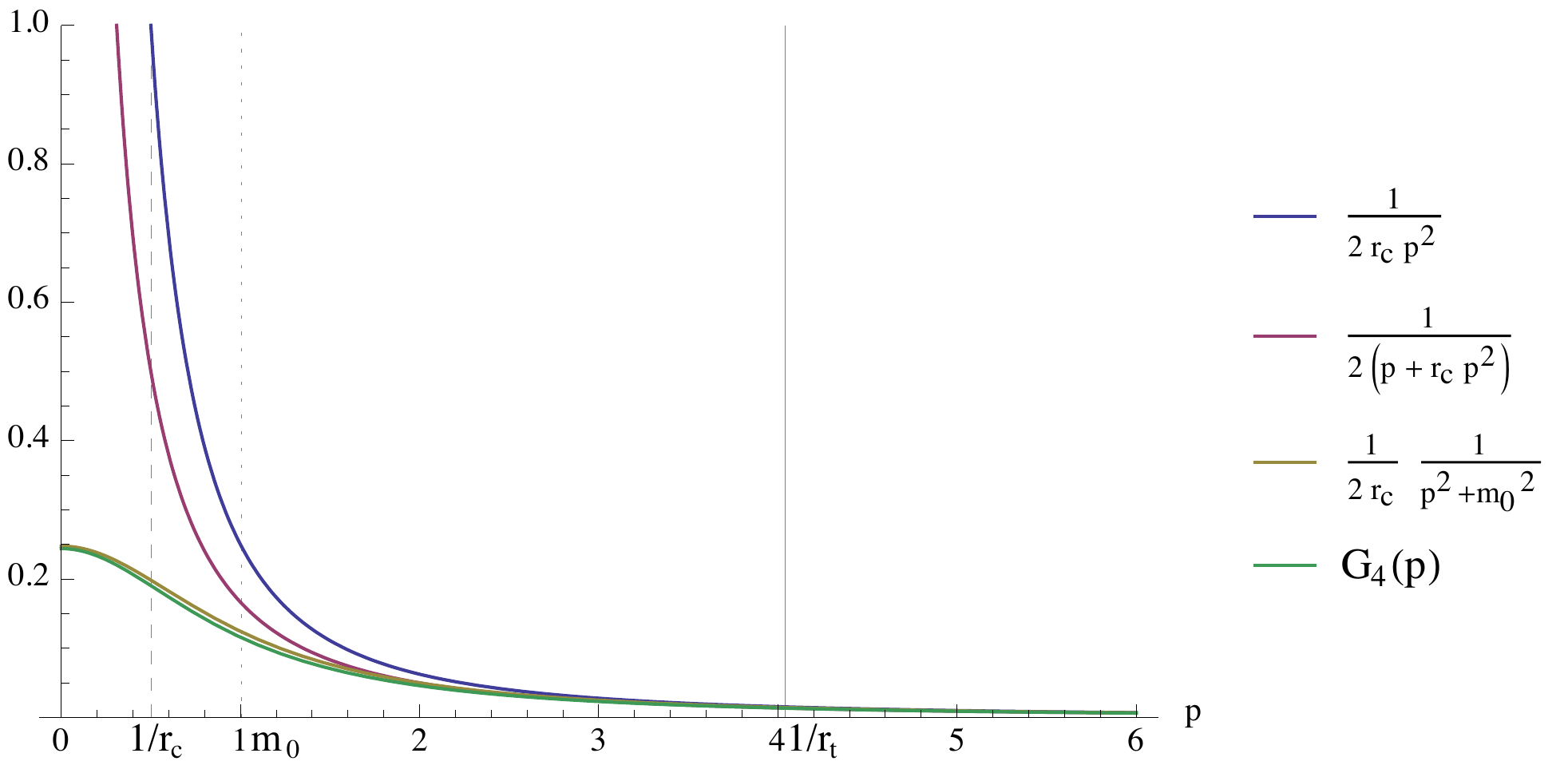}
\caption{The effective brane-to-brane propagator for $r_t<r_c$. The vertical lines are from left to right:  $1/r_c$ (dashed), $m_g$ (dotted), $1/r_t$ (solid).}\label{fig2}
\end{center}
\end{figure}

The  behavior of the brane-to-brane propagator in the two cases ($r_c<r_t$ and $r_c>r_t$)   is shown in figure \ref{fig1} and \ref{fig2}, respectively. In those figures we compare the various limiting form of the propagator (massless, massive and DGP-like) to the  result obtained numerically  in the IR-regular  example in Section \ref{concrete-1}. In  those figures,  we can see explicitly  the transitions  discussed above.

\subsection{Gravity phenomenology} \label{sec pheno}

In this subsection we will identify the phenomenologically interesting regimes which can be  acceptable in the context of the class of models described in this paper,  as a function of the model parameters. These are the regimes where the one-graviton exchange  is well approximated by a $1/q^2$ potential over the distances at which  Newtonian (or Einstein) gravity describes observation accurately\footnote{Here we limit ourselves to the linear regime. One will still have to check what happens in the non-linear regime, in particular the fate of the extra massive gravity mode and of the vDVZ discontinuity.}.

Before we analyze such  regimes, it is important to make the dependence on all the parameters of the model explicit. Following the discussion in section \ref{scales}, these include not only the brane and bulk (super)potentials which determine $\f_0$, but also the UV coupling $g_0$ which determines $A_0$, the warp factor at the brane, via equation (\ref{scales16}).

First, we  extract the explicit dependence on $A_0$ of the  expansion coefficients $d_0$ and  $d_2$ and $d_4$, entering equation (\ref{brane8})  and   given in equations (\ref{ind15}-\ref{ind15-3}). As  shown in Appendix \ref{app-smallp}, the result takes the form:
\be
d_i = e^{-A_0} {\cal D}_i (\f_0)
\ee
where the coefficients ${\cal D}_i (\f_0)$ are independent of $A_0$. Furthermore,  in Appendix \ref{app-smallp} we show that the magnitude of the coefficients ${\cal D}_i$  is controlled by the bulk curvature close to the junvtion, and we have  roughly:
\be\label{ind20}
{\cal D}_{2n}(\f_0) \approx {1\over {\cal R}_0^{2n+1}}, \qquad {\cal R}_0 \approx W_{UV}(\f_0)
\ee

We can now rewrite equations (\ref{brane8}) in a way that makes the dependence on $A_0$ explicit:
\bea
&&m_g^2 = {1\over U_0 {\cal D}_0 \left(1 - {e^{2A_0} {\cal D}_2 \over U_0 {\cal D}_0}\right)}, \quad \tilde{M}_p^2 = 2 M^3 U_0  \left(1 - {e^{2A_0} {\cal D}_2 \over U_0 {\cal D}_0}\right),  \label{ind16}\\
&& m_4^2  = e^{-2A_0}{{\cal D}_0 {\cal D}_2 - e^{-2A_0}U_0 {\cal D}_0^3 \over  {\cal D}_2^2 -  {\cal D}_4 {\cal D}_0}  \label{ind17}
\eea

Using these results  we can write the relevant scales discussed in the previous section in terms of the bulk scale at the brane and UV coupling:
\begin{itemize}
\item {\bf Transition scale} (between the large and small momentum behavior of the bulk propagator,   see equation (\ref{scales-t2})):
\be\label{ind21}
m_t \equiv {1\over r_t} = e^{-A_0}{\cal R}_0
\ee
\item {\bf DGP scale:}
\be\label{ind22}
m_c \equiv {1\over r_c} =  {2\over U_0} ;
\ee
\item {\bf Planck scale:}\\
In the DGP regime:
\be\label{ind23}
M_p^2  \approx 2 M^3 U_0   ;
\ee

In the massive gravity regime:
\be\label{ind24}
\tilde{M}_p^2  \approx 2 M^3 U_0 \left(1 + {e^{2A_0} \over U_0 {\cal R}_0}\right) ;
\ee

\item {\bf Graviton ``mass'':}
\be\label{ind25}
m_g^2 = {{\cal R}_0 \over U_0} {1 \over 1 + {{e^{2A_0} \over U_0 {\cal R}_0}}};
\ee

\item {\bf Massive gravity crossover scale}  (below which the interaction mimics a massive graviton, see equation (\ref{brane9}))
\be \label{ind26}
m_4^2 \approx e^{-2A_0} {\cal R}_0^2 + e^{-4A_0}U_0 {\cal R}_0^3 .
\ee
\end{itemize}
We will now  translate some of these parameters in the language of the 4d dual field theory:
\begin{itemize}
\item On general grounds, we expect:
\be \label{phen9a}
U_0 =  {\Lambda^2 \over M^3} u(\f_0),
\ee
 where $\Lambda$ is the UV cutoff of the theory on the brane  and  $u(\f_0)$ is a dimensionless function which is model-dependent. This is the scaling one expects for the induced correction to the 4d curvature term in the action (\ref{A3}).
\item The bulk curvature and Planck scale are expected to be related by the 5d holography relation:
\be\label{phen10}
{M \over {\cal R}_0  } \approx { N^{2/3} \over \ell_{UV} W_{UV} (\f_0)}
\ee
where $N$ is the number of degrees of freedom of the UV CFT.  Equation (\ref{phen10}) can be justified from the large-$N$ scaling  $(M \ell_{UV})^3 = N^2$ and from the fact that $W_{UV}(0) = 6/\ell_{UV}$. For a given choice of potentials, and for $N$ large but fixed, we can trust our results as long as the curvature is small in Planck units\footnote{More precisely, if we think of the gravity dual as the low-energy approximation of a full string theory setup,  we must require  that ${\cal R}_0$ be smaller than the {\em string} scale, which in perturbative string theory is parametrically smaller than $M$.}. Therefore, at the point $\f_0$ where the brane sits,  we must demand:
\be \label{phen10a}
{W_{UV}(\f_0) \over W_{UV}(0)} \equiv {1\over 6} \ell_{UV} W_{UV}(\f_0) \ll  N^{2/3}
\ee
Since the superpotential is a monotonically increasing function, this means that for any finite  $N$ there is a limit to how much we can push the brane position to large $\f_0$.

\item  As discussed at the end of  Section \ref{selfie}  (and shown in more detail in  Appendix \ref{scales}), the warp factor is related to the UV relevant coupling $g_0$ by the relation:
\be \label{phen11}
e^{A_0} =g_0^{1/\Delta_-} \,  \ell_{UV} \, e^{\bar{\cal A}(\f_0)}
\ee
where $\bar{\cal A}(\f_0)$ is given in equation (\ref{scales15}) and is generically of order one.
Thus, for fixed $\f_0$, large (small) scale factor at the interface translates into large (small) UV coupling in $AdS$ units.

\item Finally, as discussed in the introduction,  we must keep in mind that our holographic setup is supposed to be an effective description of the physics below the UV cutoff $\Lambda$, which is given by the scale of the messenger fields which couple the Standard Model to the holographic degrees of freedom. Therefore, in the discussion that follows, $r_{UV} \equiv 1/\Lambda$ will always be the short-distance cut-off.

\end{itemize}

In view of the results presented above,  in the next two subsections we will analyze three possible regions of parameter  space which make gravitational interactions phenomenologically acceptable at the observed scales (roughly from sub-millimeter scales up to cosmological scales, to stay on the safe side).  This will result in further constraints on model parameters beyond those analyzed in section 2.5-2.6, which were arising from requiring  self-tuning (plus manifest stability of the background).

\subsubsection{DGP scenario}

\begin{figure}[h!]
\begin{center}
\includegraphics[width=14cm]{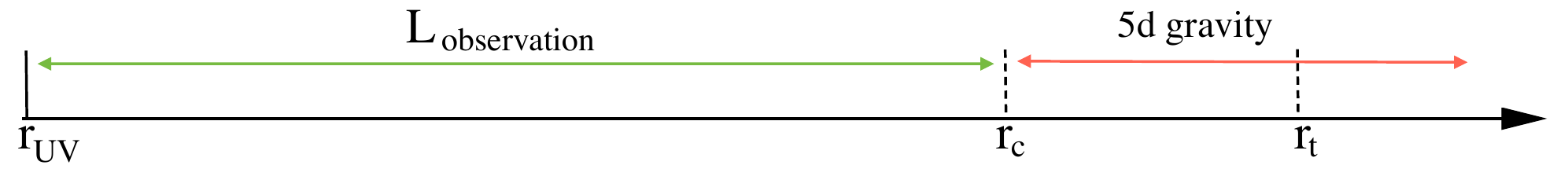}
\caption{The horizontal direction  represents the hierarchy of distance scales in the DGP scenario. The regime corresponding to the observed massless 4d gravity is represented in green. At  distances of the order $r_c$ we have a large-distance transition to the five-dimensional regime.} \label{figscales3-i}
\end{center}
\end{figure}

In  this scenario, $r_t > r_c$, and  the distance scales we  observe must all be smaller than both the transition scale $r_t$ defined in equation  (\ref{ind21}) and the DGP scale (\ref{ind22}).  We have ordinary gravity at scales:
\be \label{dgp1}
L_{\rm observation} \,\, < \, \, r_c \, \, (< \, r_t).
\ee
The situation is summarized in figure \ref{figscales3-i}.

For this scenario to be compatible with observation, if we want to be conservative (i.e. $L_{\rm observation}$ is of cosmological size), we need the dimensionless quantity:
\be\label{dgp2}
 M_p r_c = \left({M U_0 \over 2}\right)^{3/2} \approx \left(\Lambda \over M\right)^3 u^{3/2}(\f_0)
\ee
to be at least of order $10^{60}$.   This can be achieved with a large cut-off scale\footnote{In the holographic setup discussed in \cite{smgrav} this translates into a large mass for the messenger fields.}  (compared to bulk Planck scale $M$) and can be enhanced if the equilibrium position is in a region where the function $u(\f)$ is parametrically large.

The assumption $r_t > r_c$ translates into:
\be\label{dgp3}
 e^{-A_0} U_0 {\cal R}_0 \approx  {U_0 M\over N^{2/3}}\, { W_{UV}(\f_0) \over   g_0^{1/\Delta_-}} \lesssim 1,
\ee
where  we have used the relations (\ref{phen10}-\ref{phen11}). The quantity $U_0 M$ must be large by equation (\ref{dgp2}). This can be compensated by $N$ being large. Moreover we can choose  the UV coupling $g_0$ to be large as well (in $AdS$-length units)
.

In this scenario, since we ``live'' below the transition scale $r_t$, it does not matter what the other scales (related to massive gravity behavior) are, since they will be relevant  only for physics at distances $L > r_t$. Thus the first modification we observe as the scales go larger is a transition to  a 5d regime above the scale $r_c = U_0$.

At small scales on the other hand, in this scenario gravity is modified only below the short-distance cut-off $r_{UV}=1/\Lambda$.

\subsubsection{Massive gravity  scenario 1}

Suppose  we still have $r_t > r_c$, i.e.
\be
e^{-A_0} {\cal R}_0 U_0 < 1,
\ee
but that the scales we observe are beyond $r_t$, in the massive gravity regime. Then the DGP transition  is irrelevant as
\be \label{massive1}
r_c < r_t < L_{\rm observation}.
\ee
In this case, $r_t$ must be a {\em short distance} scale for gravity modification. At distances larger than $r_t$, we observe normal gravity if we are between the distance scales:
\be\label{massive2}
{1 \over m_4} < L_{\rm observation} < {1 \over m_g}
\ee
Indeed, at distances smaller than $m_4^{-1}$ there will be higher derivative corrections to the graviton propagator, whereas at distances larger than $m_g$ the gravitational interaction saturates. Therefore we need $1/m_4$ to be a microscopic distance scale  and $1/m_g$ to be a cosmological one.
\begin{figure}[h!]
\begin{center}
\includegraphics[width=14cm]{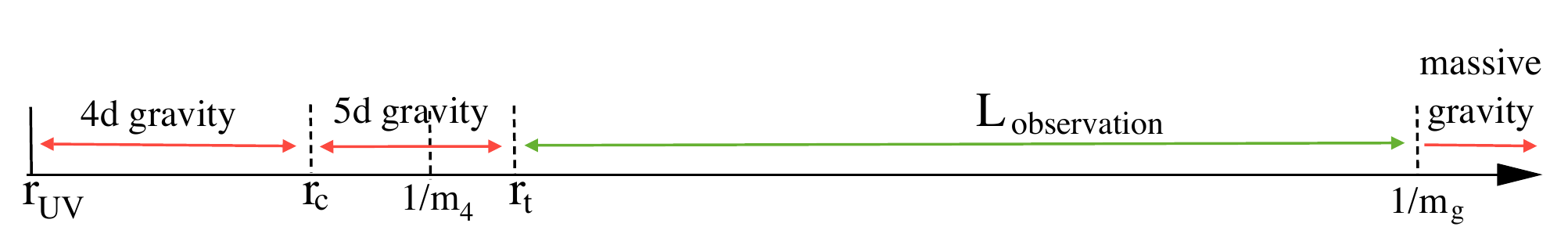}
\caption{The horizontal direction  represents the hierarchy of distance scales in the ``massive gravity 1'' scenario. The regime corresponding to the observed massless 4d gravity is represented in green. Gravity is modified at short distances at the scale $r_t$, below which it becomes five-dimensional,  and  at large  distances at the scale $1/m_g$, above which it becomes massive.} \label{figscales3-ii}
\end{center}
\end{figure}

From equation (\ref{ind26}) we notice that:
\be\label{massive3}
m_4 \approx e^{-A_0}{\cal R}_0 \left ( 1 + e^{-2A_0} U_0 {\cal R}_0\right)^{1/2} \geq m_t
\ee
Then, the massive gravity scale $m_4$ is generically at least as large as  the transition scale $m_t$,  and the left side of the inequality (\ref{massive2}) is automatically satisfied in the regime (\ref{massive1}). The situation is summarized in figure \ref{figscales3-ii}.

The ratio of the graviton ``mass'' to the 4d Planck scale is  obtained from  equations (\ref{ind24}-\ref{ind25}):
\be \label{massive3.5}
{ m_g \over \tilde{M}_p} =  {{\cal R}_0^{1/2} \over 2^{3/2} M^{3/2} U_0} {1 \over 1 + {{e^{2A_0} \over U_0 {\cal R}_0}}}  ~~~<~~~ 10^{-60}
\ee
This condition is required  if we want the transition to massive gravity to happen on  cosmological scales.

Furthermore, we have to demand that the transition scale $m_t$  be at least above the inverse (tenth of) millimeter:
\be\label{massive4}
 {m_t \over \tilde{M}_p} =  e^{-A_0}{{\cal R}_0 \over 2^{3/2} M^{3/2} U_0^{1/2}} {1 \over \left( 1 + {{e^{2A_0} \over U_0 {\cal R}}}\right)^{1/2}}  ~~~>~~~ 10^{-30}.
\ee

Notice that the denominators of equations (\ref{massive3}) and (\ref{massive4})  contains the combination:
\be \label{massive5}
e^{2A_0}/(U_0 {\cal R}_0) = e^{A_0} r_t /r_c .
\ee
Although $r_t/r_c >1$, the right hand side of the above equation  can be large or small depending on the warping, thus we have two  possibilities:
\begin{itemize}
\item  {\em large warping} i.e..
\be
e^{A_0} > {r_c\over r_t}
\ee
In this case, we can drop the ``1'' in the denominators and we obtain for the graviton mass:
\be \label{massive6}
m_g^2 \approx e^{-2A_0} {\cal R}_0 = m_t^2
\ee
As a consequence, it is impossible to satisfy equation (\ref{massive2}):  there is no room for the several orders of magnitude of massless  4d gravity we  observe.
\item {\em small  warping} i.e..
\be
e^{A_0} < {r_c\over r_t}
\ee
In this case we have
\be
{m_t^2 \over m_g^2} = e^{-2A_0} {\cal R}_0 U_0 = e^{-A_0}  r_c/r_t
\ee
Therefore, to have a large separation between $m_t$ and $m_g$ we need a very small warping. The conditions (\ref{massive3}-\ref{massive4}) simplify to:
\be \label{massive7}
 {{\cal R}_0^{1/2} \over M^{3/2} U_0} \approx \left({M \over \Lambda}\right)^2  {(\ell_{UV} W_{UV}(\f_0))^{1/2} \over u(\f_0) } {1 \over N^{1/3}}  ~~~<~~~ 10^{-60}
\ee
\be\label{massive8}
 e^{-A_0}{{\cal R}_0 \over M^{3/2} U_0^{1/2}} \approx \left({M  \over \Lambda}\right) { W_{UV}(\f_0)\over  u^{1/2}(\f_0) g_0^{1/ \Delta_-}} {1 \over N^{2/3}}  ~~~>~~~ 10^{-30}
\ee
Both conditions above can be satisfied in a technically natural way for $N$ large if for example $\Lambda/M$ and $u(\f_0)$ are also large and $g_0$ is small in $AdS$ units.
\end{itemize}

\subsubsection{Massive gravity  scenario 2}

\begin{figure}[h!]
\begin{center}
\includegraphics[width=14cm]{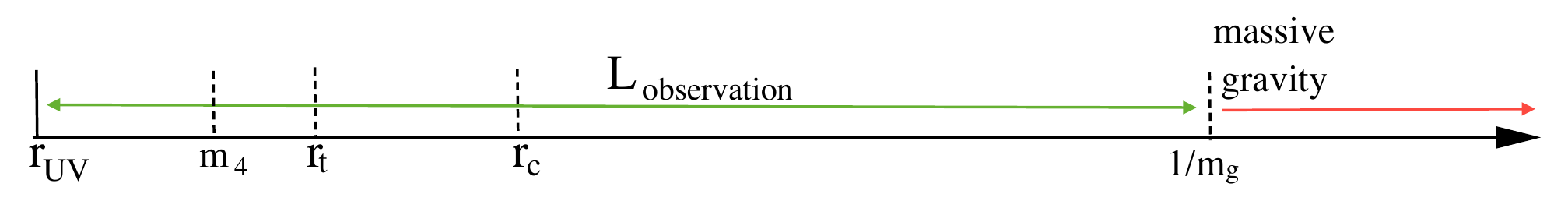}
\caption{The horizontal direction  represents the hierarchy of distance scales in the ``massive gravity 2'' scenario. Similarly to the ``massive gravity 1'' scenario (figure \protect\ref{figscales3-ii}), at distances larger than $1/m_g$ we have a transition to massive gravity. However there is no short-distance modification until  below the cut-off scale $r_{UV}$.} \label{figscales3-iii}
\end{center}
\end{figure}

Alternatively, we  can have  $r_c > r_t$, i.e.
\be\label{massive9}
e^{-A_0} {\cal R}_0 U_0 > 1.
\ee
In this case we are in  the type of scenario represented in figure \ref{fig2}: there is no DGP transition to a five-dimensional regime. Again, the  massive gravity crossover scale $m_4$ is at least as large than $m_t$, therefore at distances larger than  $r_t$ we are in the massive gravity regime. At distances shorter than $r_t$ on the other hand we are in the four-dimensional part of the DGP regime. Thus,  this scenario  reproduces the observed  gravitational interaction  at all scales satisfying:
\be\label{massive10}
r_{UV}< L_{\rm observation} < {1\over m_g},
\ee
The situation is summarized in figure \ref{figscales3-iii}.

If we want to be the theory to be manifestly ghost free (see the discussion in the section \ref{ghosts}),  we need  ${\cal R}_0 U_0 \lesssim 1$. This can be compatible with equation (\ref{massive9}) only for small warping, $e^{A_0}< 1$. From equation (\ref{ind25})  we deduce that  the graviton mass  and 4d Planck scale  are approximately:
\be
\tilde{M}_p^2 \simeq 2  M^3 U_0, \qquad m_g^2 \simeq  {{\cal R}_0 \over  U_0},
\ee
and a conservative observational constraint  that this be of cosmological scale is
\be
  {m_g \over \tilde{M}_p} =   {{\cal R}_0^{1/2} \over 2^{3/2} M^{3/2} U_0} \approx \left({M \over \Lambda}\right)^2  {(\ell_{UV} W_{UV}(\f_0))^{1/2} \over u(\f_0) } {1 \over N^{1/3}}  ~~~<~~~ 10^{-60}
\ee
i.e. the same constraint as equation (\ref{massive7}).  In this case however there is no short-distance constraint analogous to (\ref{massive8}), since gravity is ``normal'' at small distances (less than $r_t$) down to the UV cutoff.

\section{Scalar  perturbations and stability} \label{sec scalar}

We now turn to the analysis of linear perturbations in the scalar sector. The relevant modes are defined in equation (\ref{pert1}-\ref{pert3}),  in which we keep only the bulk scalar perturbations $\psi,E,\phi$, plus the brane-bending mode $\rho$ defined in equation (\ref{pert2}). We fix the gauge $\chi=B=0$ everywhere (bulk and brane).

As we saw in Section 3 (see also   Appendix \ref{app scalar} for more detail), after solving the constraints  for $E$ and $\phi$, and after eliminating the brane-bending field $\rho$, one is left with only the scalar mode $\psi$, which satisfies the bulk field equation (on each side of the brane):
\be \label{psieq}
\psi'' + \left(3{a'\over a} + 2{z'\over z}\right) \psi' + \de^\mu\de_\mu \psi = 0, \qquad z \equiv {a \f' \over a'},
\ee
and the matching conditions:
\bea
&& \left[z \psi'\right]= -\left({6\over a_0} {d U_B \over d\f}\Big|_{\f_0}\right) \Box {[z \, \psi] \over [z]} - {1\over a_0}\left( Z_0 \Box  - a_0^2 {\tilde {\cal M}}^2 \right){[ \psi] \over [z^{-1}]}, \label{sm1-sec} \\
&& \nonumber \\
&&   \left[z^2 \psi'\right] =  6 \left( 2 {U_0 \over a_0}- \left[a\over a'\right]  \right) \Box {[z \,\psi] \over [z]}  - \left({6\over a_0} {dU\over d\f}\Big|_{\f_0}\right) \Box {[ \psi] \over [z^{-1}]}\label{sm2-sec},
\eea
where $[X] \equiv X_{IR} - X_{UV}$ denotes is the discontinuity of any quantity $X$ across the brane, and $\Box = \eta^{\mu\nu}\de_\mu\de_\nu$.

One way to handle these matching conditions is to split the field $\psi$ in two parts, $\psi_{UV} = \psi(r< r_0)$ and $\psi_{IR} = \psi(r>r_0)$, and to write equations (\ref{sm1-sec}-\ref{sm2-sec}) as a matrix-like boundary condition for $\psi_{UV}$ and $\psi_{IR}$:

\be\label{scpert3}
\left(\begin{array}{l} \psi_{UV}' \\ \psi_{IR}' \end{array}\right)_{r=r_0}= \Big(\Gamma_1 + \Gamma_2 \, \de^\mu\de_\mu\Big) \left(\begin{array}{l} \psi_{UV}\\ \psi_{IR}\end{array}\right)_{r=r_0}
\ee
where $\Gamma_{1,2}$ are two $2\times 2$ matrices given explicitly in  equation (\ref{scalar15}) and which depend only on background quantities.  It is useful to  introduce a two-component wave-function  in the  whole bulk,
\be\label{scpert4}
\Psi(r) = \left(\begin{array}{l} \psi_{UV}(r) \\ \psi_{IR}(r) \end{array}\right)
\ee

Then, we can  write the problem in compact form as an asymmetric Sturm-Liouville problem with boundary conditions at $r_0$:
\bea
&&\de_r \left[ {\cal B} (r)\de_r \Psi\right] + {\cal B}(r)  \de_\mu  \de^\mu \Psi = 0 , \qquad r\neq r_0\label{sl-sec1}\\
&& \nonumber \\
&& \de_r \Psi(r_0) = \Big(\Gamma_1 + \Gamma_2\, \de^\mu\de_\mu \Big) \Psi(r_0),  \label{sl-sec2}
\eea
where we have introduced the matrix:
\be\label{scpert5}
{\cal B}(r) = \left( \begin{array}{cc} e^{2B_{UV}}\theta(r_0-r) & 0 \\0 &e^{2B_{IR}}\theta(r-r_0)\end{array}\right), \quad  e^{2B(r)} \equiv a^3 (r)z^2(r),
\ee
and $z(r)$ was defined in equation  (\ref{jun13}).  The wave-functions defined on the ``wrong'' side, i.e. $\psi_{IR}(r<r_0)$ and $\psi_{UV}(r>r_0)$,  are unphysical.

 Therefore,we have two  Sturm-Liouville problems, one on the left and one on the right,  with some generalization of Robin boundary condition which couple left and right modes at the interface\footnote{If the matrices $\Gamma_{1,2}$ were diagonal, we would have one independent Sturm-Liouville equation with  Robin boundary conditions at $r=r_0$ on each side.}.

In the following subsections we will investigate stability of the background solution under scalar perturbations, and discuss the conditions such that:

\begin{enumerate}
\item the theory does not propagate ghosts, i.e. modes with the wrong sign of the kinetic term;
\item the theory does not have tachyons around the flat solution, i.e. unstable modes that grow exponentially with time.
\end{enumerate}
The strategy we will follow will be to decompose the 5d-bulk modes into the corresponding  tower  of 4d mass eigenstates (which we assume discrete for simplicity, but this generalizes easily to a continuous spectrum), and to check for the absence of ghosts and tachyons in the usual 4-d sense. In order to do this, we have to write the effective action of the 4d modes.

\subsection{Action for scalar fluctuations}
The starting point to write the action for the 4d modes is the 5d action for scalar fluctuations. This can in principle be computed by expanding the Einstein-Dilaton action to quadratic order, using the background equations, and eliminating the redundant fields using constraints and gauge fixing. This is a very tedious calculation, but one can short-circuit it by noting that that there is a {\em unique} (up to a multiplicative constant) quadratic action whose variation gives equation (\ref{sl-sec1}) plus the boundary conditions (\ref{sl-sec2}):
\bea \label{ac1}
S_{5} = &&-{C\over 2}\int d^4x \Bigg[\int dr\, \left[\de_r \Psi^\dagger {\cal B}(r) \de_r \Psi + \de_\mu \Psi^\dagger {\cal B}(r) \de^\mu \Psi \right]  \nonumber \\  &&
 + \Psi^\dagger (r_0) \,\Sigma \Gamma_1\, \Psi (r_0) \,- \,   \de_\mu\Psi^\dagger(r_0)\,  \Sigma \Gamma_2 \, \de^\mu \Psi(r_0)
 \Bigg]
\eea
where we have introduced the matrix:
\be\label{ac2}
\Sigma \equiv \left( \begin{array}{cc} -e^{2B_{UV}(r_0)} & 0 \\0 & e^{2B_{IR}(r_0)}\end{array}\right)
\ee
Varying the action (\ref{ac1}) gives both the bulk field equations (\ref{sl-sec1})  and the matching conditions (\ref{sl-sec2})  (from integrations by parts plus the variation of the localized terms).

The value of the prefactor $C$ in equation (\ref{ac1})  is irrelevant for now\footnote{We will be concerned with the value of $C$ when we couple the scalar mode to a source in a later subsection.}, but its sign is crucial to decide whether there are ghosts in the model. However, it is well  known that the  Einstein-Dilaton theory we started with, described by the action  (\ref{A2}-\ref{A3}),  has no {\em bulk} ghost scalar modes, and  the one physical bulk scalar perturbation is healthy.  This knowledge  fixes the  sign of the bulk kinetic term to be the correct one, therefore we conclude that $C$ must be positive. In the rest of this section  we will set $C=1$.

We will now evaluate the action on a  ``Kaluza-Klein'' mode with 4d mass eigenvalue $m^2$,  of the form:
\be\label{ac3}
\Psi(r, x^\mu) = \Psi(r) \phi(x) ,
\ee
where the radial wave-function $\Psi$ satisfies (for $r\neq r_0$):
\be \label{ac4}
-{\cal B}^{-1} {d\over dr}\left({\cal B}(r) {d\Psi(r) \over dr}\right) = m^2 \Psi(r) , \qquad r\neq r_0,
\ee
plus the boundary condition (\ref{sl-sec2}). Inserting the KK ansatz (\ref{ac3}) in the action, and using  the boundary conditions, we arrive at the effective  action for the four-dimensional mode $\phi(x)$:
\be \label{ac5}
S_4 = -{1\over 2}{\cal N} \int d^4x \, \left(\de^\mu \phi \de_\mu \phi + m^2 \phi^2\right),
\ee
where
\be\label{ac6}
{\cal N} = \int_{r<r_0} dr\,  e^{2B_{UV}} \psi_{UV}^2 + \int_{r>r_0} dr\,  e^{2B_{IR}} \psi_{IR}^2 -  \Psi^\dagger (r_0) \,\Sigma \Gamma_2\, \Psi (r_0)
\ee
The action (\ref{ac5}) describes a four-dimensional scalar mode with mass $m$, and for it not to be neither a ghost nor a tachyon we must require that the two conditions hold simultaneously:
\be\label{ac7}
i) \, {\cal N} >0, \qquad ii)\, m^2 \geq 0.
\ee

\subsection{No ghosts} \label{ghosts}

We first consider the condition ${\cal N}>0$. The radial integrals in (\ref{ac6}) are manifestly positive, so the only constraint comes from the localized term.   Using the explicit form of the matrix $\Gamma_2$ in equation (\ref{scalar15}), after some algebra the condition ${\cal N}>0$ becomes:
\be\label{ac7.5}
 0< \int dr\,  \Psi^\dagger {\cal B} \Psi  \, +\,  \left(\begin{array}{cc}  {[z \psi] \over [z]} & -{[\psi]\over [1/z]} \end{array}\right)  {\cal K } \left(\begin{array}{c}  {[z \psi] \over [z]} \\ -{[\psi]\over [1/z]} \end{array}\right),
\ee
where we have defined:
\be\label{ac8}
{\cal K} \equiv  a_0^2\left( \begin{array}{cc} \tau_0& -6{dU_B\over d\f}\Big|_{\f_0} \\ -6{dU_B\over d\f}\Big|_{\f_0}  &  Z_0\end{array}\right) , \quad \tau_0\equiv 12\left(3{W_B\over W_{IR}W_{UV}}\Big|_{\f_0} - U_0\right).
\ee
Therefore, it is sufficient that the matrix ${\cal K}$ in equation (\ref{ac7}) have positive eigenvalues for the no-ghost condition to be satisfied. The eigenvalues are given by:
\be\label{ac9}
\lambda_{\pm} = a_0^2 \left[{Z_0+\tau_0 \over 2} \pm \sqrt{{\left(Z_0+\tau_0\right)^2 \over 4} - \left(\tau_0Z_0 - 36 \left({dU_B \over  d\f}\Big|_{\f_0} \right)^2\right)} \right].
\ee
They are both real since the matrix ${\cal K}$ is symmetric. In addition, they are both positive if  both conditions below are met:
\be\label{ac10}
\tau_0>0, \qquad Z_0 \tau_0 > 36 \left({dU_B \over  d\f}\Big|_{\f_0} \right)^2.
\ee
Recall that for the coupling of the induced gravity term, $U_0>0$, for the spin-2 modes not to be ghost-like at short distances. Note that for the scalar perturbations this term contributes with the wrong sign and this agrees with the observation made in \cite{padilla}.

The first condition in equation (\ref{ac10}) implies,  among other things,   $W_B(\f_0)>0$, i.e.  the brane  has positive tension. This  is not surprising,  as negative tension branes that are allowed to fluctuate usually lead to ghost-like modes or tachyons \cite{fax1}. Similarly, the second condition demands that $Z_0>0$,  meaning that the scalar field brane kinetic term  should have the correct sign.

Notice that (\ref{ac10}) is not a functional constraint, i.e. it does not need to be satisfied for arbitrary value of $\f$: it only needs to hold at the stabilized brane position.

The relations (\ref{ac10})  are  useful sufficient conditions for the absence of ghosts. They are not necessary, since even if they are violated, equation (\ref{ac7.5})  may still hold thanks to the positive contribution to the bulk term. However this has to be checked by performing the fluctuation analysis. On the other hand, the relations (\ref{ac10})  are very simple and depend only on background quantities.

\subsection{No tachyons}
We now consider a solution of equations (\ref{sl-sec1}-\ref{sl-sec2}) which is a 4d mass eigenstate, i.e. $\de_\mu\de^\mu \Psi = m^2 \Psi$. This implies that $\Psi$ satisfies the radial eigenstate equation  (\ref{ac4}), with  eigenvalue $m^2$. The model has tachyonic instabilities if  the radial operator (\ref{ac4}) has negative eigenvalues.

We now  multiply both sides of equation (\ref{ac4}) by $\Psi^\dagger$ and integrate over the radial direction:
\be
m^2  \left[\int_{r<r_0} e^{2B_{UV}}\psi^2_{UV} + \int_{r>r_0} e^{2B_{IR}}\psi^2_{IR}\right]  = - \int_{r<r_0}  \psi_{UV}(e^{2B_{UV}} \psi'_{UV})' - \int_{r>r_0}  \psi_{IR}(e^{2B_{IR}} \psi'_{IR})'.
\ee
Integrating by parts and using the boundary conditions, as well as the mass shell condition $\de_\mu \de^\mu = m^2$, we find:
\be
m^2 \int {\cal B} \,\psi^2 = \int (\Psi')^\dagger{\cal B}(\Psi')  + \Psi^\dagger(r_0) \Sigma \left (\Gamma_1 + m^2 \Gamma_2\right) \Psi(r_0).
\ee
The first term on the r.h.s. is positive, therefore (recall the definition of ${\cal N}$ in equation (\ref{ac6})):
\be
m^2 {\cal N} - \Psi^\dagger(r_0) \Sigma \Gamma_1  \Psi(r_0) \geq 0.
\ee
Using the explicit form  of $\Gamma_1$ from equation (\ref{scalar15}) and the definition (\ref{ac2}),  we find:
\be
\Sigma \Gamma_1 = a^4_0 {\tilde{{\cal M}}^2  \over [1/z]^2} \left( \begin{array}{rr} 1& -1 \\ -1  & 1\end{array}\right),
\ee
which leads to  a lower  bound on the eigenvalues:
\be
m^2 {\cal N} \geq a^4 \tilde{{\cal M}}^2 \, [\psi]^2/[1/z]^2.
\ee
This   implies that,  if there are no-ghosts (${\cal N}>0$), then the  absence of tachyonic instabilities is guaranteed if:
\be
\tilde{\cal  M}^2 \equiv {d^2 W_B \over d\f^2}\Big|_{\f_0} - \left[{d^2 W \over d\f^2}\right]^{IR}_{UV} \geq 0.
\ee
This  is a  ``positive mass squared'' condition for the effective brane  mass.

\subsection{Scalar-mediated interaction}

In this section we derive the interaction between two brane sources mediated by the exchange of the scalar modes, at the linearized level.

The action including localized sources can be obtained by adding to equation (\ref{ac1}) the  linearized version of the brane-matter action (\ref{ind3-ii}), keeping only scalar modes. The corresponding sources  at linear order are:
\be \label{stress-tensor-x}
T_{\mu\nu} = -{2\over \sqrt{\gamma}} {\delta S_m \over \gamma^{\mu\nu}},  \quad O =   {\delta S_m \over \delta \f}.
\ee

We will assume   $T^{\mu\nu}$ is conserved, therefore it does not couple to the $E$-mode in the decomposition  (\ref{pert3}). Then,  we keep only the  metric perturbation  $\psi$, which  will couple to the trace of the stress tensor
  $T \equiv \eta^{\mu\nu}T_{\mu\nu}$. 

The resulting action is:
\bea \label{actot}
S = &&-{M^3\over 2}\int d^4x \Bigg[\int dr\, \left[\de_r \Psi^\dagger {\cal B}(r) \de_r \Psi + \de_\mu \Psi^\dagger {\cal B}(r) \de^\mu \Psi \right]  \nonumber \\  &&
 + \Psi^\dagger (r_0) \,\Sigma \Gamma_1\, \Psi (r_0) \,- \,   \de_\mu\Psi^\dagger(r_0)\,  \Sigma \Gamma_2 \, \de^\mu \Psi(r_0)
 \Bigg]   \nonumber \\
&& + \int d^4x\, e^{4A_0}\hat{\Psi}^\dagger {\cal T}
\eea
where we have introduced the vectors:
\be\label{scforce2-i-app}
\hat{\Psi} = \left(\begin{array}{c} \hat{\psi}(r_0) \\ \hat{\chi} (r_0)\end{array} \right), \qquad  {\cal T} =\left(\begin{array}{c} e^{-2A_0}T \\  O\end{array} \right).
\ee
We have fixed the  overall multiplicative constant $C$ in the quadratic part of the action  (\ref{actot}) by demanding that the bulk term matches the  Einstein action expanded to quadratic order.  The  coefficient  of the last  term can be checked by recalling the definition (\ref{stress-tensor-x}) of the brane stress tensor and noting that the linearized action involving only  a trace perturbation $h_{\mu\nu} = 2 e^{2A}\eta_{\mu\nu}\psi$ (cfr. the definitions (\ref{pert1}-\ref{pert3}))  is  
\be \label{coeff-linear}
\delta S_m = -\int \sqrt{-\gamma}{1\over 2}h^{\mu\nu}T_{\mu\nu} = \int e^{2A_0} \psi T. 
\ee
In writing the last line in  equation (\ref{actot}), one has to   take into account that the brane is at $r=r_0 + \rho$:  expanding around the equilibrium position, one finds that only the gauge-invariant, continuous combinations $\hat{\psi} = \psi(r_0) + A'(r_0)\rho$ and $\hat{\chi} = \chi(r_0) + \bar{\f}(r_0)\rho$ enter.

We now rewrite the sources in terms of  $\Psi \equiv (\psi_{UV}, \psi_{IR}) $.   Using equations  (\ref{scalar13.5-sec}-\ref{scalar13.6-sec}), we find:
\be \label{scforce3-i-app}
\hat{\Psi} = P \Psi, \qquad P \equiv - {z_{IR} z_{UV}\over [z]}\left( \begin{array}{cc} {1\over z_{IR}}&-{1\over z_{UV}}   \\ 1 & 1\end{array}\right).
\ee
The source term  in equation (\ref{actot})  becomes then:
\be \label{scforce4-app}
S_{source}  =  \int d^4x e^{4A_0} \, \Psi^\dagger {\cal J} \qquad {\cal J} = \left(\begin{array}{c} J_{UV} \\ J_{IR}\end{array} \right) \equiv P^\dagger {\cal T}.
\ee

The bulk equation  is still given by  (\ref{sl-sec1}), but the matching condition (\ref{sl-sec2})  is now modified by the addition of the sources:
\bea
&&\de_r \left[ {\cal B} (r)\de_r \Psi\right] + {\cal B}(r)  \de_\mu  \de^\mu \Psi = 0 , \qquad r\neq r_0\label{sl-sec1-2}\\
&& \nonumber \\
&& \de_r \Psi(r_0) = \Gamma \Psi(r_0) + \Sigma^{-1}{{\cal J} \over M^3},   \qquad  \Gamma \equiv \Gamma_1 + \Gamma_2\, \de^\mu\de_\mu . \label{sl-sec2-2}
\eea
where the matrix $\Sigma$ is given in equation (\ref{ac2}).

To compute the interaction between sources on the brane,  we will write the bulk-to-brane Green's function associated to equations (\ref{sl-sec1-2}-\ref{sl-sec2-2}), i.e. we look for a solution (in momentum space) of the form:
\be\label{scpsi}
\Psi(r,p) = {\cal G}(r,r_0;p){{\cal J}(p)\over M^3},
\ee
where $ {\cal G}(r,r_0;p)$ is a $2\times 2$ matrix propagator satisfying:
\bea
&&-\de_r \left[ {\cal B} (r)\de_r {\cal G}(r,r_0;p)\right] + p^2 {\cal B}(r)   {\cal G}(r,r_0;p) = 0 , \qquad r\neq r_0\label{scalarG1}\\
&& \nonumber \\
&& \de_r {\cal G}(r,r_0;p) \Big|_{r_0}= \Gamma {\cal G}(r_0,r_0;p) + \Sigma^{-1}
 \label{scalarG2}
\eea

To solve equations  (\ref{scalarG1}-\ref{scalarG2}), we take ${\cal G}$  of the form:
\be \label{scforce7}
{\cal G}(r,r_0;p) = {\cal D}(r;p) {\cal H}(p), \qquad {\cal D} = \left( \begin{array}{cc} D_{UV}(r) & 0    \\ 0 &  D_{IR}(r) \end{array}\right).
\ee
where ${\cal H}$ is a constant $2\times 2$ matrix  and the diagonal matrix  ${\cal D}(r;p)$   satisfies the bulk equation:
\be \label{scforce8}
-{\cal B} \de_r^2 {\cal D} - \de_r{\cal B} \de_r {\cal D} + p^2 {\cal B}  {\cal D} =
0, \qquad r \neq r_0,
\ee
with normalizable boundary conditions in the UV and in the IR.
Since the upper and lower part of ${\cal D}$ are independent functions on $r<r_0$ and $r>r_0$ respectively, we are free to choose any (non-vanishing) boundary condition at $r_0$ for each of them. A convenient choice is:
\be \label{scforce8-i}
{\cal D}'(r_0) = -\Sigma^{-1}
\ee
where $\Sigma$ is defined in equation (\ref{ac2}).

With the ansatz (\ref{scforce7}-\ref{scforce8}), ${\cal H}$ must satisfy  the  algebraic  matrix equation:
\be\label{scforce9}
\Big[ {\cal D}'(r_0;p) - \Gamma {\cal D}(r_0;p)\Big] {\cal H}(p)  = \Sigma^{-1}
\ee
which can be solved by inverting the left hand side. Then, with the boundary condition (\ref{scforce8-i}), the brane-to-brane scalar propagator ${\cal G}(p) = {\cal D}(r_0;p) H(p)$ is:
\be\label{scforce10}
{\cal G}(p) = - \Big[ \Sigma \Gamma +  \left({\cal D}\right)^{-1}(r_0;p)\Big]^{-1}.
\ee
This is the matrix version of equation (\ref{ind8}), the matrix $\Sigma \Gamma$ in the denominator being the quadratic form governing   the brane-localized terms.

By inserting the solution (\ref{scpsi}) back in the action we obtain  the  interaction mediated by scalar modes:
\be \label{scforce11}
S_{int} = -{1\over 2}\int {d^4 q \over (2\pi)^4} {\cal T}^\dagger (q)  G_s(q)  {\cal T}(-q),
\ee
where:
\be\label{scforce12}
G_s(q) \equiv {e^{4A_0}\over 2M^3} \left[ P^{\dagger -1}\Sigma \left(\Gamma_1 + e^{2A_0}
q^2 \Gamma_2\right) P^{-1} + P^{\dagger -1}\ {\cal D}^{-1}(r_0;q)  P^{-1}
\right]^{-1}.
\ee
The above expression  is now expressed  in terms of the physical momentum observed on the brane, $q = e^{-A_0}p$, and we have rotated back to the basis of stress-tensor and dilaton charge sources, (\ref{scforce2-i-app}).

We can simplify equation (\ref{scforce12}) by noticing that the first term in the parenthesis  is nothing but the matrix $\Sigma\Gamma$, expressed in terms of the basis $(\hat{\psi}(r_0), \hat{\chi}(r_0))$, and we can read-off its expression from  equations (\ref{gi1}) and (\ref{gi2}):
\be \label{scforce13}
P^{\dagger -1}\Sigma \left(\Gamma_1 + e^{2A_0}
q^2 \Gamma_2\right) P^{-1} =  e^{4A_0}\tilde{{\cal M}}^2\left( \begin{array}{ccc} 0& &  0  \\  0 &  &  2\end{array}\right) + e^{4A_0} \left( \begin{array}{cc} \tau_0& -6{dU_B\over d\f} \\ -6{dU_B\over d\f}\Big|_{r_0}  &  Z_0\end{array}\right) q^2.
\ee

In order for the model to be viable, one must carefully examine the strength and the range of the coupling to matter, and compare it with current constraints on fifth forces and violations of the equivalence principle. We will postpone this discussion  to further work, where we will explore in more detail the phenomenology (and viability) of concrete realisations of the framework. Here, we limit ourselves to observe the following important features:
\begin{enumerate}
\item The scalar modes that couple to the dilaton have a mass controlled by $\tilde{M}^2$, which is determined by the second derivatives of the bulk and brane potentials at the interface. This can be large as it is expected to scale as $\Lambda^4$. On the other hand, the modes which couple to the trace of the stress tensor, and which come from the gravitational sector, have a mass which is controlled by the ${\cal D}^{-1}(r_0,0)$, as in the case of the spin 2.
\item Similarly, the normalization of the propagator (i.e. the strength of the coupling) is  roughly controlled by the two eigenvalues of the second term in equation (\ref{scforce13}), given explicitly in equation (\ref{ac9}). If these are large, i.e. if both $\tau_B(r_0), Z_B(r_0) \gg U_B(r_0)$, these modes may have weaker couplings to matter than the spin-2 modes.
\end{enumerate}
For a full phenomenological discussion one must provide a specific model of the  coupling of the dilaton to brane matter, diagonalize the propagator matrix and take into account the mixing between the modes. This goes beyond the scope of the present work.

\subsection{On the presence of the vDVZ problem}

We will not attempt here a complete discussion of the phenomenology related to the scalar mode exchange. However, we close off by briefly discussing a point which is worth pointing out: the possibility that the exchange of the light scalar modes (those which correspond to the zero eigenvalue of the mass matrix appearing in equation (\ref{scforce13}),  and couple to the trace of the stress tensor)  may naturally  cancel that of the  longitudinal component of the tensor modes at scales where the interaction looks effectively four-dimensional. If that is the case, the interaction will be completely similar to the exchange of massless gravitons with only two helicities, i.e. the van Dam-Veltman-Zakharov problem \cite{VDVZ} would  be absent at the linear order.

Consider the total interaction between two brane stress tensors, to which both the tensors and scalars contribute, over distances larger than the inverse mass of the ``heavy'' scalar modes, corresponding to the non-zero eigenvalue in equation (\ref{scforce13}), with  $\tilde{{\cal M}} \approx \Lambda$. Beyond  this scale, the scalar contribution reduces effectively to the exchange of the light modes and we can ignore the effect of the scalar charge $O$ of the source, defined in equation (\ref{scforce2-i-app}). Although this is not completely equivalent, we can  simplify the discussion by simply setting the scalar source $O=0$ in equation (\ref{scforce2-i-app}). This leaves the coupling to the stress tensor trace (the upper component of ${\cal T}$) in equation (\ref{scforce11}).

Now suppose that the momenta are in the ``DGP'' regime, where the ``bulk'' contributions (i.e. the last term in equation (\ref{scforce12}) for the scalar, and the ``$1$'' in the denominator of equation (\ref{brane5}) for the tensor) are negligible. Then, the total exchange between two stress tensors, mediated by the tensor and the light scalar modes, gives the approximate potential:
\be\label{vdvz1}
{\cal V}(q)  \simeq -{1\over q^2}\left[{1\over 4M^3 U_0}\left(T_{\mu\nu}(q)T^{\mu\nu}(-q) -{1\over 3}T^\mu_\mu(q) T^\nu_\nu(-q)\right)  + {1\over 2 M^3 \tau_0} T^\mu_\mu (q)T^\nu_\nu(-q)\right].
\ee
where we recall the definition (\ref{ac8}):
\be\label{vdvz2}
\tau_0 =  12\left(3{W_B\over W_{IR}W_{UV}}\Big|_{\f_0} - U_0\right).
\ee
Now suppose that:
\be\label{vdvz3}
{W_B\over W_{IR}W_{UV}}\Big|_{\f_0} \ll U_0, \quad \Rightarrow \quad \tau_0 \simeq -12U_0.
\ee
Then equation (\ref{vdvz1}) becomes approximately:
\be\label{vdvz4}
{\cal V}(q) \simeq  -{1\over q^2}\left[{1\over 2M_p^2}\left(T_{\mu\nu}(q)T^{\mu\nu}(-q) -{1\over 2}T^\mu_\mu(q) T^\nu_\nu(-q)\right)\right], \qquad M_p^2 = 2 M^3 U_0,
\ee
i.e. the scalar mode changes the $1/3$ into $1/2$ in the tensor structure, which becomes that of a massless graviton with two transverse polarizations.
We have already seen this mechanism in brane-world models supporting {\em localized} massless gravitons, in which the brane-bending mode cancels the extra longitudinal polarization of the would-be five-dimensional graviton zero mode \cite{gata}. The novelty here is that the same mechanism is reproduced at the level of quasi-localized resonances, in the regime where the DGP mechanism is at work.

The left-over interaction from the light scalar, i.e. the first subleading term in the approximation (\ref{vdvz3}), takes the form:
\be\label{vdvz5}
\delta V(q) \simeq  {1\over 2 M_{eff}^2} {T^\mu_\mu(q) T^\nu_\nu(-q) \over q^2}, \qquad M_{eff}^2 = 2 M_p^2 \left({W_{UV}W_{IR}\over -W_B}\right)_{\f_0}\!\!\!\!U_0. 
\ee
 If $W_B(\f_0)<0$, this  mismatch can be seen as the exchange of a healthy (i.e. non-ghostlike) light scalar with a coupling much weaker than gravity since, by equation (\ref{vdvz3}), $M_eff\gg M_p$.  Depending on the scales in the model, this can be made invisible in precision tests of the equivalence principle.

The price to pay in this situation is that we have to relax the sufficient condition $\tau_0>0$ which would automatically make the model ghost-free: although there are manifestly no ghosts in the high momentum regime, one still has to check that this situation persists at {\em all} momenta, both in the 5d and in the massive gravity regime, when the tensor and scalar modes decouple because of the differences in the tensor and scalar  bulk propagators.

The absence of ghosts is equivalent to the requirement  that the  quantity (\ref{ac6}) is positive for all modes. This question can be addressed using a spectral representation for the propagator, and checking for violation of positivity of the corresponding spectral density\footnote{In the recent work \cite{newpadilla}, this formalism was used to  constrain self-tuning models. In fact, the reasoning  in \cite{newpadilla} applies to theories featuring  ``degravitation'' of the cosmological constant, in which the coupling of (effective) gravitons to vacuum energy vanishes, and a change in vacuum energy has no effect. This is stronger than self-tuning, and  it is not the case in our framework, in which a change in the vacuum energy {\em does} have an effect (it changes  the background solution) but it does not contribute to the 4d curvature.}. This leads to the conclusion that, if we cancel the linearized vDVZ discontinuity at large momenta to reproduce equation (\ref{vdvz4}), this necessarily introduces ghost-like modes\footnote{We thank Massimo Porrati for an illuminating discussion on this point.}.  Indeed, consider the spectral representation for the effective 4d Euclidean scalar propagator:
\be \label{vdvz6}
G_4(q)  = \int_0^{+\infty} ds\,  {\rho(s) \over q^2 + s}.
\ee
The spectral density $\rho(s)$ is essentially given by the quantity ${\cal N}^{-1}$, computed for $m^2=s$, in equation  (\ref{ac6}).  Stability requires $\rho(s)$ to be non-negative for all $s>0$. On the other hand,  to cancel the vDVZ discontinuity, in the relevant momentum range where the propagator  behaves as $1/q^2$ we must have:
\be  \label{vdvz7}
G_4(q) \simeq -{1\over 24}{1\over M^3 U_0}{1\over  q^2}
\ee
which is incompatible with equation (\ref{vdvz6}) in which $\rho(s)$ is non-negative.

Therefore, the resolution of the  vDVZ problem must be sought at the  nonlinear level. We leave this for future work.

\section*{Acknowledgements}
\addcontentsline{toc}{section}{Acknowledgements}

We would like to thank Antonio Amariti, Davide Forcella, David Langlois, Antonio H. Padilla, Massimo Porrati, Daniele Steer,  and Lukas Witkowski for discussions. C.C. and F.N. thank the Crete Center for Theoretical Physics for hospitality during the initial stages of this work.
This work was supported in part  by the Advanced ERC grant SM-grav, No 669288.

\newpage
\appendix

\begin{appendix}
\renewcommand{\theequation}{\thesection.\arabic{equation}}
\addcontentsline{toc}{section}{Appendices}
\section*{APPENDIX}

\section{The different types of junctions} \label{types-app}

In this Appendix, we  analyze the different possible qualitative behavior at the intersection, depending on the sign and the size of the brane potential (and its derivative) at $\f_0$.
As we have seen in section \ref{selfie}, once the IR solution $W_{IR}$ and the brane potential $W_{B}$ are fixed,  the interface position $\f_0$ and the UV superpotential $W_{UV}$ are determined by the two equations (\ref{sol1}).

We begin with some preliminary considerations.
\begin{enumerate}
\item
First,  we have to fix some discrete ambiguities. Notice that the superpotential equation (\ref{superpot-i}) is invariant under $W\to -W$, thus there is a two-fold degeneracy of solutions. We fix this ambiguity by choosing $W_{IR} >0$. Since there is the the flow equations have the symmetry $(u,W)\to (-u,-W)$,  fixing the positie sign of $W$  implies by equation  (\ref{FE11}) that the coordinate $u$ increases as $A(u)$ decreases,  i.e. that $u$ increases towards the IR.

\item Notice that the matching conditions (\ref{FE4}) are written assuming the same direction of the normal on both sides of the interface, therefore the direction of $u$ does not change as we cross the brane.

\item The function $W(\f(u))$ is a monotonically increasing function of $u$:
\be\label{all1}
{d\over du } W(\f(u)) = \dot{\f}W'(\f(u)) = (\dot{\f})^2
\ee
Therefore in crossing the interface from the IR to the UV, we have to continue the solution in the direction where $W$  decreases, because  this is the same direction in which $u$ decreases. This is explained in figure \ref{interface}: the arrows indicate the direction of increasing $u$ (which is the same as increasing $W$), and at the interface the arrows must point away from the brane on one of the solutions  and towards the brane on the other.

\begin{figure}[h!]
\begin{center}
\includegraphics[width=7cm]{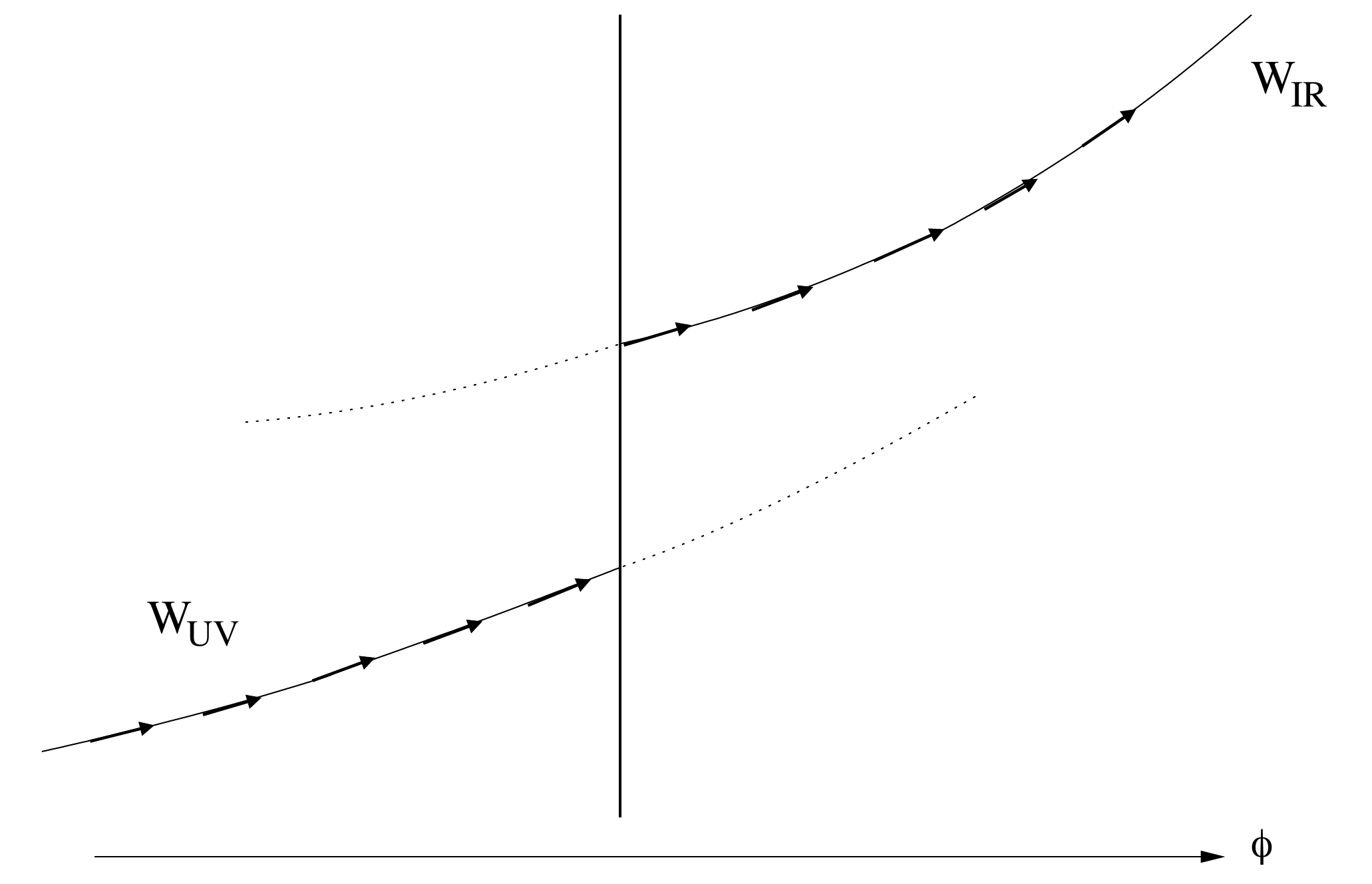} \hspace{0.5cm} \includegraphics[width=7cm]{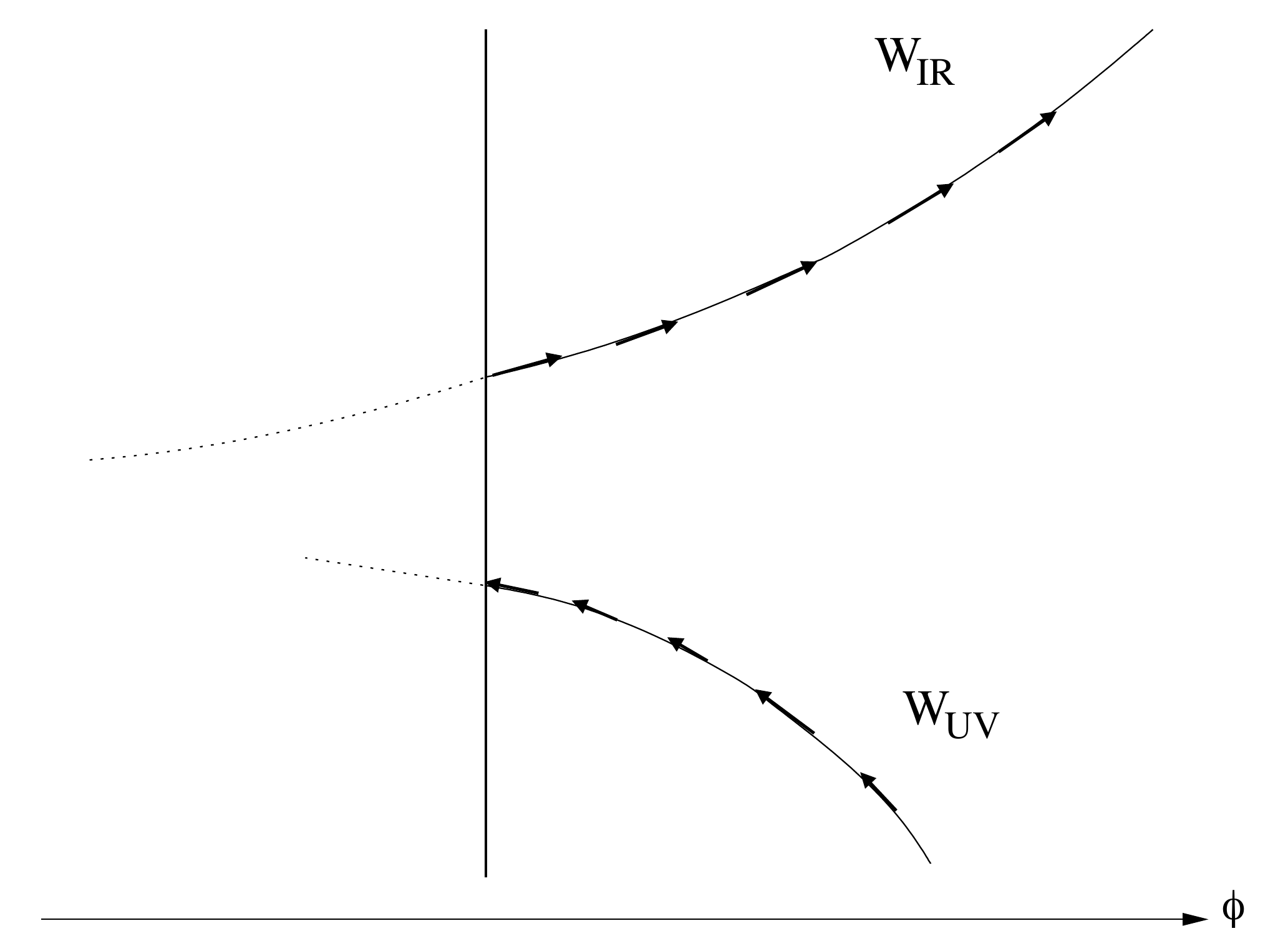} \\
(a) \hspace{7cm} (b)
\caption{The figures above show the allowed ways of joining two solutions  of the superpotential equation at the interface. The arrows indicate the direction of increasing $u$, which coincides with the direction of increasing $W$. The junction must be  such  that the arrows on the IR solution point away from the brane, whereas those on the UV solution point towards the brane. In terms of $\f$, this condition can be realized on opposite sides of the point $\f=\f_0$  when $W'_{UV}>0$ (left figure),  or on the same side, when  $W'_{UV}<0$ (right figure). The dotted lines indicate how the solutions would continue past the interface.} \label{interface}
\end{center}
\end{figure}

\item Finally, notice that in the $(\f, W(\f))$ plane there is a {\em forbidden region},  where no solution to the superpotential equation (\ref{superpot-i}) exists: it is the region where $W'$ becomes imaginary.  Indeed, for $W'(\f_0)$ to be real, $W(\f_0)$  must satisfy:
\be \label{all2}
|W(\f_0)| > B(\f_0), \qquad B(\f) \equiv {\sqrt{-2V(\f)} \over Q}.
\ee
where $Q = \sqrt{d/2(d-1)}$. This condition in particular must hold for $W_{UV}(\f_0) = W_{IR}(\f_0) - W_B(\f_0)$. This implies that there can be   no solutions to the matching condition  such that $ |W_{IR}(\f_0) - W_B(\f_0)| < B(\f_0)$. This fact will be used later.

\end{enumerate}
We will now analyse the full geometry we  obtain depending on the value of $W_B$ and its derivative at the interface. We must distinguish three cases.
\begin{itemize}
\item[A1.] $ W_B(\f_0) < 0$

In this case, $W_{UV}(\f_0) > W_{IR}(\f_0) >0$ and  the structure of the full solution is as shown in figure \ref{caseA}. The sign of $W'_{UV}$ is fixed by the sign of $W'_{IR} - W'_B$ at the interface. If $W'(\f_0) >0$, the solution can connect  directly to the UV at $\f=0$ (figure \ref{caseA} (a) ). On the other hand, if $W'_{UV} < 0$, then by the junction rules shown in figure \ref{interface} we must follow $W_{UV}$ for {\em increasing} $\f$ (i.e. on the same side of the interface in field space (figure \ref{caseA} (b)). As shown in the figure,  the solution will eventually reach the curve $B(\f)$ where it can be glued continuously (as described in \cite{multirg}) with a solution with $W' >0$, which will in turn flow to the UV fixed point.

\item[A2.]  $0< W_B(\f_0) < W_{IR}(\f_0) - B(\f_0)$

In this case, we have $B(\f_0)  < W_{UV}(\f_0) < W_{IR}(\f_0)$,  and depending on the sign of $W'$ the structure is  essentially the same as in case $A1$, except that the UV superpotential starts lower than the IR superpotential (see figure \ref{caseB}).

\begin{figure}[h!]
\begin{center}
\includegraphics[width=7.3cm]{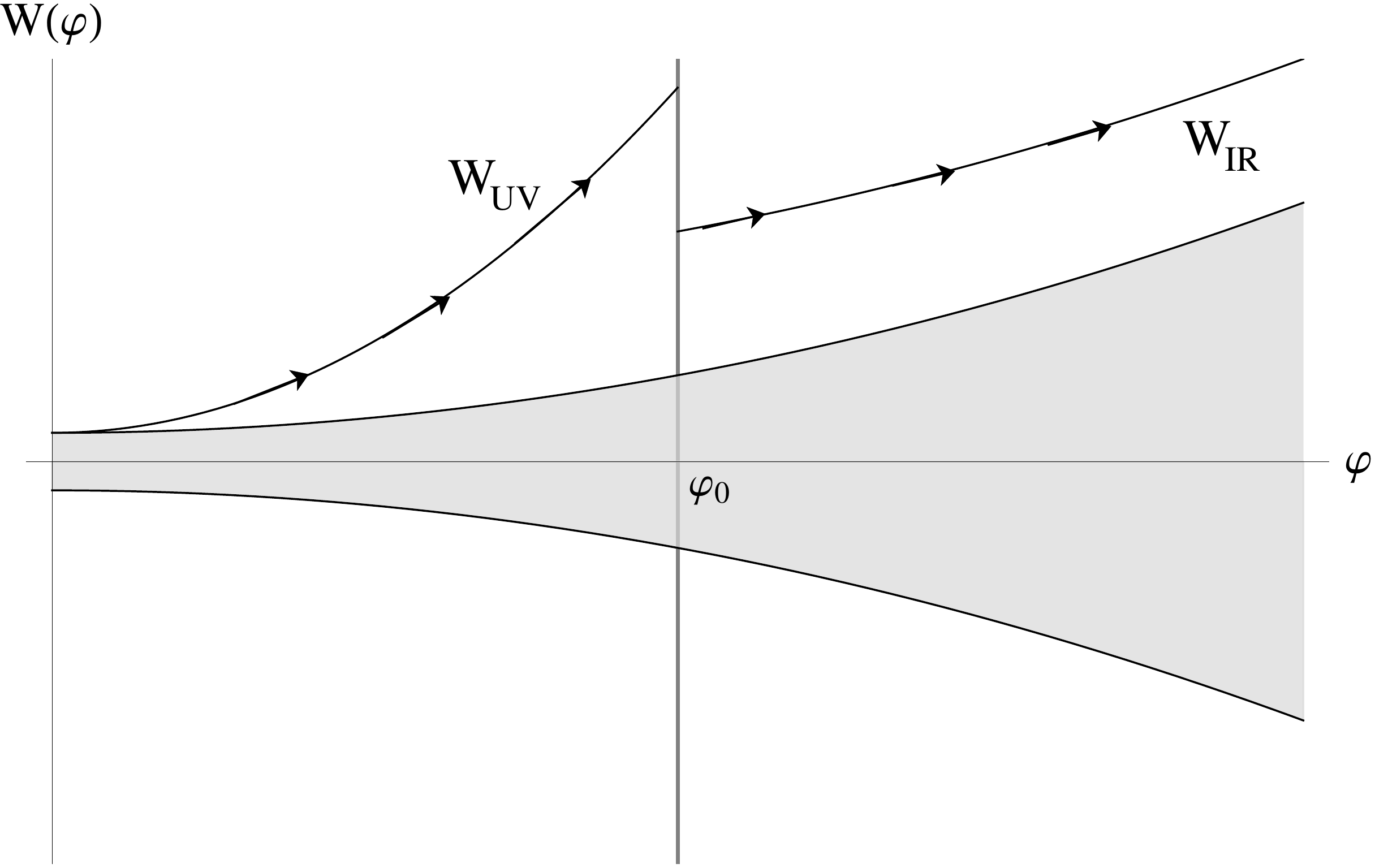} \includegraphics[width=7.3cm]{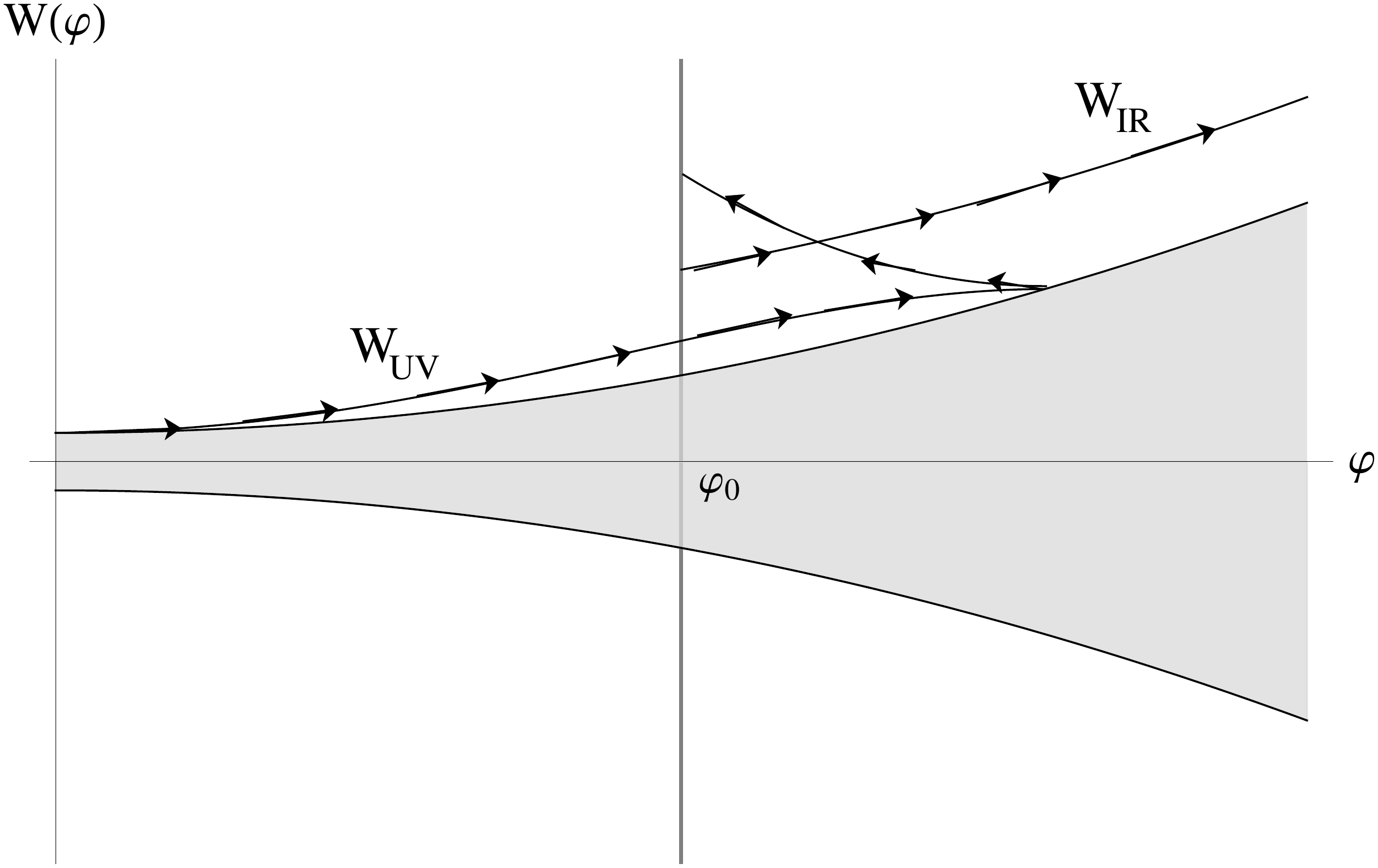} \\
(a) \hspace{7cm} (b)
\caption{Sketch of the global behavior of the superpotential   in case $A1$, when $W_B <0$. The left figure corresponds to the case $W'_{UV}(\f_0)>0$, whereas the right figure corresponds to $W'_{UV}(\f_0)<0$. In the former case,  the UV solution connects directly to the fixed point at $\f=0$. In the latter, it continues past $\f_0$ and goes to the UV fixed point  after a bounce at some $\f > \f_0$. The shaded area corresponds to the forbidden region $-B(\f) < W(\f) < B(\f)$, where $B(\f)$ is defined in equation (\protect\ref{all2}).} \label{caseA}
\end{center}
\end{figure}

\begin{figure}[h!]
\begin{center}
\includegraphics[width=7.3cm]{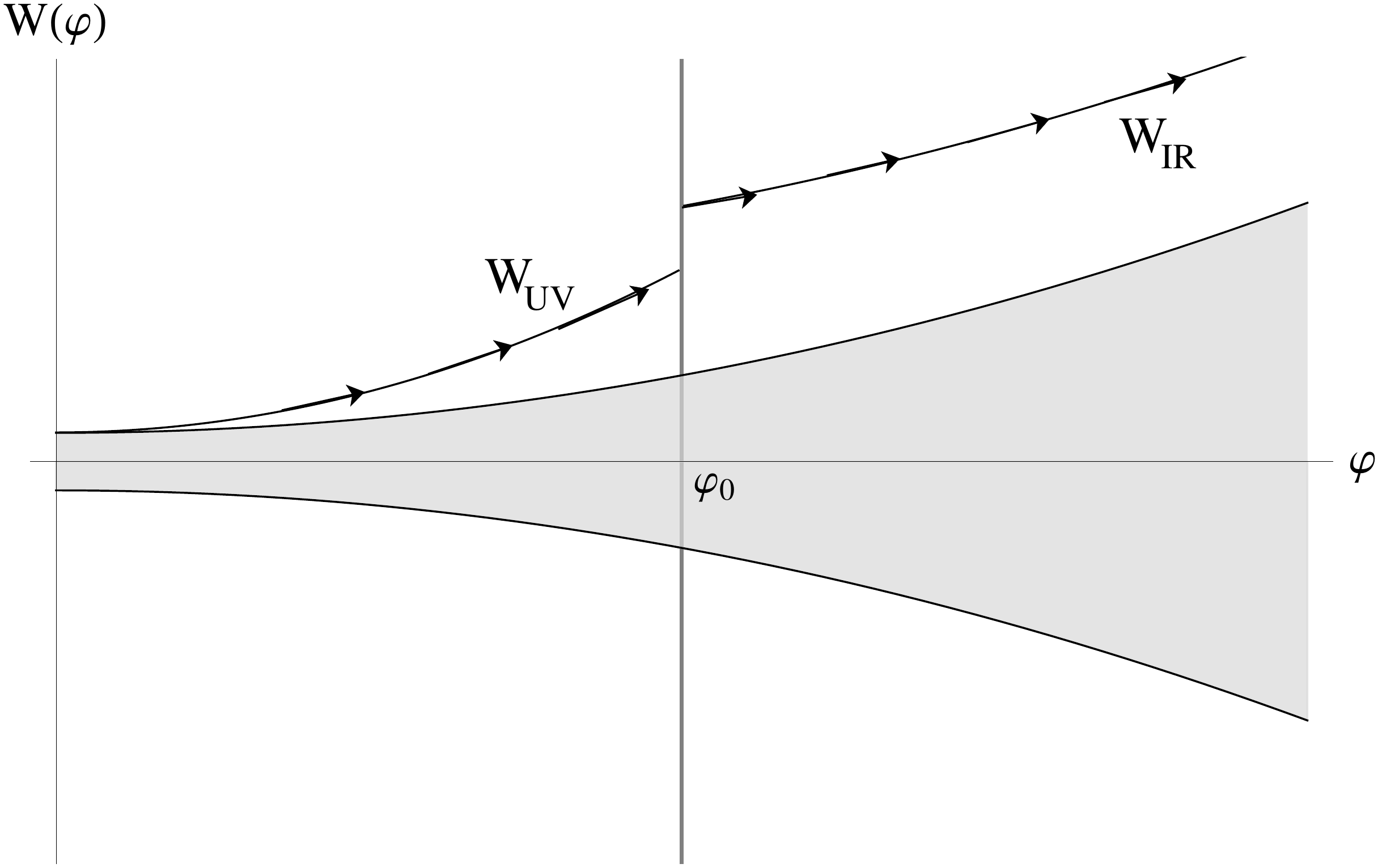} \includegraphics[width=7.3cm]{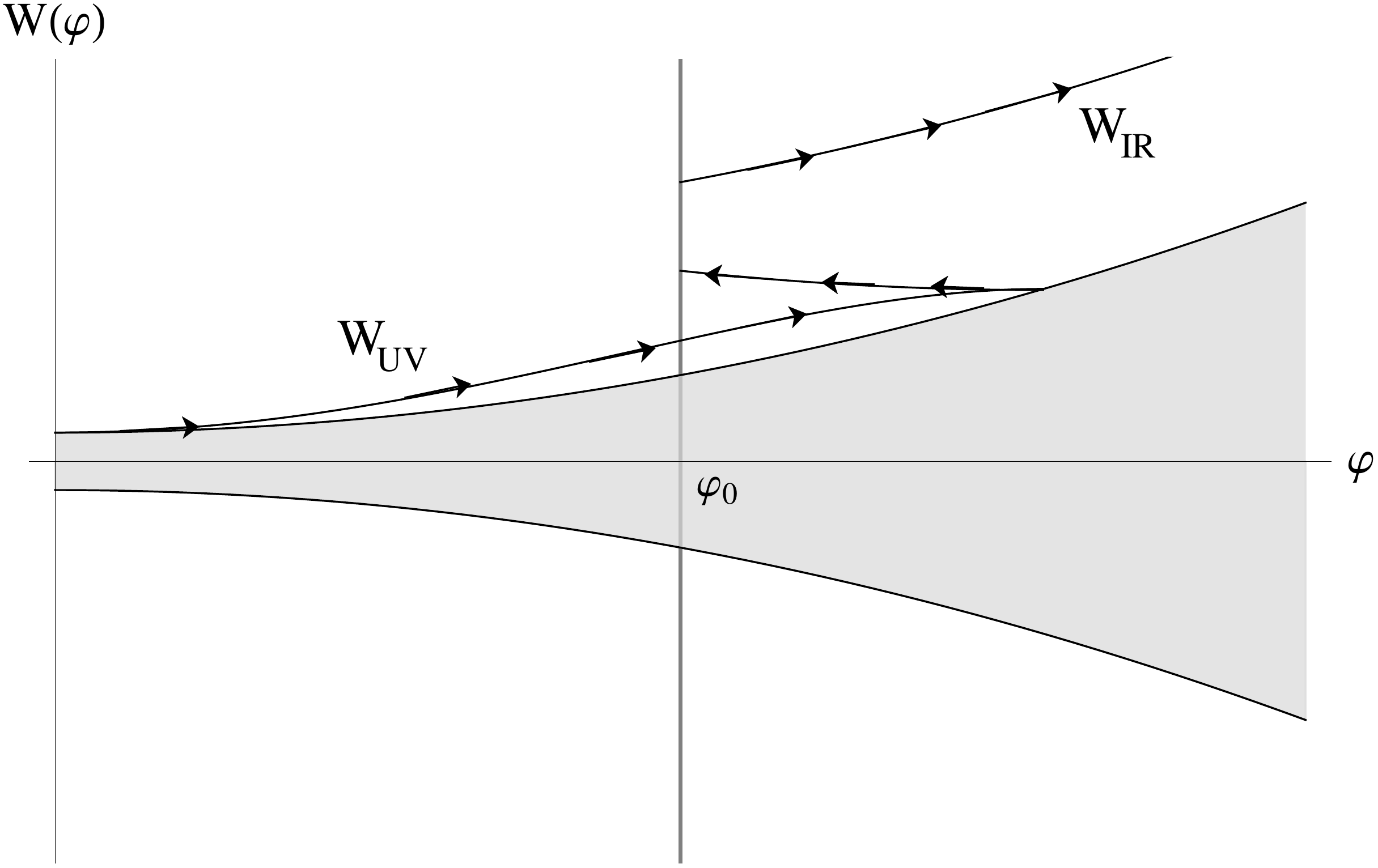} \\
(a) \hspace{7cm} (b)
\caption{The behavior of the full solution in case $A2$, i.e. $W_B(\f_0)>0$ and $W_{UV}(\f_0) >B(\f_0)$. The left and right figures correspond to $W_{UV}'(\f_0)>0$ and $W_{UV}'(\f_0)<0$. In the latter case, like in  the analog situation in case $A1$ (figure \protect\ref{caseA} (b)), the solution bounces at $\f > \f_0$ before reaching the interface from the right.} \label{caseB}
\end{center}
\end{figure}

\item[B.]  $ W_B (\f_0)> W_{IR}(\f_0) + B(\f_0)$

In  this case $W_{UV}(\f_0)$ is negative. The two possible  behaviors across the interface corresponding to either sign of $W'_{IR}(\f_0)-W'_B(\f_0)$ are represented in figure \ref{caseC}. Notice that as we cross the interface into the ``UV'' region, we are forced to follow the superpotential to more and more negative values. However, now $\dot{A}(u) = -W > 0$, thus the scale factor actually increases with $u$, and the maximum value is attained {\em at the interface}: thus this solution does not connect to an asymptotic $AdS$ boundary, but it connects two finite-volume  regions with asymptotically vanishing scale factor (i.e. two regions of the IR type). For generic brane potential,  the solution  will be singular, i.e. the interface will {\em not} connect two IR-acceptable solutions. For this to happen, we need a fine-tuning of the bulk and brane potentials, i.e. we need:
\be
W_B(\f_0) = 2 W_{IR}(\f_0).
\ee
This fine-tuning  is the asymmetric version of the similar fine-tuning of the brane tension to the bulk cosmological constant in the  one-brane  Randall-Sundrum setup.

\begin{figure}[h!]
\begin{center}
\includegraphics[width=7.3cm]{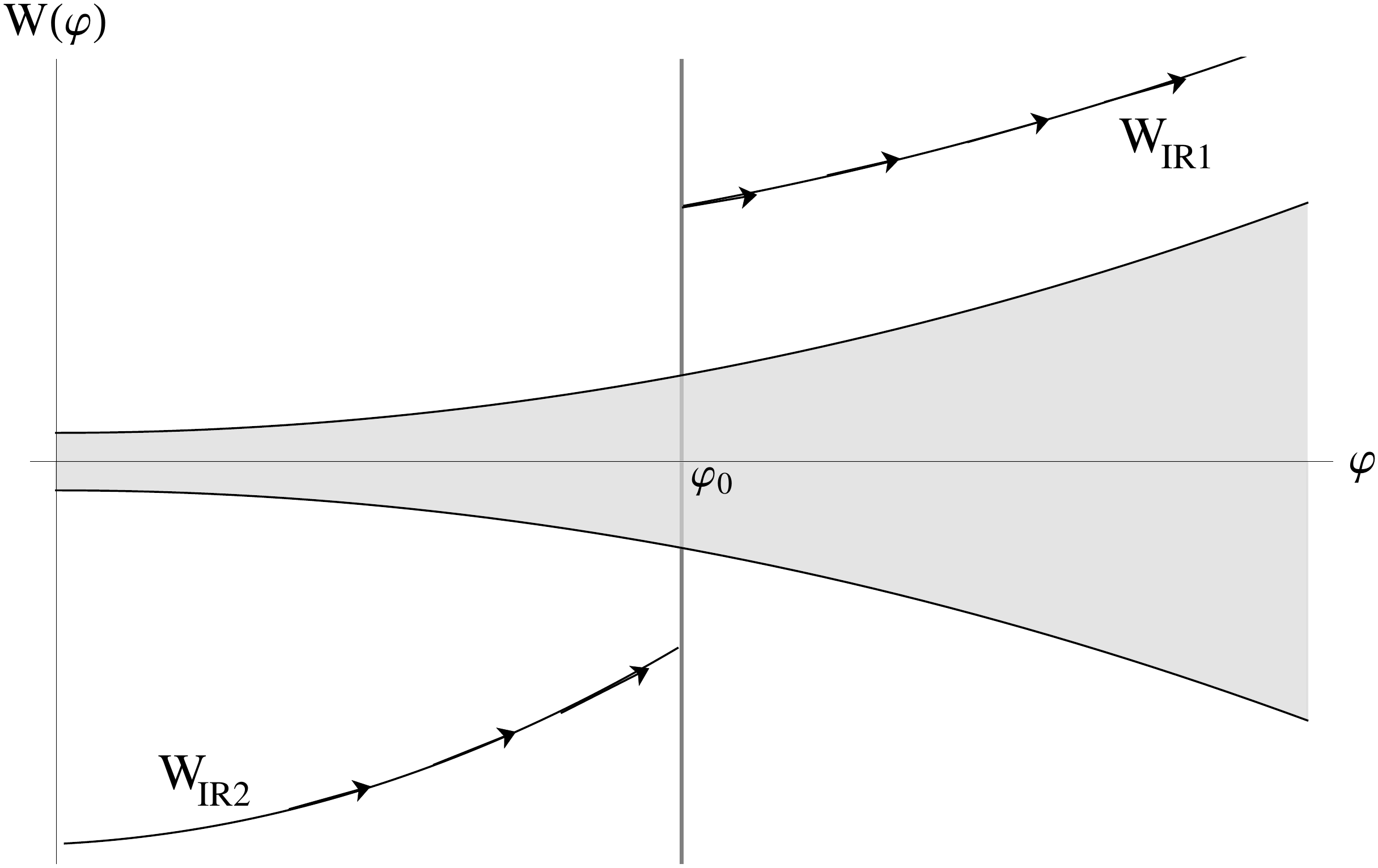} \includegraphics[width=7.3cm]{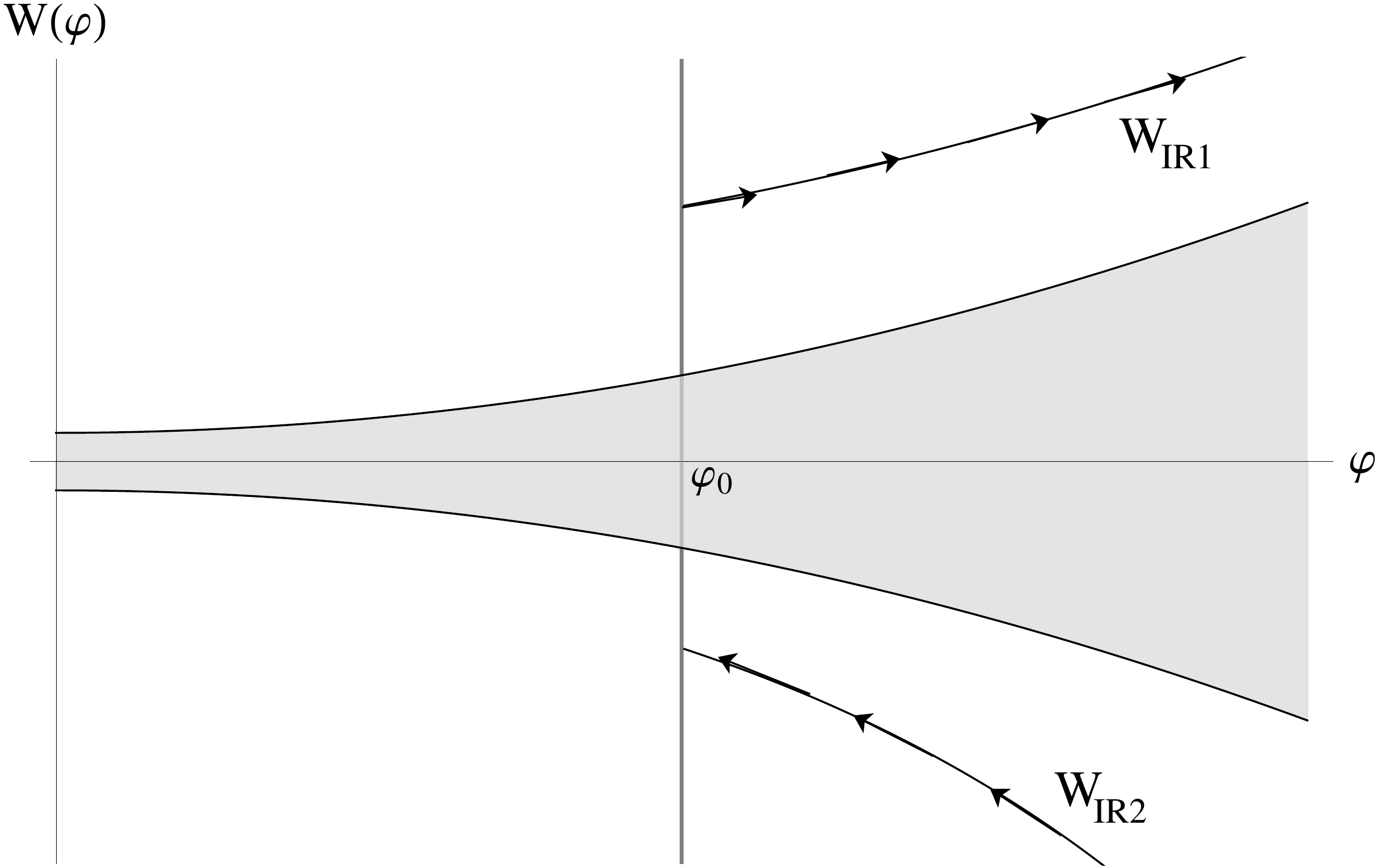} \\
(a) \hspace{7cm} (b)
\caption{The behavior of the superpotential in case B, when $W_{IR}(\f_0)-W_B(\f_0)< -B(\f_0)$. Contrary to cases A1 and A2, in this case the solution does not connect to the UV on either side of the interface.  Rather, the interface joins two ``IR'' (i.e. finite-volume)  regions (hence both branches are labeled IR in the figures).  Without fine tuning,  at least one of them will have an unacceptable singularity.} \label{caseC}
\end{center}
\end{figure}
\end{itemize}

Based on the discussion above, solutions of type $B$  do not have an asymptotically AdS large-volume region (thus the holographic dual is not UV-complete) and they do not realize the self-tuning mechanism. This leaves cases $A1$ and $A2$. Following the discussion  in section 5, if  $W_B > 0$ the model is manifestly ghost-free. Thus, case $A2$ is safe, whereas in case $A1$ one still has to check the absence of ghosts explicitly.

\section{The holographic parameters and the integration constants} \label{scales}

In this Appendix we show how choosing  the integration constants for the  metric  at the interface is equivalent to fixing the UV data at the $AdS$ boundary.

To this end it is convenient to rewrite the metric in   conformal coordinates:
\be \label{conformal}
ds^2 = e^{A(r)}\left(dr^2 + \eta_{\mu\nu}dx^\mu dx^\nu\right), \qquad \f=\f(r), \qquad du = e^{A(r)} dr.
\ee
In these coordinates, (\ref{FE10}-\ref{FE11}) become:
\be\label{scales5b}
A_{UV}' = - {e^{A_{UV}} \over 2(d-1)} W_{UV}, \qquad  \f_{UV}' = e^{A_{UV}} {d W_{UV} \over d\f}, \quad ' \equiv {d\over dr}.
\ee
and similarly in the IR.
We have to choose initial conditions for  equations (\ref{scales5b}), and this can be done in two equivalent ways:
\begin{enumerate}
\item {\em At the interface:} There, the value $\f=\f_0$ is fixed by the superpotential matching equations. Then, the free integration constants of equations  (\ref{scales5b})  are $r_0$ (the position of the interface) and  $A_0\equiv A(r_0)$.
\item {\em At the AdS  boundary.} This makes the holographic interpretation of the integration constant transparent, since in holographic theories,  fixing the near-boundary behavior of the  metric and scalar fields fixes  the geometry and the couplings of the dual field theory in the UV.  Conventionally, the boundary is set at $r=0$. This  fixes one of the two integration constants.  Then,  the asymptotic form of the metric and scalar field are:
\be \label{scales5c}
\exp A(r)  =  {\ell_{UV}\over r}\left(1 + O(r^2)\right), \qquad \f(r) = g_0 r^{\Delta_-}(1 + O(r)) , \quad r \to 0,
\ee
where $\Delta_- = d-\Delta$ and  $\ell_{UV}^2 = d(d-1)/V(\f=0)$. The quantity  $g_0$ appears as the second   integration constant, and  it represents the value of the relevant coupling deforming the UV CFT.
\end{enumerate}

The two ways of fixing the integration constants   are equivalent, and we will show below how one can translate from one to the other. In particular, we will show how, once  the equilibrium position $\f_0$ is fixed,   the  choice of the UV coupling  $g_0$,  determines both $r_0$ and $A_0$.  This  allows to translate the dependence on $A_0$ into a dependence on $g_0$, which is a physical parameter of the UV theory.

We start by integrating equations (\ref{scales5b}), on the UV side, with respect to $\f$:
\be \label{scales6}
A_{UV}(\f) = A_0 - {1\over 2(d-1)} \int_{\f_0}^{\f}  {W_{UV} \over dW_{UV}/d\f},  \quad r(\f) = r_0 +  \int_{\f_0}^{\f}  d\tilde\f {e^{-A_{UV}(\tilde\f)} \over dW_{UV}/d\tilde\f},  \quad \f < \f_0,
\ee
where $A_0$ and $r_0$ are  arbitrary integration constants.
Given $A_0$ and $r_0$, one can in principle invert the  relation between $r$ and $\f$  and obtain $A_{UV}(r), \f_{UV}(r)$.

With equations (\ref{scales6}), we have  fixed the solution completely by choosing  the integration constants $(A_0,r_0)$ at the brane. The boundary of $AdS$ in this solution corresponds to setting\footnote{We supposed that the UV AdS fixed-point to be at a maximum of the potential situated at $\f=0$, see Section \ref{interlude}.} $\f = 0$ in equations  (\ref{scales6}). If we want to adhere to the usual conventions in which the boundary is at $r=0$ in conformal coordinates, then we must choose:

\be \label{scales7}
r_0 =  \int^{\f_0}_{0} d\tilde\f {e^{-A_{UV}(\tilde\f)} \over dW_{UV}/d\tilde\f}
\ee
This equation  fixes the (coordinate) distance from the boundary to the brane, in terms of the bulk metric. With this determination, equation (\ref{scales6}) becomes:
\be\label{scales7a}
r(\f) = \int_0^{\f}d\tilde\f {e^{-A_{UV}(\tilde\f)} \over W_{UV}'(\tilde\f)}.
\ee

We can extract the dependence on $A_0$ of the result (\ref{scales7})   by using the expression  for $A(\f)$   in  equation (\ref{scales6}):
\be \label{scales8}
r_0 = e^{-A_0} \int_{0}^{\f_0} {d\tilde{\f}\over W_{UV}'(\tilde\f)} \exp \left(-{1\over 2(d-1)}\int_{\f_0}^{\tilde{\f}} {W_{UV}(\tilde\f)\over dW_{UV}'(\tilde\f)} \right).
\ee

To connect $A_0$ with the UV boundary data $g_0$ appearing in equation (\ref{scales5c}),  we need the asymptotic behavior of $\f$ close to the boundary. This  can be read-off by taking the $\f \to 0$ limit of equations (\ref{scales6}). Close to $\f =0$,  the superpotential $W_{UV}$ has the form characteristic of a UV fixed point\footnote{In special cases the flow maybe driven by a vev and $\Delta_{-}\to \Delta_{+}$.},
 \be \label{scales9}
W_{UV} \simeq {2(d-1) \over \ell_{UV}} + {\Delta_- \over 2} \f^2 + \ldots , \qquad \ell_{UV} \equiv \sqrt{d(d-1)\over V(0)}.
\ee
Then, from (\ref{scales6}), the scale factor behaves as:
\be \label{scales10}
A(\f) \sim {1\over \Delta_-} \log \f \, + \bar{A} + O(\f), \qquad \f\to 0,
\ee
where $\bar{A}$ is a constant, defined by:
\be\label{scales11}
\bar{A} =
 \lim_{\f \to 0}\left[ A_{UV} (\f) + \, {1\over \Delta_-} \log \f\right].
\ee
Using equation (\ref{scales9}) and (\ref{scales11})  we can write   $r(\f)$ for  $\f\to 0$ from equation (\ref{scales7a}):
\be \label{scales12}
r(\f) \simeq {\ell_{UV} \over \Delta_-} e^{-\bar{A}} \int_0^\f  d\tilde\f {\tilde\f^{1/\Delta_-}\over \tilde\f} =  \ell_{UV} e^{-\bar{A}}\f^{1/\Delta_-},
\ee
which we can invert for $\f(r)$ close to $r=0$:
\be\label{scales13}
\f(r) \simeq  g_0 r^{\Delta_-} \qquad g_0 \equiv \left(e^{\bar{A}} \over \ell_{UV}\right)^{\Delta_-}
\ee

We want to relate the constant $\bar{A}$ defined in equation (\ref{scales11}) to the integration constant $A_0$ defined at the brane, appearing in equation (\ref{scales6}). This can be done by writing the limit in equation (\ref{scales11}) as:
\be\label{scales14}
\bar{A} = \lim_{\f \to 0}\left[A_0 - {1\over 2(d-1)} \int_{\f_0}^{\f}  {W_{UV} \over W_{UV}'} + \, {1\over \Delta_-}\int_{\f_0}^\f {1\over \f} + {1\over \Delta_-} \log \f_0\right]
\ee
Since the above expression is finite, we can put the second and third term under the same integral sign and take the limit by replacing $\f$ with zero: in doing so, we  find the desired relation between $A_0$ and $\bar{A}$:
\be\label{scales15}
\bar{A} = A_0 + \bar{{\cal A}} (\f_0), \qquad \bar{{\cal A}} (\f_0) \equiv {1\over \Delta_-} \log \f_0 +  {1\over 2(d-1)} \int^{\f_0}_{0} \left( {W_{UV} \over W_{UV}'} - {2(d-1) \over \Delta_- \f}\right).
\ee

Notice that $\bar{\cal A}(\f_0)$ depends only on quantities appearing in the action, which determine both $\f_0$ and $W_{UV}$ via the matching conditions plus IR regularity.

Using equation (\ref{scales15}) and (\ref{scales13}) we can finally relate $A_0$ to the UV data $g_0$ and the equilibrium position:
\be\label{scales16}
g_0 = e^{\Delta_- A_0} \left(e^{ \bar{{\cal A}}(\f_0)} \over \ell_{UV}\right)^{\Delta_-}.
\ee
In holography, AdS boundary conditions  are specified by fixing $g_0$: by equation (\ref{scales16}) this fixes $A_0$ (and thus completely fixes the geometry), since the quantity  in the parenthesis is determined dynamically from the matching equations at the interface. The coordinate position of the brane is also fixed by equation (\ref{scales8}).

\section{Avoiding Weinberg's no-go  theorem} \label{nogo}

Any claim to have a working self-adjustment mechanism for the cosmological constant has to be confronted with Weinberg's no-go theorem \cite{Weinberg:1988cp} (see also \cite{padilla} for an updated review and discussion). Below, we review Weinberg's theorem and we show how our framework avoids it.

Weinberg's argument starts from the following assumptions. Consider a model based  on an action of the form:
\be \label{wein0}
S[\phi_i, \gamma_{\mu\nu}] = \int d^4 x {\cal L}(\phi_i, \gamma_{\mu\nu}),
\ee
where the Lagrangian density ${\cal L}$ depends on the four-dimensional metric $\gamma_{\mu\nu}$ plus  other fields generically denoted by $\phi_i$ (which may be  scalars or tensors with respect to the four-dimensional Lorentz group).

Now suppose the  field equations that one obtains stemming from the action (\ref{wein0})  have a solution which preserves rigid space-time translations $x^\mu \to x^\mu + \alpha^\mu$.  The field equations in this case  reduce to:
\be \label{wein1}
{\de {\cal L} \over \de \gamma_{\mu\nu}} = 0 , \qquad  {\de {\cal L} \over \de \phi_i} = 0.
\ee
We will suppose that the two equations above hold independently\footnote{Weinberg also considers  the case when the two equations are proportional to each other, i.e. when
$$
\gamma^{\mu\nu}{\de {\cal L} \over \de \gamma^{\mu\nu}} = \sum_i f(\phi_i) {\de {\cal L} \over \de \phi_i}
$$
 This is not the case in the model under consideration, and   will not be of interest here.}.
On such a solution,  the fields are constant and diffeomorphism invariance is broken to rigid space-time $GL(4)$ transformations, under which:
\be\label{wein2}
x^\mu \to M^\mu_{\,\nu} x^\nu, \quad \gamma_{\mu\nu} \to \gamma_{\rho\sigma}M^\rho_{\,\mu}M^\sigma_{\,\nu}, \quad {\cal L} \to {\cal L} \det M.
\ee
Using these transformations properties one can easily show that, under an infinitesimal $GL(4)$ transformation $M = 1 + \delta M$:
\be\label{wein3}
\delta  {\cal L} = Tr\delta M {\cal L} =  {\de {\cal L}  \over \de \phi_i} \delta \phi_i +  {\de {\cal L}  \over \de \gamma_{\mu\nu}} \left(\delta M_{\mu\nu} + \delta M_{\nu\mu}\right)
\ee
We now solve  the field equations for the fields $\phi_i$  and set  their values on-shell:
\be\label{wein4}
{\de {\cal L}  \over \de \phi_i}  = 0 \qquad \Rightarrow \qquad \phi_i = \bar{\phi}_i,
\ee
Then,  equation (\ref{wein3}) implies:
\be \label{wein5}
{\de {\cal L} \over \de  \gamma_{\mu\nu}} = {1\over 2}\gamma^{\mu\nu}{\cal L} \qquad  \Rightarrow \qquad {\cal L} = \sqrt{g} V_{eff}(\bar{\phi}_i)
\ee
where $V_{eff}(\phi_i)$ is some function that depends on the fields $\phi_i$ only.  Finally,  the metric field equation is:
\be \label{wein5-i}
{\delta {\cal L} \over \de \gamma_{\mu\nu}} = {1\over 2}\sqrt{\gamma}\gamma^{\mu\nu} V_{eff}(\bar{\phi}_i) = 0
\ee
 which generically will not be satisfied unless the parameters in the  Lagrangian obey one relation, i.e. they are fine-tuned.

Now we will reproduce this line of reasoning in the holographic setup and show where the loophole lies. First, we have to bring the problem in the same form as in Weinberg's argument. To this end, we will reduce the problem of extremizing the original action (\ref{A1}) to a purely 4d problem with an effective action of the same form as in equation (\ref{wein0}),   where the only remaining dynamical variables are the induced metric and scalar field on the interface, $\gamma_{\mu\nu}(u_0), \phi (u_0)$.  To do this we first extremize the bulk action,    on each side of the interface,  but  without imposing the matching condition.

On each side, we take  an ansatz of the form:
\be\label{wein6}
ds^2 =  du^2 + e^{A(r)}\gamma^{(0)}_{\mu\nu}dx^\mu dx^\mu , \qquad \phi = \phi(u)
\ee
where $\gamma^{(0)}_{\mu\nu}$ is invariant under space-time translations (this is a slight generalization of the solution (\ref{FE7}) ) and coincides with the metric of the UV dual CFT (cfr. equation (\ref{int4-iii}), therefore it is fixed by the UV boundary condition, as we discussed in subsection \ref{int-UV}.

One can show that the action, evaluated on such solutions, is a total derivative:
\be
S_{on-shell}  = -\int d^d x \sqrt{\gamma^{(0)}} \int du\,  {\de \over \de u} \Big[ e^{4A(u)} W(\phi(u)) \Big].
\ee
Using this result, the on-shell action reduces to the sum of   three boundary terms:
\begin{itemize}
\item Two finite boundary terms coming  from each side of the interface,
 \be \label{wein7}
S_{UV} = -  e^{4A(u_0)}\int d^4x \sqrt{\gamma^{(0)}} W_{UV}(\phi_0) , \quad
S_{IR}= e^{4A(u_0)}\int d^4x\sqrt{\gamma^{(0)}}  W_{IR}(\phi_0).
 \ee
where $W_{UV}$ and $W_{IR}$ are the solutions of the superpotential equation on each side (in particular, as we have discussed in section \ref{interlude}, $W_{IR}$ is fixed by regularity).
To write  equation (\ref{wein7}) we have assumed continuity of the metric and the scalar fields.
\item A divergent  boundary term $S_{UV}$ coming from the boundary of AdS, which can be renormalized by supplementing the original action with appropriate counterterms. The counterterms {\em do not depend on the solution (\ref{wein6}) }. After renormalization is carried out, one is left with \cite{papa}:
\be \label{wein8}
S_0 = \int d^4x \sqrt{\gamma^{(0)}} C_{UV} g_0^{d/d-\Delta} =  \int d^4x \sqrt{\gamma^{(0)}} \< O \> g_0,
\ee
where in the second equality we have used equation (\ref{int5}).  Notice that this contribution does not depend on $A_0$ nor $\phi_0$.
One may have expected also a boundary term from the far IR, but this always vanishes if the solution is IR-regular or has an acceptable IR singularity.
\end{itemize}

So far we have ``integrated out'' the bulk but we have not yet solved the field equations for the metric and scalar field at the interface.  The effective 4d action for these variables is the sum of the terms in equations (\ref{wein7}) and (\ref{wein8}), plus  the world-volume action (\ref{A3}):
\bea \label{wein9}
&& S_{eff}[A_0,\f_0; C_{UV}] = \int d^4x \sqrt{\gamma^{(0)}} C_{UV} g_0^{d/d-\Delta} + \nonumber \\
&&  + \int d^4x \sqrt{\gamma^{(0)}}  e^{4A_0} \Big[W_{IR}(\f_0) - W_{UV} (\f_0; C_{UV})- W_{B}(\f_0)\Big]
\eea
This action depends on the dynamical variables $(\f(u_0), A(u_0))$; on the {\em fixed} quantities $g_0$ and $\gamma^{(0)}_{\mu\nu}$ which are part of the definition of the UV CFT;  and on the extra free parameter $C_{UV}$.  Notice that we should not vary the effective action with respect to  $g_0$ nor $\gamma^{(0)}_{\mu\nu}$ nor $C_{UV}$ (in particular the first line in equation (\ref{wein9}) is a constant, independent of the dynamical variables.

Extremizing the action with respect to the dynamical variables $(\f_0, A_0)$ gives back the matching conditions, (\ref{match1}-\ref{match2}), as  expected.

We can now compare the action (\ref{wein9}) with the one assumed in the no-go theorem, (\ref{wein0}). First,  notice that the $A_0$ equation of motion is  essentially the same  as (\ref{wein5-i}):
\be \label{wein10}
V_{eff}(A_0,\f_0; C_{UV}) \equiv  e^{4A_0} \Big[W_{IR}(\f_0) - W_{UV} (\f_0; C_{UV})- W_{B}(\f_0)\Big] =0.
\ee
Contrary to equation (\ref{wein5-i}) however,  this equation determines the extra parameter $C_{UV}$ (which does not appear in the full definition of the model, neither in the bulk nor on the brane nor on the boundary) and does not require fine-tuning between the model parameters.     {\em This is where the no-go theorem fails:} it assumed that the action depends only on dynamical variables, determined by their own field equations, and that there are no extra free parameters. This is true for weakly coupled field theories. Here however the quantity $C_{UV}$ is not a dynamical variable but it  is determined in a different way: on the gravity side, by insisting that the UV solution, through the matching conditions at the brane, glues correctly to the fixed IR-regular solution; in  the dual field theory language, it is the strong coupling dynamics which determines the value of the  VEV of the operator in the UV. These are  affected also by the low energy degrees of  freedom. Indeed, it is natural that the presence of the brane-world degrees of freedom   at intermediate energies affect the UV value of the VEVs and the running of couplings, but not the bare UV  coupling $g_0$.

\section{Linearized bulk equations and matching conditions} \label{app pert}

In this appendix we derive the perturbed  equations and matching conditions  for the tensor and scalar modes. We restrict to the physically interesting case of a five-dimensional bulk, i.e. from now on we set   $d=4$.   We use  conformal coordinates in the bulk, such that the unperturbed metric and scalar field are:
\be \label{app1}
ds^2 = a(r)\left(dr^2 + \eta_{\mu\nu} dx^\mu dx^\nu \right),  \qquad \f = \bar{\f}(r),
\ee
where $\eta_{\mu\nu} = diag(-,+,+,+)$. We denote derivatives with respect to $r$ by a prime.

The background Einstein equations are, in these coordinates:
\be\label{app2}
-a^2 V(\fb) = 3\left( 2 {a^{'2}\over a^2} + {a''\over a} \right) \, , \quad (\fb')^2 =  6\left(2 {a^{'2}\over a^2} - {a''\over a}  \right),
\ee
or in terms of the superpotential:
\be\label{app3}
a' = -{a^2 W \over 6}, \qquad \fb' = a {dW \over d\f}.
\ee
The brane is located at the equilibrium position $r_0$. All quantities with a subscript $0$ are evaluated at $r_0$ (e.g. $a_0 \equiv a(r_0)$ etc).

We write  the perturbed 5-d metric and scalar field as:
\bea
&&ds^2 = a^2(r)\left[(1+2\phi) dr^2 + 2 A_\mu dx^\mu dr  + (\eta_{\mu\nu} + h_{\mu\nu}) dx^\mu, dx^\nu\right],  \label{app4-1} \\
&& \f = \fb(r) + \chi, \label{app4-3}
\eea
where the quantities $\phi, A_\mu, h_{\mu\nu}, \chi$ are functions of $r,x^\mu$ and will be treated as small perturbations around the $r$-dependent homogeneous background.  We further decompose the metric perturbations  in a scalar-tensor decomposition\footnote{We set to zero the transverse vector modes $A^T_\mu$ and $V^T_\mu$  appearing in the general decomposition (\ref{pert3}),  since there is no physical vector in the bulk, and these modes decouple.}:
\be
h_{\mu\nu} =  2\eta_{\mu\nu}\, \psi +2 \de_\mu\de_\nu E + \hat{h}_{\mu\nu}, \quad A_\mu = \de_\mu B\label{app4-2}
\ee
where the tensor perturbation $\hat{h}_{\mu\nu}$ is transverse and traceless: $\de^\mu\hat{h}_{\mu\nu} = h^\mu_{\,\mu} = 0$. Unless explicitly stated, all indices are raised and lowered with the flat Minkowski metric $\eta_{\mu\nu}$.

\subsection{Perturbed bulk  equations}
In  the bulk, the system contains  one tensor perturbation $\hat{h}_{\mu\nu}$ and (before gauge-fixing) five scalar perturbations ($\psi,\phi,B, E, \chi$). The components of the linearized Einstein tensor are:
\bea
&&G_{rr}^{(1)}  =  12{a'\over a} \psi' + 3\de^\mu\de_\mu \psi - 3{a'\over a}\de^\mu\de_\mu(B-E'), \quad G_{r\mu}^{(1)} =  3{a'\over a} \de_\mu\phi -  3\de_\mu \psi'  \label{app5}\\
&&G_{\mu\nu}^{(1)} = -{1\over2} \left[a^{-3}\left(a^3 \hat{h}'_{\mu\nu}\right)' + \de^\rho \de_\rho \hat{h}_{\mu\nu} \right] +  3 \eta_{\mu\nu}\left[\psi'' + 3{a'\over a} \psi' - {a'\over a} \phi' + 2 {a''\over a} \psi -  2 {a''\over a} \phi\right] + \nonumber \\
&& \qquad - \de_\mu\de_\nu\left[ 2 \psi + \phi - (B-E')' - 3{a'\over a}(B-E')\right].  \label{app6}
\eea
The linearized Einstein equations are then:
\be \label{app7}
G_{ab}^{(1)} = M^{-3}\left({1\over {\sqrt{g}}}{\delta S_{bulk}[g,\phi]  \over \delta{g^{ab}}}\right)^{(1)},
\ee
where the right hand side is the linearized matter stress tensor obtained from the variation of the matter bulk action in equation (\ref{A2}).   At linear order, Einstein equations do not couple tensor and scalar modes and we can discuss the two sectors separately.

\paragraph{Tensor modes}  Since there are is no tensor-like matter, the transverse-traceless part of the right hand side of equation (\ref{app7}) is identically zero (this can be easily checked explicitly). Therefore,   the linearized field equation for tensor modes  $\hat{h}_{\mu\nu}$ is obtained by setting to zero the first square bracket  in equation (\ref{app6}), and it reads:
\be \label{app8}
\de_r \left( a^3 \de_r \hat{h}_{\mu\nu}\right) + a^3 \de^\rho\de_\rho   \hat{h}_{\mu\nu} = 0 .
\ee

\paragraph{Scalar modes}
Keeping only scalar modes,  the perturbed Einstein equations (\ref{app7}) are, to linear order\footnote{To be  precise, equations (\ref{rmu}) and (\ref{munotnu}) hold when acted upon by $\de_\mu$ and $\de_\mu\de_\nu$, respectively, as one can see by comparing them with equations (\ref{app5}) and (\ref{app6}). Here we restrict to  perturbations that  can be expanded in Fourier modes  with finite momentum $p_\mu\neq 0$. Then, equations (\ref{rmu}) and (\ref{munotnu}) hold as well.}:
\bea
& (rr) & \quad  4{a'\over a} \psi' + \de^\mu\de_\mu \psi - {a'\over a}\de^\mu\de_\mu(B-E') \nonumber\\
&& \quad  = {1\over 6} \fb'\, \chi' -{a^2\over 6} {dV\over d\f} \chi - {a^2\over 3} V(\fb) \phi,  \label{rr}\\
& (r\mu) & \quad  {a'\over a} \phi -  \psi' =  {1\over 6} \fb' \chi \label{rmu}\\
& (\mu\neq \nu) &\quad  2 \psi + \phi - (B-E')' - 3{a'\over a}(B-E') =0 \label{munotnu} \\
& (\mu=\nu) & \quad  \psi'' + 3{a'\over a} \psi' - {a'\over a} \phi' + 2 {a''\over a} \psi -  2 {a''\over a} \phi =  \nonumber \\
&& \quad \quad = {(\fb')^2 \over 6} \phi - {1\over 3}\left({(\fb')^2 \over 2} + a^2 V\right) \psi - {\fb' \over 6} \chi' - {a^2 \over 6}  {dV\over d\f} \chi \label{munu}
\eea
where the right hand sides are the explicit form of  the linearized matter stress tensor appearing in equation (\ref{app7}).
We  also have the perturbed Klein-Gordon equation (which is not independent of equations (\ref{rr}-\ref{munu}) , but it can be useful to work with):
\bea
(KG) && 0 = a^{-3}\left(a^3\chi'\right)' + \de^\mu\de_\mu \chi - a^2 {d^2 V\over d\f^2}\chi  \nonumber \\ &&
\qquad - 2 a^2 {d V\over d \f} \phi -  \bar{\f}' \phi' + 4 \bar\f'\psi' -  \bar\f' \de^\mu\de_\mu (B - E')  \label{kg}
\eea

 These equations contain  five scalar perturbations, but we can impose two scalar gauge conditions plus two scalar constraints (this will be discussed in detail in Appendix \ref{app scalar}). These leave one physical scalar bulk fluctuation, which can be taken to be the gauge-invariant combination:
\be\label{app9}
\zeta (r,x^\mu) \equiv \psi(r,x^\mu) - {1\over z(r)} \chi(r,x^\mu), \qquad z \equiv  {a \bar\f'\over a' }
\ee
From   equations (\ref{rr}-\ref{munu}) one can obtain  a single second order equation for the $\zeta(r,x^\mu)$, which reads:
\be\label{app10}
\de_r \left(a^3 z^2 \de_r \zeta\right) + a^3 z^2 \de^\mu\de_\mu \zeta = 0.
\ee
Regardless of the gauge fixing, one can arrive at  equation (\ref{app9})  by solving   equation (\ref{rmu})  for $\phi$ and  equation (\ref{rr}) for  $\de^\mu\de_\mu (B-E')$ in favor of $\chi$ and $\psi$, and inserting their expression in equation  (\ref{kg}).

Equations (\ref{app8}) and (\ref{app10}) describe the full system of linearized perturbations in the bulk.

\subsection{Brane perturbations and linearized junction conditions}

In order to write the linearized Israel matching conditions (\ref{FE5}-\ref{FE6}) we need to write the perturbed induced metric, normal vector, and extrinsic curvature, to linear order, in terms of the metric perturbations (\ref{app4-1}-\ref{app4-2}),  plus the brane-bending mode $\rho(x^\nu)$. The latter is  defined by perturbing the embedding equation:
\be
r(x^\mu) = r_0 + \rho (x^\mu),
\ee
where $r_0$ is the unperturbed equilibrium position.

The normal vector $n^A$ and induced metric $\gamma_{AB}\equiv g_{AB} - n_A n_B$ are, to first order in perturbations:
\bea
&& n^A = a^{-1}(r_0+ \rho)\left(1-\phi, -A_\mu - \de_\mu\rho\right), \label{pert 6-1}\\
&&  \gamma_{AB}dX^A dX^B = a^2(r_0+\rho)\left[(A_\mu + \de_\mu \rho)dr dx^\mu + (\eta_{\mu\nu} + h_{\mu\nu})dx^\mu dx^\nu\right],\label{pert 6-2}.
\eea
 It is convenient to explicitly expand to linear order in $\rho$ the prefactor $a(r_0 + \rho)$ in equation (\ref{pert 6-2}),  and to write the perturbed induced metric as:
\be\label{pert13}
\gamma_{\mu\nu} = a_0^2 \left(\eta_{\mu\nu} + \tilde{h}_{\mu\nu}\right),  \quad  \tilde{h}_{\mu\nu} \equiv h_{\mu\nu} + 2{a'_0\over a_0} \eta_{\mu\nu}\rho .
\ee
The scalar field perturbation at the (perturbed) brane position is:
\be \label{pert14}
\f(r(x^\mu))  = \bar{\f}_0 + \chi + \bar\f'_0 \rho
\ee
In equations (\ref{pert13}-\ref{pert14}) all quantities are evaluated at  $r_0$, the unperturbed equilibrium position.

From equation  (\ref{pert13}-\ref{pert14}) we can deduce the continuity conditions (\ref{FE3}) to linear order:
\be \label{pert7-1}
\Big[h_{\mu\nu} + 2 \eta_{\mu\nu}{a'_0\over a_0} \rho\Big]^{IR}_{UV} = 0, \qquad \Big[\bar\f_0' \rho+ \chi\Big]^{IR}_{UV} = 0.
\ee

Notice that the bulk metric and scalar field perturbations  are not continuous at the brane, unless one chooses a gauge where $\rho=0$. This is not the most convenient choice to deal with bulk perturbations, however. We will come back to the gauge fixing problem in Appendix \ref{app scalar} when we discuss in detail the matching  conditions in the scalar sector.

The linearized junction conditions are given by (\ref{FE5}-\ref{FE6}).   On the the right hand side, in addition to  the brane  action $S_{brane}$ in equation (\ref{A3}), we allow the possibility of some localized matter:
\be \label{pert8-1}
S_{loc} = S_{brane}[\gamma, \f] + S_m, \qquad S_m \equiv \int d^4 x \sqrt{\gamma} {\cal L}_m (\psi_i, \f_0)
\ee
The localized matter fields $\psi_i$  (which may include the Standard Model fields), are taken to be   trivial in the vacuum. We  assume the matter fields are  minimally coupled to the induced metric but   may have a direct coupling to the dilaton $\f$ evaluated on the brane. The localized matter stress tensor is defined as:
\be \label{pert9-1}
T_{\mu\nu} =  -{2\over \sqrt{\gamma}}{\delta S_m \over \delta \gamma^{\mu\nu}}
\ee
We also define the ``dilaton charge operator'' $O$ of the localized matter as:
\be\label{pert10-1}
O = {1\over \sqrt{\gamma}}{\delta S_m \over \delta \f}.
\ee

With these ingredients, the perturbed  matching conditions are derived by linearizing both sides of the two equations:
\bea
&& \left[\Big(K_{\mu\nu} - \gamma_{\mu\nu} K\Big)^{(1)}\right]^{IR}_{UV} = \left({1\over {\sqrt{\g}}}{\delta S_{brane}[\g,\f]  \over \delta{\g^{\mu\nu}}}\right)^{(1)} -{1\over 2 M^3} T_{\mu\nu}, \label{pert11-1} \\
&& \left[\Big(n^a \de_a \f \Big)^{(1)}\right]^{IR}_{UV}  =  -\left({1\over {\sqrt{\g}}}{\delta S_{brane}[\g,\f]  \over \delta{\f}}\right)^{(1)} - {1\over M^3} O \chi. \label{pert12-1}
\eea

In order to proceed, we need the components of the  extrinsic curvature, 
\be\label{kappaAB}
K_{AB} = \nabla_{\left(A\right.} n_{\left.B\right)} - {1\over 2}n^C\nabla_C \left(n_A n_B\right)
\ee 
They are, to linear order in the perturbations,
\bea\label{pert7}
&& K_{rr}^{(1)} = 0, \quad K_{r\mu}^{(1)} =  a'\left(A_\mu + \de_\mu \rho\right) + a\left(\de_\mu \phi + {a'\over a} \de_\mu\rho\right) , \\ &&K_{\mu\nu}^{(1)} = a'(r_0+\rho)\left[(1-\phi)\eta_{\mu\nu} + h_{\mu\nu}\right] +  a(r_0)\left[ {1\over 2} h'_{\mu\nu} - \de_{\left(\mu\right. } (A_{\left.\nu \right)} + \de_\nu \rho)\right].
\eea

Using the scalar-tensor decomposition (\ref{app4-2}), the left hand sides of the matching conditions (\ref{pert11-1}-\ref{pert12-1}) are, to linear order in the perturbations:
\bea \label{pert9}
\Big(K_{\mu\nu} - \gamma_{\mu\nu} K\Big)^{(1)} && = -3 a'_0 \left[\left(1-\phi+ {a''_0 \over a'_0}\rho + 2 \psi \right)\eta_{\mu\nu} + 2 \de_\mu\de_\nu E + \hat{h}_{\mu\nu}\right] + \nonumber  \\ && \!\!\!\!\!+  a_0 \left[{1\over 2} \hat{h}'_{\mu\nu} - 3\eta_{\mu\nu}\psi' + (\de_\mu\de_\nu  - \eta_{\mu\nu}\de^\rho \de_{\rho}) \left(E'-B -\rho \right)\right],
\eea
\be \label{pert10}
\Big(n^a \de_a \f \Big)^{(1)} = a_0^{-1}\left[\f'_0 + \chi' + \f''_0 \rho - {a_0'\over a_0}\f'_0 \rho - \f'_0 \phi \right].
\ee

The right hand sides of equations (\ref{pert11-1}-\ref{pert12-1}) are obtained by linearizing the expressions on the right hand side of equations (\ref{FE5}-\ref{FE6}). For this, we need the linearized expressions of the brane Ricci tensor for the induced metric in equation (\ref{pert13}):
\be\label{pert14.5}
R^{(\gamma)}_{\mu\nu} = -{1\over 2} \de^\rho \de_\rho \tilde{h}_{\mu\nu} - {1\over 2} \de_\mu \de_\nu \tilde{h}^{\rho}_{\,\rho} + \de^\rho \de_{\left(\mu\right.} \tilde{h}_{\left.\nu\right)\rho}, \quad \tilde{h}_{\mu\nu} \equiv h_{\mu\nu} + 2{a'_0\over a_0} \eta_{\mu\nu}\rho .
\ee
Decomposing $h_{\mu\nu}$ in tensor and scalar components as in equation (\ref{app4-2}),  the above expression becomes:
\be \label{pert15}
R^{(\gamma)}_{\mu\nu} = -{1\over 2} \de^\rho \de_\rho \hat{h}_{\mu\nu}  - \left(2 \de_\mu\de_\nu + \eta_{\mu\nu}\de^\rho\de_\rho\right) \left( \psi + {a'_0\over a_0} \rho\right)
\ee
whence:
\bea
&& R^{(\g)} = -{6\over a_0^2}\de^\rho\de_\rho \left(\psi +{a'_0\over a_0} \rho\right)  , \label{pert16}\\
&& G_{\mu\nu}^{(\g)} =  -{1\over 2} \de^\rho \de_\rho \hat{h}_{\mu\nu} - 2\left(\de_\mu\de_\nu - \eta_{\mu\nu}\de^\rho\de_\rho\right) \left( \psi + {a'_0\over a_0} \rho\right). \label{pert17}
\eea
Notice that the longitudinal component $E$ of the metric perturbation drops out of the Ricci tensor.

We can finally  obtain, to linear order in the perturbations, the expressions on right hand sides of equations (\ref{pert11-1}-\ref{pert12-1}):
\bea
\left({1\over {\sqrt{\g}}}{\delta S_{brane}[\g,\f] \over \delta{\g^{\mu\nu}}}\right)^{(1)}&& = {1\over 2}a_0^2 W_0\left.\bigg[\eta_{\mu\nu} + \hat{h}_{\mu\nu} + 2 \de_\mu\de_\nu E + \right. \nonumber \\
&& \left. \qquad \qquad \;\;+ 2 \eta_{\mu\nu}\left(\psi + {a_0'\over a_0}\rho\right)+ \eta_{\mu\nu}{W_0'\over W_0}\left(\chi + \bar\f'_0 \rho\right) \right] +  \nonumber  \\
&& - U_0\left[{1\over 2} \de^\rho \de_\rho \hat{h}_{\mu\nu} + 2 \left(\de_\mu\de_\nu - \eta_{\mu\nu}\de^\rho\de_\rho\right) \left( \psi + {a'_0\over a_0} \rho\right)\right] +\nonumber \\
&&  - \left({dU_B\over d\f} \right)_0\left(\de_\mu\de_\nu - \eta_{\mu\nu}\de^\rho\de_\rho\right)\left(\chi + \bar\f'_0 \rho\right), \label{pert18}\\
&& \quad \nonumber \\
&& \quad \nonumber \\
\left({1\over {\sqrt{\g}}}{\delta S_{brane}[\g,\f]  \over \delta{\f}}\right)^{(1)} &&= \left({dW_B\over d\f}\right)_0 + \left({d^2W_B\over d\f^2}\right)_0 \left(\chi + \bar\f'_0 \rho\right) + \nonumber \\
&&\!\!\!\!\!\!  - {Z_0\over a_0^2}\de^\mu\de_\mu\left(\chi + \bar\f'_0\rho\right)+   {6\over a_0^2}\left({dU_B\over d\f} \right)_0 \de^\mu\de_\mu  \left( \psi + {a'_0\over a_0} \rho\right) \label{pert19}
\eea
The brane matter stress tensor and dilaton charge appear as inhomogeneous  source terms  in equations (\ref{pert11-1}-\ref{pert12-1}).

In the following two subsections we will decompose the junction conditions in their tensor and scalar components, respectively.

\subsection{Tensor  junction conditions}  \label{app tensor}

Since tensor and scalar modes are decoupled at linear order, to study the tensor modes it is enough to set all the scalar modes to zero in the equations found in the previous subsection:
\be \label{ten1}
\phi = B = \psi =E = \rho = 0,  \quad h_{\mu\nu} = \hat{h}_{\mu\nu}.
\ee

The continuity equation  across the interface, equation (\ref{pert7-1}) becomes simply
\be\label{ten2}
\left[\hat{h}_{\mu\nu}\right]^{IR}_{UV} = 0,
\ee
i.e. tensor modes are continuous across the interface.

The tensor (i.e. transverse and traceless) part of the  second junction conditions is  found by imposing (\ref{ten1})  in  equations (\ref{pert11-1}) and (\ref{pert18}) , and moreover by keeping only  the transverse traceless component of the matter stress tensor, defined by:
\be\label{ten3}
\hat{T}_{\mu\nu} = T_{\mu\nu} - {1\over 3}\eta_{\mu\nu} T +  {1\over 3}{\de_\mu\de_\nu\over \de^2} T +  {1\over 3}\eta_{\mu\nu}{\de^\rho\de^\sigma\over \de^2} T_{\rho\sigma} - {2\over \de^2} \de_{\left(\mu\right.} \de^\rho T_{\left.\nu\right)\rho} + {2\over 3} {\de_\mu\de_\nu\over \de^2}{\de^\rho\de^\sigma\over \de^2} T_{\rho\sigma}
\ee
where $T \equiv T^\mu_{\:\mu}$ and $\de^2 = \de^\mu\de_\mu$.  We will assume the matter stress tensor to be conserved, in which case the expression above reduces to the first three terms only:
\be\label{ten4}
\hat{T}_{\mu\nu} = T_{\mu\nu} - {1\over 3}\eta_{\mu\nu} T +  {1\over 3}{\de_\mu\de_\nu\over \de^2} T.
\ee
Setting all modes to zero except  $\hat{h}_{\mu\nu}$ in equations (\ref{pert9}) and (\ref{pert18})  and replacing $T_{\mu\nu}$ by $\hat{T}_{\mu\nu}$,  equation (\ref{pert11-1}) becomes:
\be\label{ten5}
 \left[-3a_0'\hat{h}_{\mu\nu} +   {1\over 2} a_0\hat{h}'_{\mu\nu}\right]^{IR}_{UV} = {1\over 2}a_0^2 W_0\hat{h}_{\mu\nu}  - {1\over 2} U_0 \de^\rho \de_\rho \hat{h}_{\mu\nu} - {1\over 2M^3}\hat{T}_{\mu\nu}.
\ee
Notice that the first term on each side cancel thanks to  the continuity of $\hat{h}_{\mu\nu}$ and  to the background matching condition since, in conformal coordinates and for $d=4$,  $a' = -a^2 W/6$,  and $[W]^{IR}_{UV} = W_0$   by equation (\ref{match1}). This leaves the simple jump condition for the first  derivative (plus the source term):
\be
a_0 \left[\hat{h}'_{\mu\nu}\right]^{IR}_{UV} =  -  U_0 \de^\rho \de_\rho \hat{h}_{\mu\nu} -{1\over M^3}\hat{T}_{\mu\nu}.
\ee

\subsection{Scalar  junction conditions} \label{app scalar}

The relevant modes are defined in equation (\ref{pert1}), in which we keep only the bulk scalar modes,
\be
\phi, \quad \chi,  \quad  A_\mu =  \de_\mu B,  \quad h_{\mu\nu} = 2\psi \eta_{\mu\nu} + 2\de_\mu\de_\nu E,  \label{scalar1}
\ee
plus the brane-bending mode $\rho(x)$ defined in equation (\ref{pert2}). Unlike the tensor modes, these fields are not gauge-invariant. Rather, they transform as follows under an infinitesimal scalar coordinate transformation $(\delta r,\delta x^\mu) = (\xi^5,  g^{\mu\nu}\de_\nu\xi)$:
\bea
&& \delta \psi = - {a'\over a} \xi^5 \qquad \delta \phi =-(\xi^5)'  - {a'\over a} \xi^5 \nonumber \\
&& \nonumber \\
&& \delta B = -\xi' - \xi^5 \qquad \delta E = - \xi  \label{scalar2}  \\
&&  \nonumber \\
&& \delta \chi = - \fb' \xi^5, \qquad \delta \rho  = \xi^5(r_0,x). \nonumber
\eea
It is convenient to partially fix the gauge:
\be\label{gaugefix1}
B=0
\ee
 by an appropriate shift  $\xi(x,r)$. This leaves a residual  gauge freedom with parameters $\xi^5(x,r)$ and $r$-independent $\xi(x^\mu)$:  a $\xi^5$-transformation can be compensated by an appropriate $\xi(r,x)$ to leave the condition $B=0$ unchanged, and only $\xi'$ affects $B$. Therefore  we are still free to do radial gauge-transformations and $r$-independent space-time diffeomorphisms and keep this gauge choice.

In this gauge, setting the brane sources to zero\footnote{The localized sources  will be added back at the end of this section},   the first matching conditions (\ref{FE3}) become:
\be\label{scalar3}
\Big[a^2(r_0+\rho) \left(2 \psi \eta_{\mu\nu} + 2\de_\mu\de_\nu E\right)\Big]^{UV}_{IR} = 0,   \qquad \Big[\fb(r_0+\rho) + \chi\Big]^{IR}_{UV} =0
\ee
Expanding the scale factor and the background scalar field profile, these are equivalent to the following continuity conditions:
\be \label{scalar4}
\Big[\hat{\psi}\Big]^{UV}_{IR} = 0, \quad \Big[\hat{\chi}\Big]^{UV}_{IR} = 0, \qquad \Big[E\Big]^{IR}_{UV} =0
\ee
where we have defined the new {\em bulk} perturbations:
\be\label{scalar5}
\hat{\psi}(r,x) = \psi + A'(r) \rho(x),  \quad \hat{\chi}(r,x) = \chi + \fb'(r) \rho(x),
\ee
where $A'=a'/a$.

The gauge-invariant scalar perturbation (\ref{pert3a}) has the same expression in terms of these new continues variables:
\be
\zeta = \hat{\psi} - {A' \over \fb'} \hat{\chi}.
\ee
In general however $\zeta(r,x)$ is not continuous across the brane, since the background quantity $A'/\fb'$ jumps:
\be\label{scalar6}
\Big[\zeta\Big]^{UV}_{IR} = \left[{A' \over \fb'} \right]^{UV}_{IR} \, \hat{\chi} (r_0)
\ee
Notice that this equation is gauge-invariant since, under a gauge transformation:
\be\label{scalar7}
\delta \hat{\chi}(r,x) = - \fb'(r) \left[\xi^5(r,x) - \xi^5(r_0,x)\right],
\ee
therefore $\hat{\chi}(r_0)$ on the right hand side of equation  (\ref{scalar6})  is invariant.

It is convenient to fix the remaining gauge freedom by imposing:
\be \label{chigauge}
\chi(r,x)=0.
\ee
To do this, one needs  different diffeomorphisms on the left and on the right of the brane, since $\fb'$ differs on both sides. The continuity  for $\hat{\chi}$ then  becomes the condition:
\be\label{scalar8}
\rho_{UV} (x) \fb_{UV}'(r_0) = \rho_{IR} (x) \fb_{IR}'(r_0)
\ee
i.e. the brane profile looks different from the left and from the right. This is not a problem, since equation (\ref{scalar8}) tells us how to connect the two sides given the background scalar field profile.

In the gauge (\ref{gaugefix1}-\ref{chigauge}) we have:
\be\label{scalar9}
\zeta = \psi = \hat{\psi} - A' \rho,  \qquad \hat{\chi}(r_0) = \fb' (r_0)\rho.
\ee
This makes it simple to solve for $\phi$ using the bulk constraint equation  (in particular, the $r\mu$-component of the perturbed Einstein equation (\ref{rmu}):
\be
\phi = {a\over a'} \psi' =  {a\over a'}\hat{\psi}' + \left({a'\over a} - {a''\over a}\right)\rho
\ee
where it is understood that this relation holds both on the UV and IR sides.

In the gauge $\chi = B =0$, we can write the second matching conditions (\ref{pert11-1}-\ref{pert12-1}), using equations (\ref{pert9}-\ref{pert10}) and (\ref{pert18}-\ref{pert19}):
\bea
&&\!\!\!\!\!\!\!\!\!\!\Bigg[-3 a' \left(2\hat{\psi} \,\eta_{\mu\nu} + 2\de_\mu\de_\nu E\right) 
+ {1\over 2} a_0 (\fb')^2 \rho\, \eta_{\mu\nu} + a_0\left(\de_\mu\de_\nu - \eta_{\mu\nu}\de^\sigma\de_\sigma\right) \left(E' -\rho\right) \Bigg]_{UV}^{IR}= \nonumber \\
= &&  {a^2(r_0)\over 2} W_B(\f_0) \left(2\eta_{\mu\nu} \hat{\psi} + 2\de_\mu\de_\nu E\right)_{r_0} \ + \ {a^2(r_0)\over 2}  {dW_B \over d\varphi}\Big|_{\f_0} \,  \eta_{\mu\nu}\, \fb'_0 \rho \nonumber \\
&-&
  2 U_B(\f_0) \left(\de_\mu\de_\nu - \eta_{\mu\nu}\de^\sigma\de_\sigma\right) \hat{\psi}  - {dU_B\over d\f}\Big|_{\f_0}\left(\de_\mu\de_\nu - \eta_{\mu\nu}\de^\rho\de_\rho\right)\left(\bar\f'_0 \rho\right)\,\,\,, \label{pertmatch1}\\
&& \nonumber \\
&&\Bigg[{\fb' \over a'} \hat{\psi}' +\left({(\fb')^2 \over 6 a'} - {\fb'' \over a \fb'} \right) \fb'\rho\Bigg]_{UV}^{IR}= \nonumber \\
&=&
\ -{d^2 W_B \over d\f^2}\Big|_{\f_0}\fb'\rho \  +\  {Z_B(\f_0) \over a^2} \fb' \de^\sigma \de_\sigma \rho \ -\  {6 \over a^2} {dU_B \over d\f}\Big|_{\f_0} \de^\sigma \de_\sigma\hat{\psi} \label{pertmatch2}
\eea
Using the background matching conditions  (\ref{match1}) and (\ref{match2}), as well as the definitions  (\ref{FE10}-\ref{FE11}) in conformal coordinates,
\be
{a'\over a^2}  = -{1\over 2(d-1)} W, \qquad \fb' = a {dW \over d\f},
\ee
one can see that the first two terms on each side of equation  (\ref{pertmatch1}) cancel each other, and we are left with an equation that fixes the matching condition for $E'(r,x)$:
\be\label{scalar10}
\Bigg[E'  - \rho\Bigg]_{UV}^{IR} = -2 {U_0 \over a_0} \, \hat{\psi}(r_0) - {1\over a_0}\left({dU_B\over d\f} \right)_0\bar\f'_0 \rho
\ee

Equation (\ref{pertmatch2}) fixes the discontinuity of $\hat{\psi}'$. It is convenient to write the equations for $\hat{\psi}$ and $\rho$  in the form:
 \bea
&& \Big[\hat{\psi}\Big]_{UV}^{IR} = 0 \,; \qquad \qquad\Big[\fb' \rho\Big]_{UV}^{IR} = 0 \, ; \label{scalar11-1}\\
&& \Bigg[{\fb'a \over a'} \hat{\psi}'\Bigg]_{UV}^{IR} = \Bigg[\left({Z_B(\f_0) \over a}\de^\mu\de_\mu - {\cal M}_b^2 \right) \fb'\rho - {6 \over a} {dU_B \over d\f}(\f_0) \de^\mu\de_\mu \hat{\psi}\Bigg]_{r_0} \label{scalar11-2}
\eea
where we have defined the brane mass:
\be\label{scalar12}
{\cal M}_b^2 \equiv a(r_0) {d^2 W_b \over d\f^2}\Big|_{\f_0} + \Bigg[\left({(\fb')^2 \over 6} {a\over a'} - {\fb'' \over  \fb'} \right) \Bigg]_{UV}^{IR}.
\ee
Using the background Einstein's equations (\ref{app3}) this can also be written as:
\be
{\cal M}_b^2 =  \left[{a'\over a} - {a''\over a'}\right]_{UV}^{IR} +  a \left({d^2 W_B \over d\f^2} - \left[ {d^2 W \over d\f^2 }\right]^{IR}_{UV} \right),
\ee
(to see this, use (\ref{app2}) and  take a radial derivative of eq. (\ref{app3}) to write $\fb'' = a d^2W/d\f + a'/a$).\\

We can eliminate $E$ from  equation (\ref{scalar10})  by acting with $\de^\mu\de_\mu$ on both sides and using Einstein's equations  (\ref{rr}) and (\ref{rmu}), with $\chi=0$:
\be
\Box E' = -{a\over a'} \left[\Box \psi + {a\over a'}\left(2{a'^2\over a^2} -  {a''\over a}\right) \psi' \right].
\ee
Notice that the combination multiplying $\psi'$ can be written as $(a/a')(\fb')^2/6$ using (\ref{app2}).

The bulk equation (\ref{pert4-2}) for $\zeta$ ($\equiv \psi$ in this gauge) on both sides of the brane is:
\be
\psi'' + \left(3{a'\over a} + 2{z'\over z}\right) \psi' + \de^\mu\de_\mu \psi = 0,
\ee
where $z = \fb'a/a'$. We can also write  it in terms of $\hat{\psi}$ using (\ref{scalar9}).

To summarize, we arrive at the following equations and matching conditions, either in terms of $\psi$:
\bea
&&
\psi'' + \left(3{a'\over a} + 2{z'\over z}\right) \psi' + \de^\mu\de_\mu \psi = 0, \label{b1}\\
&& \nonumber \\
&&\Big[\psi\Big]_{UV}^{IR} = -\Big[{a'\over a \fb'}\Big]_{UV}^{IR} \fb' \rho , \qquad \Big[ \fb' \rho \Big]_{UV}^{IR} = 0\, ; \label{m1}\\
&& \nonumber \\
&&
\Big[{a^2 \over a^{'2}} {\fb^{'2} \over 6} \psi'\Big]_{UV}^{IR} =  \left( {2U_0 \over a} - \Big[{a \over a'}\Big]_{UV}^{IR}\right) \Box \left(\psi + {a'\over a}  \rho\right) + {1\over a_0}\left({dU_B\over d\f} \right)_0\bar\f'\Box \rho; \label{m2}\\
&& \nonumber \\
&&
\Big[{a \fb'\over a'} \psi'\Big]_{UV}^{IR} =- {6\over a_0} \left({dU_B\over d\f} \right)_0 \Box\left(\psi + {a'\over a} \rho\right) + \left( {Z_B(\f_0) \over a}\Box - \tilde{{\cal M}_b}^2 \right) \fb'\rho\, ; \label{m3}\\
&& \nonumber\\
&& \Box \equiv \de^\mu \de_\mu, \quad z \equiv {a\fb' \over a'} , \quad  \tilde{{\cal M}_b}^2 = a \left({d^2 W_B \over d\f^2} - \left[ {d^2 W \over d\f^2 }\right]^{IR}_{UV} \right)\, . \label{def1}
\eea

Notice that these equations  have  6 free parameters: 4 in the bulk    (two integration constants for equation (\ref{b1}) in the UV, and  two in the IR) and two brane  parameters ($\rho$ on each side). From these 6 we can subtract one: a rescaling   of the solution, which is not a true parameter since the system is homogeneous in $(\rho,\psi)$. There is a total of 4 matching conditions, plus 2 normalizability conditions if the IR is confining, or only one if it is not.
 Therefore,in the confining case, we should find a quantization condition for the mass spectrum, whereas in the non-confining case the spectrum is continuous and the solution unique given the energy. The goal will be to show that such solutions exist only for positive values of $m^2$, defined as the eigenvalue of $\Box$.  To see this, one must go to the Schrodinger formulation.

To put the matching conditions (\ref{m1}-\ref{m3}) in a more  useful  form, it is convenient to eliminate $\rho_{L,R}$ altogether using  equations (\ref{m1}):
\be\label{scalar13}
\left[{a'\over a} \rho\right] = -[\psi] , \qquad [\fb'\rho] = 0
\ee
These can be solved to express the continuous quantities  $\hat{\psi}(r_0)$ and $\fb'\rho$ which appear on the r.h.s. of (\ref{m2}-\ref{m3}) in terms of $\psi_{UV,IR}$ only ($[x] \equiv x_{IR} - x_{UV}$ for any quantity $x$):
\be\label{scalar13.5}
\hat{\psi}(r_0) = {[z\, \psi] \over [z]} , \qquad \fb'(r_0) \rho = -{[\psi] \over [1/z]}.
\ee
Using the above identifications, equations (\ref{m2}-\ref{m3}) become relation between the left and right functions and their derivatives:
\bea
&& \left[z \psi'\right]= -{6\over a_0} {d U_B \over d\f}\Big|_{\f_0} \Box {[z \, \psi] \over [z]} - {1\over a_0}\left( Z_0 \Box  - a^2_0 {\tilde {\cal M}}^2 \right){[ \psi] \over [z^{-1}]} \label{sm1} \\
&&  \left[z^2 \psi'\right] =  6 \left( 2 {U_0\over a_0}- \left[a\over a'\right]  \right) \Box {[z \,\psi] \over [z]} \label{sm2} - {6\over a_0} {dU_B\over d\f}\Big|_{\f_0} \Box {[ \psi] \over [z^{-1}]}
\eea
Since the left hand side is in general non-degenerate, these equations can be solved to give $\psi'_L$ and $\psi'_R$ as linear combinations of $\psi_L$ and $\psi_R$, i.e. one  can put (\ref{sm1}-\ref{sm2}) in the general form:
\be \label{scalar14}
\left(\begin{array}{c} \psi_{UV}'(r_0)  \\ \psi_{IR}'(r_0) \end{array} \right) =  \left(\Gamma_1 + \Gamma_2 \de^\mu\de_\mu \right) \left(\begin{array}{c} \psi_{UV}(r_0)  \\ \psi_{IR}(r_0) \end{array}\right)
\ee
where the matrices   $\Gamma_1$ and $\Gamma_2$ are given by:
\bea \label{scalar15}
\Gamma_1 = &&{a_0 \tilde{{\cal M}}^2\over [z]^2} \left( \begin{array}{cc} -z_{IR}^2 & z_{IR}^2 \\ -z_{UV}^2 & z_{UV}^2\end{array}\right) , \nonumber \\
&& \\
 \Gamma_2 = && {1\over [z]^2 a_0} \left( \begin{array}{cc}
-12z_{IR}{d U_B \over d\f}\Big|_{\f_0}  + \tau_0+ Z_0 z_{IR}^2 &
  6z_{IR}\left({z_{IR}\over z_{UV}}+1\right){d U_B \over d\f}\Big|_{\f_0} - \tau_0{z_{IR} \over z_{UV}} -  Z_0 z_{IR}^2 \\
- 6z_{UV}\left({z_{UV}\over z_{IR}}+1\right){d U_B \over d\f}\Big|_{\f_0}  + \tau_0{z_{UV} \over z_{IR}}+ Z_0 z_{UV}^2 &
12 z_{UV}{d U_B \over d\f}\Big|_{\f_0} - \tau_0 - Z_0 z_{UV}^2
\end{array}\right)   \nonumber
\eea
where
\be \label{scalar15-a}
\tilde{{\cal  M}}^2 =  {d^2 W_B \over d\f^2}\Big|_{\f_0} - \left[{d^2 W \over d\f^2}\right] , \quad  \tau_ 0 = 12\left(3{W_B\over W_{IR}W_{UV}}\Big|_{\f_0} -  U_0\right).
\ee

\subsection{Gauge-invariant action for scalar modes}
Here we show that the action for the scalar perturbation, equation (\ref{ac1}),   can be written in a gauge-invariant form. To this end, we show that the action depends  solely  of the gauge-invariant  bulk variable $\zeta$ and gauge-invariant brane variables $\hat{\psi}(r_0), \hat{\chi}(r_0)$.

First, notice that equation (\ref{ac1}) was obtained in the gauge $\chi=0$ in the bulk. In this gauge, the scalar quantity $\psi$ coincides with the  gauge-invariant variable $\zeta$ (see equation (\ref{pert3a})).  Therefore, the bulk part of the action (first line in equation (\ref{ac1}))   can be written in a manifestly gauge-invariant fashion by replacing the 2-component object  $\Psi$ with $Z \equiv (\zeta_{IR}, \zeta_{UV})$.

Next, we consider the localized terms in the second line of equation (\ref{ac1}). Using the expressions for $\Gamma_1$ and $\Gamma_2$ in equation (\ref{scalar15})  and after some tedious algebra we obtain, for the first localized term:
\be\label{gi1}
 \Psi^\dagger (r_0) \,\Sigma \Gamma_1\, \Psi (r_0) =   a^4_0 \tilde{{\cal M}}^2  \left(\begin{array}{cc}  {[z \psi] \over [z]} & -{[\psi]\over [1/z]} \end{array}\right)    \left( \begin{array}{ccc} 0& \quad  & 0 \\ &\quad  & \\  0  & \quad  &  2\end{array}\right) \left(\begin{array}{c}  {[z \psi] \over [z]} \\ -{[\psi]\over [1/z]} \end{array}\right),
\ee
and for the second:
\be\label{gi2}
\de_\mu\Psi^\dagger(r_0)\,  \Sigma \Gamma_2 \, \de^\mu \Psi(r_0) =  a^2_0 \de_\mu \left(\begin{array}{cc}  {[z \psi] \over [z]} & -{[\psi]\over [1/z]} \end{array}\right)  \left( \begin{array}{cc} \tau_0& -6{dU_B\over d\f}\Big|_{\f_0} \\ -6{dU_B\over d\f}\Big|_{\f_0}  &  Z_0\end{array}\right)   \de^\mu \left(\begin{array}{c}  {[z \psi] \over [z]} \\ -{[\psi]\over [1/z]} \end{array}\right)
\ee
where $\tau_0$ was defined in equation (\ref{scalar15-a}).

From equation (\ref{scalar13}) we observe that the components of the  2-vectors entering the above matrix products coincide, in our gauge $\chi=0$, with the gauge-invariant combinations:
\be \label{gi3}
-{[\psi]\over [1/z]} = \hat{\chi}(r_0), \qquad  {[z \psi] \over [z]} = \hat{\psi} (r_0)
\ee
i.e. the gauge-invariant dilaton and metric trace on the brane.

\section{The bulk propagator for tensor modes} \label{app-bulk}
The bulk propagator $D(p,r)$ is  defined by equation (\ref{ind6}). It   must satisfy normalizability  conditions at the asymptotic $AdS$ boundary (UV) and in the deep interior (IR).
Here, normalizability is to be understood as square-integrability  with respect to the appropriate integration measure, i.e.
\be
\int e^{(d-1)A} |\Psi |^2 < \infty.
\ee

The bulk propagator  $D(p,r)$ can then be written in terms of normalizable UV and IR eigenfunctions of the radial operator $\de_r e^{(d-1)A}\de_r$, with ``energy'' determined by $p^2$:
\be\label{ind9}
D(p,r) = \left\{\begin{array}{ll} \Psi_{UV}^{(p)}(r)  & \quad  r<r_0 \\ & \\ \Psi_{IR}^{(p)}(r)  & \quad r>r_0 \end{array} \right.
\ee
where  $\Psi_{UV}$ and $\Psi_{IR}$ satisfy the equations:
\bea
&& \left[\de_r e^{(d-1)A_{UV}(r)} \de_r   - e^{(d-1)A_{UV}(r)} p^2 \right] \Psi_{UV}^{(p)} = 0  \label{ind9-i}  \\
&& \nonumber \\
&&  \left[\de_r e^{(d-1)A_{IR}(r)} \de_r   - e^{(d-1)A_{IR}(r)} p^2 \right] \Psi_{IR}^{(p)} = 0  \label{ind9-iii}
\eea
and the matching conditions:
\bea
&& \Psi_{IR}^{(p)}(r_0) = \Psi_{UV}^{(p)}(r_0)   \label{ind9-iv}\\
&& \nonumber \\
&&\left[\de_r \Psi^{(p)}_{IR} - \de_r \Psi^{(p)}_{UV}\right]_{r_0} = -1 \label{ind9-ii}
\eea
The matching conditions (\ref{ind9-iv}-\ref{ind9-ii}) follow by integrating  equation (\ref{ind6}) on a small interval across the interface.

The mode functions in equation (\ref{ind9})  are normalizable in the UV and IR,  respectively. The solution therefore  has four integration constants  and four  conditions (two normalizability conditions plus two matching conditions) that fix the wave-functions uniquely (notice that the system is not homogeneous, and does not have a rescaling freedom).

\subsection{Large-$p$ behavior} \label{app-largep}
 At large Euclidean $p^2$,  we can approximate the bulk equations as in flat space, neglecting the  derivatives of $A(r)$,
\be \label{ind10}
\de_r^2 \Psi^{(p)}(r)= p^2 \Psi^{(p)}(r).
\ee
For small $r$,  the $AdS$ boundary  acts as an infinite barrier and imposes a vanishing wave-function at $r=0$ (this is equivalent to  normalizability in the UV). In  the interior, assuming the IR is reached as  $r\to +\infty$\footnote{This is the case for example when the IR geometry asymptotes to an $AdS$ interior. A full classification of the possible IR geometries in a general Einstein-dilaton theory, can be found in Section 4 of \cite{multirg}.}, the solutions for positive $p^2$ are real exponentials,  and for normalizability we require the solution to be vanishing as $r\to +\infty$.

The solution satisfying appropriate boundary conditions  (vanishing in the IR and for $r\to 0$) and matching condition at $r_0$ is:
\be\label{ind11}
\Psi_{IR}^{(p)}=  {\sinh p r_0 \over p}e^{-p r},  \quad \Psi_{UV}^{(p)} ={e^{-p r_0}\over p }\sinh p r,    \qquad p \equiv \sqrt{p^2}
\ee
For large $p r_0$, we observe that:
\be \label{ind12}
D(p,r_0) \simeq {1\over 2p},  \quad \quad  p r_0 \gg 1
\ee
like in flat space.

\subsection{Perturbation expansion for small-$p$ } \label{app-smallp}

For small $p$,  the bulk propagator   has the form of an expansion in $p^2$:
\be \label{ind14}
D(r_0, p) = d_0 + p^2 d_2 + p^4 d_4 +  \ldots
\ee
where the coefficients $d_i$ can be computed perturbatively in $p^2$ solving equation (\ref{ind6}) iteratively. We concentrate on the case $d=4$.
\begin{itemize}
\item $O(p^0)$ \\
Setting $p=0$ in equations (\ref{ind9-i}) and (\ref{ind9-iii}), we can integrate them immediately and find:
\be\label{sp8}
\Psi_{UV}^{(0)} = C_{1,UV}^{(0)} \int^r_0 e^{-3A_{UV}(r')} dr' +  C_{2,UV}^{(0)}, \qquad \Psi_{IR}^{(0)} =  C_{1,IR}^{(0)}\int^r_0 e^{-3A_{UV}(r')} dr' + C_{2,IR}^{(0)}
\ee
Normalizability implies $C_{2,UV}^{(0)} =  C_{1,IR}^{(0)}= 0$. The matching conditions  (\ref{ind9-iv}-\ref{ind9-ii}) determine the values:
\be\label{sp9}
 C_{1,UV}^{(0)} = e^{3A_0} , \qquad C_{2,IR}^{(0)}  =  e^{3A_0}\int^{r_0}_0 e^{-3A_{UV}(r')} dr'
\ee
Therefore, to lowest order in small $p$:
\be\label{sp9a}
\Psi_{UV}^{(0)}(r) = e^{3A_0} \int^r_0 e^{-3A_{UV}(r')} dr' , \quad \Psi_{IR}^{(0)}(r) = e^{3A_0} \int^{r_0}_0 e^{-3A_{UV}(r')} dr',
\ee
and we find:
\be\label{sp10}
D(0,r_0) = d_0 =  e^{3A_0}\int^{r_0}_0 e^{-3A_{UV}(r')} dr'
\ee
\item $O(p^2)$ \\
To the next order, we write:
\be \label{sp11}
\Psi_{UV}^{(p)} \simeq  \Psi_{UV}^{(0)} + p^2  \Psi_{UV}^{(2)}, \quad \Psi_{IR}^{(p)} \simeq  \Psi_{IR}^{(0)} + p^2  \Psi_{IR}^{(2)}
\ee
The corrections  to the wave-functions at order  $p^2$  satisfy the equations:
\bea
&& \de_r\left( e^{3A_{UV}(r)} \de_r  \Psi_{UV}^{(2)}\right) =  e^{3A_{UV}(r)} \Psi_{UV}^{(0)} \qquad r< r_0  \label{sp12}  \\
&& \nonumber \\
&& \de_r \left(e^{3A_{IR}(r)} \de_r   \Psi_{IR}^{(2)}\right) = e^{3A_{IR}(r)} \Psi_{IR}^{(0)} \qquad r> r_0 .  \label{sp13}
\eea
The  matching conditions for $\Psi^{(2)}$ are:
\be\label{sp14}
\Psi_{IR}^{(2)}(r_0) = \Psi_{UV}^{(2)}(r_0) , \qquad \left(\de_r\Psi_{IR}^{(2)}\right)(r_0) = \left(\de_r \Psi_{UV}^{(2)}\right)(r_0)
\ee
as follows from equations (\ref{ind9-iv}-\ref{ind9-ii}) and from the matching conditions at order $p^0$.
Integrating twice equations (\ref{sp12}-\ref{sp13}), the general solution  with normalizable  homogeneous parts   are:
\bea
\Psi_{UV}^{(2)}(r) = && \int_0^r dr' e^{-3A_{UV}(r')} \int_0^{r'} dr'' \Psi^{(0)}_{UV}(r'')   e^{3A_{UV}(r'')} + \nonumber \\
&&+ \, C_{UV}^{(2)}\int_0^r dr' e^{-3A_{UV}(r')}  \label{sp15}\\
\Psi_{IR}^{(2)}(r) = &&\int_{r_0}^r dr' e^{-3A_{IR}(r')} \int_{r_0}^{r'} dr'' \Psi^{(0)}_{IR}(r'')   e^{3A_{IR}(r'')}+ \nonumber \\
&& + \, C_{IR}^{(2)}\int_0^r dr' e^{-3A_{UV}(r')} \label{sp16}
\eea
Imposing the continuity conditions (\ref{sp14}) at $r=r_0$ we find:
\bea \label{sp17}
&& C_{UV}^{(2)} = - \int_{0}^{r_0} dr' \Psi^{(0)}_{UV}(r') e^{3A_{UV}(r')},  \label{sp17}   \\ &&  C_{IR}^{(2)} = \int_0^{r_0} dr' e^{-3A_{UV}(r')}\left[C_{UV}^{(2)} + \int_{0}^{r'} dr'' \Psi^{(0)}_{UV}(r'') e^{3A_{UV}(r'')}\right]. \label{sp17-2}
\eea
Inserting this result into equation (\ref{sp15}) we find:
\be \label{sp18a}
\Psi_{UV}^{(2)}(r) = - \int_0^r dr' e^{-3A_{UV}(r')} \int_{r'}^{r_0} dr'' \Psi^{(0)}_{UV}(r'')   e^{3A_{UV}(r'')}.
\ee
Recall that $d_2 = \Psi_{UV}^{(2)}(r_0)$: evaluating equation (\ref{sp18a}) at $r=r_0$ and using equation (\ref{sp9a}), we obtain:
\be\label{sp18}
d_2  = - e^{3A_0} \int_0^{r_0} dr'  e^{-3A_{UV}(r')}\int_{r'}^{r_0} dr'' e^{3A_{UV}(r'')}    \int_0^{r''} dr'''e^{-3A_{UV}(r''')}.
\ee
\item  $O(p^{2n})$
One can continue the above procedure iteratively: the wave-functions at order $2n$ satisfy the equations
\bea
&& \de_r\left( e^{3A_{UV}(r)} \de_r  \Psi_{UV}^{(2n)}\right) =  e^{3A_{UV}(r)} \Psi_{UV}^{(2n-2)} \qquad r< r_0  \label{sp19}  \\
&& \nonumber \\
&& \de_r \left(e^{3A_{IR}(r)} \de_r   \Psi_{IR}^{(2n)}\right) = e^{3A_{IR}(r)} \Psi_{IR}^{(2n-2)} \qquad r> r_0 .  \label{sp20}
\eea
and must be continuous, with continuous derivative, at $r_0$. This system of equation is identical to the one we have solved at order $p^2$, and the solution  is  as follows:
\be \label{sp21}
\Psi_{UV}^{(2n)}(r) = - \int_0^r dr' e^{-3A_{UV}(r')} \int_{r'}^{r_0} dr'' \Psi^{(2n-2)}_{UV}(r'')   e^{3A_{UV}(r'')}.
\ee
The coefficient $d_{2n}$ is obtained by evaluating the above expression at $r_0$ and consists of $2n+1$ alternating integrals:
\bea\label{sp22}
d_{2n} = && (-)^n e^{3A_0} \int_0^{r_0} dr_1  e^{-3A_{UV}(r_1)}\int_{r_1}^{r_0} dr_2 e^{3A_{UV}(r_2)}  \int_0^{r_2} dr_3  e^{-3A_{UV}(r_3)} \ldots \nonumber \\
&& \ldots \int_{r_{2n-1}}^{r_0} dr_{2n}  e^{3A_{UV}(r_{2n})} \int_0^{r_{2n}} dr_{2n+1}  e^{-3A_{UV}(r_{2n+1})} .
\eea
\end{itemize}
We will now extract the explicit dependence on $A_0$ of the  expansion coefficients $d_{2n}$. This can be achieved by writing $A_{UV}$ as a function of $\f$ as in equation  (\ref{scales6}),
\be
A_{UV}(\f) = A_0 + {\cal A}_{UV}(\f_0,\f) , \quad {\cal A}_{UV}(\f_0,\f) \equiv  -{1\over 2(d-1)} \int_{\f_0}^{\f}  {W_{UV} \over dW_{UV}/d\f}
\ee
 and by changing variables to $\f$  in all the integrals   (\ref{ind15}-\ref{ind15-3}), using the identity (valid for $0<\f<\f_0$ and $0< r< r_0$):
\be
{d\f \over dr} = e^{A_{UV}(r)} {dW_{UV} \over d \f}.
\ee
The result takes the form:
\be
d_i = e^{-A_0} {\cal D}_i (\f_0)
\ee
where the coefficients $ {\cal D}_i (\f_0)$ are independent of $A_0$ but depend only on the superpotentials and the equilibrium position $\f_0$:\\

 \be\label{ind15a}
{\cal D}_0 (\f_0)=  \int_0^{\f_0} d\f'  {e^{-3 {\cal A}_{UV}(\f_0,\f')} \over W'_{UV}(\f')};
\ee
\\

\be\label{ind15b}
{\cal D}_2 (\f_0) =  -  \int_0^{\f_0} d\f' {e^{-3 {\cal A}_{UV}(\f_0,\f')} \over W'_{UV}(\f')} \int_{\f'}^{\f_0} d\f''{ e^{3{\cal A}_{UV}(\f_0,\f'')} \over W'_{UV}(\f'')} \int_0^{\f'''} d\f''' {e^{-3 {\cal A}_{UV}(\f_0,\f''')} \over W'_{UV}(\f''')}
\ee
and similarly for ${\cal D}_4(\f_0)$.

Therefore, the expansion coefficients of the bulk propagator at low momenta all scale as $e^{-A_0}$ times a function that depends only on the bulk  potentials.

Notice that, for fixed $\f_0$,  the exponential $e^{-{\cal A}_{UV}}$ appearing in the integrals, is bounded between zero and one, and:
\be\label{ind18}
e^{-{\cal A}_{UV}(\f_0,\f)} \to \left\{\begin{array}{ll} 0 & \quad\f \to 0 \\ 1 & \quad \f\to \f_0. \end{array}\right. ,
\ee
As a consequence, the scale controlling ${\cal D}_i$ is approximately the bulk curvature  ${\cal R}$ at the interface, encoded in the superpotential factors in the denominators:
\be \label{ind19}
{\cal R}_0 \approx W_{UV}(\f_0),
\ee
and we have,   roughly:
\be\label{ind20-app}
{\cal D}_{2n}(\f_0) \approx {1\over {\cal R}_0^{2n+1}},
\ee

\subsection{Regularity of the small-$p$ expansion} \label{secreg}
Here we discuss if and at which order the expansion in $p^2$ used in equation (\ref{ind14}) may break down.

First, consider the case when the bulk theory has a confining IR. In this case the spectrum of normalizable  eigenmodes is discrete, and the bulk Green's function $D(p,r)$ can be expanded in terms of the  eigenfunctions of the bulk radial operator:
\be\label{reg1}
D(p,r)  = \sum_n  {f_n \over p^2 + m_n^2}
\ee
where $f_n$ are some constants and  $m_n$ are the  ``eigenvalues''  for the radial kinetic operator, that is:
\be \label{reg2}
\de_r e^{3A} \de_r h_n(r) +  e^{3A(r)}m_n^2 h_n(r) = 0
\ee
In this case, it is clear from equation (\ref{reg1}) that the small momentum expansion is regular.

Things are more subtle if the bulk theory has  no gap, but rather it has a continuous spectrum starting at $m=0$. This is the case either if  the theory reaches a  conformal fixed point in the IR, or if $\f$ reaches infinity but the superpotential grows slower than $\exp \gamma \f$ with  $\gamma^2 < 1/6$ \cite{superp}. In both cases, the IR is reached as the conformal coordinate $r\to +\infty$, where the scale factor behaves as:
\be \label{reg3}
e^{A_{IR}(r)} \sim {1\over r^{z}} , \quad r\to +\infty, \qquad  z\geq 1
\ee
The constant $z$ is related to the steepness of the bulk potential, with $z=1$ corresponding to $AdS$ asymptotics in the interior (thus to the case of a  conformal IR fixed point). For more details, the reader is referred to \cite{superp}.

The small-$p$ behavior is expected to be governed by the far IR of the theory, i.e. by the behavior of the geometry as  $r\to \infty$. In  this region, the bulk wave equation equation  simplifies to:
\be\label{reg4}
h''(r) - {3z \over r} h'(r) - p^2 h(r) = 0
\ee
This approximation  is valid in the asymptotic region where the metric can be approximated by (\ref{reg3}), and is independent of the value of $p$.
The solution of equation (\ref{reg4}) which is  normalizable at infinity is:
\be\label{reg5}
\Psi_{IR}(r) = c_{IR}(p) r^{1+3z\over 2} K_{1+3z\over 2}(pr)
\ee
where $K$ is the modified Bessel function which is exponentially vanishing  at infinity, and $c_{IR}(p)$ is for the moment unknown.

For fixed large $r$, but for $p \ll 1/r$,  we can also expand equation (\ref{reg5}) for small argument:
\be\label{reg6}
\Psi_{IR}(r) \simeq c_{IR}(p)\left\{ p^{-{1+3z\over 2}} \left[1 + \alpha_1 r^2 p^2 + O(r^4p^4)\right] + \beta  p^{3z+1\over2} r^{1+3z} \left[ 1 +  \alpha_2 r^2 p^2 + O(r^4p^4)\right]\right\},
\ee
where  $\alpha_1, \alpha_2$ and $\beta$ are some {\em fixed} constants arising from the expansion of the Bessel function.

We can compare equation (\ref{reg6})   with the small-$p$ expansion of the IR wave-function $\Psi_{IR}$ given in equation  (\ref{sp11}).  To lowest order in this expansion  the normalizable IR wave-function is a constant (see equation (\ref{sp9}) :
\be \label{reg6a}
\Psi_{IR}^{(0)} = C^{(0)}.
\ee
This is consistent with the fixed $r$,  $p\to 0$ limit of equation (\ref{reg6}) if the momentum dependence in $c_{IR}(p)$ is fixed to be:
\be \label{reg7}
c_{IR}(p) =  C^{(0)} p^{3z+1\over 2}.
\ee
Inserting this expression back in equation (\ref{reg6}) we find, at small $p$ and large $r$:
\be\label{reg8}
\Psi_{IR}(r)  \simeq C_0 \left\{ \left(1 + \alpha_1 r^2 p^2 + O(r^4p^4)\right) + \beta (pr)^{1+3z} \left(1+ \alpha_2 p^2 r^2 + O(r^4p^4) \right) \right\}
\ee
The only source of non-analyticity in the above expression is the $p^{1+3z}$ prefactor. Thus, we have a regular expansion in $p$ at least  up to the order $1+3z \geq 4$.  The larger is $z$ (and the faster the scale factor vanishes), the further the non-analytic terms arise in the expansion. The coefficients $d_{2n}$ are well-defined and finite as long as $2n < 1+3z$.  The earliest the expansion can fail is at $2n=4$  for $z=1$, with the  appearance of  terms  $~p^4\log p$ which are familiar for massless fields in asymptotically $AdS$ space-times.

Notice that we may evade the above argument, and have a singular limit  of $D(p,r_0)$ as $p\to 0$, only if we somehow lose the constant solution (\ref{reg6a}). This is the case, for example, in the Randall-Sundrum type matching: if we impose $Z_2$ symmetry at the brane, then the matching condition (\ref{ind9-ii}) to lowest order in $p$ becomes $(\de_r \Psi_{IR}^{(0)})(r_0) = -1/2$ which is {\em not} obeyed by the constant solution. This signals a singularity of the propagator as $p\to 0$, which indeed turns out to be the massless pole associated to the graviton zero mode in this theory. However, in our case we can only find the  solution (\ref{sp9})   to the matching conditions, thus the expansion makes sense up to order $1+3z$.

\end{appendix}


\end{document}